\documentclass[article,onecolumn,superscriptaddress,nofootinbib,floatfix,onecolumn]{revtex4}

\usepackage{amsmath, latexsym, amssymb, hyperref, graphicx, color, multirow, makecell, diagbox}
\usepackage{tabularx}
\usepackage[T1]{fontenc}

\usepackage{dcolumn}
\usepackage{bm}
\usepackage{xspace}
\usepackage{comment}

\usepackage{listings}
\lstdefinestyle{mystyle}{mathescape,basicstyle=\small\ttfamily,frame=leftline,aboveskip=4mm,belowskip=4mm,xleftmargin=20pt,framexleftmargin=10pt,numbers=none,framerule=2pt,abovecaptionskip=0.0mm,belowcaptionskip=3.5mm}
\usepackage{float}
\usepackage{epsfig}
\usepackage{longtable}
\usepackage{rotating}
\usepackage{ulem}
\usepackage{amssymb}
\usepackage{amsmath}
\usepackage{txfonts}
\usepackage{upgreek}
\usepackage{mathtools}
\usepackage{tabularx}
\usepackage{booktabs}
\usepackage{appendix}

\usepackage{soul} % Striking text
\usepackage[capitalize]{cleveref}
\definecolor{nicered}{rgb}{.7,.1,.1}
\definecolor{nicegreen}{rgb}{.1,.5,.1}
\definecolor{darkblue}{rgb}{0,0,.5}
\hypersetup{colorlinks, citecolor=nicegreen,linkcolor=nicered, urlcolor=darkblue}
\usepackage{multirow}
\usepackage{slashed}
\usepackage{verbatim}

\def\ttt#1{\texttt{\small #1}}

\providecommand{\ttbar}{t\overline{t}}

\newcommand{\pp}{p-p}
\newcommand{\ppbar}{p-$\overline{\mathrm{p}}$}
\providecommand{\pA}{p-A}
\providecommand{\pPb}{p-Pb}
\providecommand{\AaAa}{A-A}
\providecommand{\PbPb}{Pb-Pb}

\newcommand{\ABgagaX}{A \,B\,$\xrightarrow{\gaga}$ A\, $\ellell$ \,B}
\newcommand{\sigmagagaX}{\sigma_{\gaga\to \ellell}}

\newcommand{\epem}{\mathrm{e}^+\mathrm{e}^-}
\providecommand{\lele}{\ell\ell}
\newcommand{\ellell}{\ell^+\ell^-}
\newcommand{\mm}{\mu\mu}
\newcommand{\mumu}{\mu^+\mu^-}
\newcommand{\tautau}{\tau^+\tau^-}
\newcommand{\tata}{\tau\tau}
\newcommand{\gaga}{\gamma\gamma}

\newcommand{\sqrts}{\sqrt{s}}

\newcommand{\snn}{s_{_\text{NN}}}
\newcommand{\sqrtsnn}{\sqrt{s_{_\text{NN}}}}

\newcommand{\pT}{p_\text{T}}
\newcommand{\pTmumu}{p_\text{T}^{\mm}}
\newcommand{\kt}{k_{\perp}}

\newcommand{\helaconia}{\textsc{HELAC-Onia}}
\newcommand{\madgraph}{\textsc{MadGraph5\_aMC@NLO}}
\newcommand{\mgshort}{\textsc{MG5\_aMC}}
\newcommand{\gammaUPC}{\ttt{gamma-UPC}}
\newcommand{\starlight}{\textsc{Starlight}}
\newcommand{\superchic}{\textsc{Superchic}}

\newcommand{\lhe}{\textsc{lhe}}

\usepackage{xspace}

\newcommand*{\ie}{i.e.,\@\xspace} 
\newcommand*{\cm}{c.m.\@\xspace}

\begin{document}

\title{Dimuon and ditau production in photon-photon collisions at next-to-leading order in QED}
%\title{NLO QED corrections to dilepton production in ultraperipheral proton and nuclear collisions}
%\title{Exclusive dimuon and ditau production at the LHC: the need of NLO QED corrections}
%\title{Dimuon and ditau production in high-energy photon fusion at NLO QED accuracy}
%\title{Next-to-leading-order QED corrections to dimuon and ditau production in photon-photon collisions}

\author{Hua-Sheng~Shao}\email{huasheng.shao@lpthe.jussieu.fr}
\affiliation{Laboratoire de Physique Th\'eorique et Hautes Energies (LPTHE), UMR 7589,\\ Sorbonne Universit\'e et CNRS, 4 place Jussieu, 75252 Paris Cedex 05, France}
\author{David~d'Enterria}\email{david.d'enterria@cern.ch}
\affiliation{CERN, EP Department, CH-1211 Geneva, Switzerland}

\begin{abstract}
\noindent 
Next-to-leading-order (NLO) quantum electrodynamics (QED) corrections to the production of muon and tau pairs in photon-photon collisions, $\gaga\to\mumu,\tautau$, are calculated in the equivalent photon approximation. We mostly consider $\gaga$ processes in ultraperipheral collisions of hadrons at the LHC, but the $\gaga\to\tautau$ process in $\epem$ collisions at LEP is also discussed. The NLO terms are found to modify the total fiducial cross sections by up to 5\%, increasing the tails of the dilepton acoplanarity and transverse momentum distributions, and depleting by up to 15\% the yields at high masses, with respect to the leading-order predictions including the very small virtuality of the colliding photons. At the LHC, the calculations obtained with the charge form factor for protons and lead ions including the NLO QED corrections improve the data--theory agreement for all measured differential distributions, and prove an indispensable ingredient for the extraction of precision quantities in photon-photon processes, such as the anomalous magnetic moment of the tau lepton. 
\end{abstract}

\maketitle

%\vspace{-0.6cm}

\section{Introduction}

The study of photon-photon ($\gaga$) interactions at multi-GeV energies was first realized in the laboratory in $\epem$ collisions at DESY PETRA in the 1980s~\cite{Morgan:1994ip} and at CERN LEP in the 1990s~\cite{Whalley:2001mk,Przybycien:2008zz}, and has received a significant experimental and theoretical boost in collisions with hadron beams in the last twenty years thanks to the large $\gamma$ energies and luminosities accessible at the BNL Relativistic Heavy-Ion Collider (RHIC)~\cite{Bertulani:2005ru} and CERN Large Hadron Collider (LHC)~\cite{Baltz:2007kq,deFavereaudeJeneret:2009db}. In the equivalent photon approximation (EPA)~\cite{vonWeizsacker:1934nji,Williams:1934ad}, the electric field created by a charge accelerated at high energies can be interpreted as a flux of quasireal photons whose energies $E_{\gamma}$ and number densities $N_{\gamma}$ grow proportionally to the Lorentz relativistic factor ($E_{\gamma}\propto \gamma_L$) and squared charge ($N_{\gamma}\propto Z^2$) of the beam particles, respectively~\cite{Brodsky:1971ud,Budnev:1975poe}. At the LHC, one can exploit the availability of multi-TeV beams (leading to boosts of $\gamma_L = E_\text{beam}/m_\mathrm{p,N}\approx 7500,\,3000$ in collisions with protons and ions\footnote{For proton and nucleon masses: $m_\mathrm{p,N} = 0.9315, 0.9383$~GeV, respectively. Natural units, $\hbar=c=1$, are used throughout the paper.} at nucleon-nucleon center-of-mass energies (\cm) of $\sqrtsnn = 14,\,5.5$~TeV, respectively) and large electric charges (up to $Z=82$ for Pb ions) to produce a very large number of different processes via photon-photon fusion~\cite{Shao:2022cly}.\\

Most of the studies of $\gaga$ processes have been carried out in proton-proton (\pp), proton-nucleus (\pA), and nucleus-nucleus (\AaAa) ``ultraperipheral'' collisions (UPCs) at impact parameters larger than twice their transverse radii, \ie\ without overlap of their transverse matter profiles, and with the protons and/or ions coherently emitting quasireal photons. Since the hadrons remain intact after their electromagnetic interaction, the UPCs feature extremely clean topologies characterized by just an exclusive final state produced in an otherwise empty detector. The requirement of coherent emission from the hadron charge distribution imposes very low maximum virtualities of the emitted photon: $Q_\text{max}^2 \approx 1/R^{2}$, where $R$ is the charge radius, \ie\ $Q^2\lesssim~0.08$~GeV$^2$ for protons ($R_\mathrm{p}\approx 0.7$~fm), and $Q^2\lesssim 8\cdot10^{-4}$~GeV$^2$ for Pb nuclei ($R_\mathrm{Pb}\approx 7$~fm).
%, with the outgoing protons (or zero-degree neutrons in the case of ions) detectable with near-beam detectors far forward/backward from the interaction point.
More recently, $\gaga$ collisions have also been observed in nonexclusive final states where the hadrons overlap and produce many additional particles in parton-parton interactions concomitant with the two-photon interaction~\cite{STAR:2018ldd,ATLAS:2018pfw,ALICE:2022hvk}, as well as in high-luminosity \pp\ runs with large number of overlapping collisions (pileup) in the same bunch crossing~\cite{CMS:2024skm}. To identify $\gaga\to\mathrm{X}$ collisions in both exclusive and nonexclusive events, one exploits the fact that the colliding photons have very small virtualities and, therefore, that the final state is basically produced at rest in the transverse plane, \ie\ $p_\mathrm{T}^{X}\approx 0$, and thus its decay products are back-to-back in azimuthal angle $\phi$, %($\Delta\phi=\phi_1-\phi_2\approx\pi$), 
\ie\ have a very small acoplanarity $A_\phi = 1-|\Delta\phi|/\pi \approx 0$. In the case of \pp\ collisions, one can in addition detect one or both surviving proton(s) in dedicated very forward detectors inside the LHC beamline~\cite{FP420RD:2008jqg,CMS:2014sdw,Tasevsky:2015xya,CMS:2022hly}, provided that the $\gaga$ collision produces a system with large enough masses $m_{\gaga}\gtrsim 100$~GeV~\cite{Bruce:2018yzs,CMS:2018uvs,ATLAS:2020mve}.\\

%At particle colliders, each bunch crossing yields photon-photon collisions that can produce pairs of charged fermions or bosons, or a central neutral system X with overall charge-conjugation symmetry $C = + 1$. Since the colliding quasireal photons have very small transverse momenta, on average $\kt\approx \gamma_L/R\approx 30$ and 200~MeV for proton and lead beams, respectively, the central system is produced at rest in the transverse plane, and its decay products are back-to-back in azimuth angle $\phi$, \ie\ acoplanarity $A_\phi = \Delta_\phi = ||-\pi \approx 0$. High-energy $\gaga$ collisions were first studied at $\epem$ colliders~\cite{X}, but the availability of hadronic beams with multi-TeV energies, very large luminosities, and/or electric charges as large as $Z=82$ (for lead ions) at the Large Hadron Collider (LHC)~\cite{Baltz:2007kq,dEnterria:2008puz,deFavereaudeJeneret:2009db}, as well as previously at RHIC~\cite{}, has allowed to study $\gaga$ collisions up to invariant masses hitherto experimentally unexplored. 
%Although high-energy photon-photon processes have been studied in $\epem$ and $e$-p collisions since more than thirty years ago~\cite{Vermaseren:1982cz,Schuler:1997ex,Uehara:1996bgt}, as well as in the last twenty years with heavy ions at the Relativistic Heavy Ion Collider (RHIC)~\cite{Bertulani:2005ru}, 

\begin{figure}[htpb!]
\centering
\includegraphics[width=0.99\textwidth]{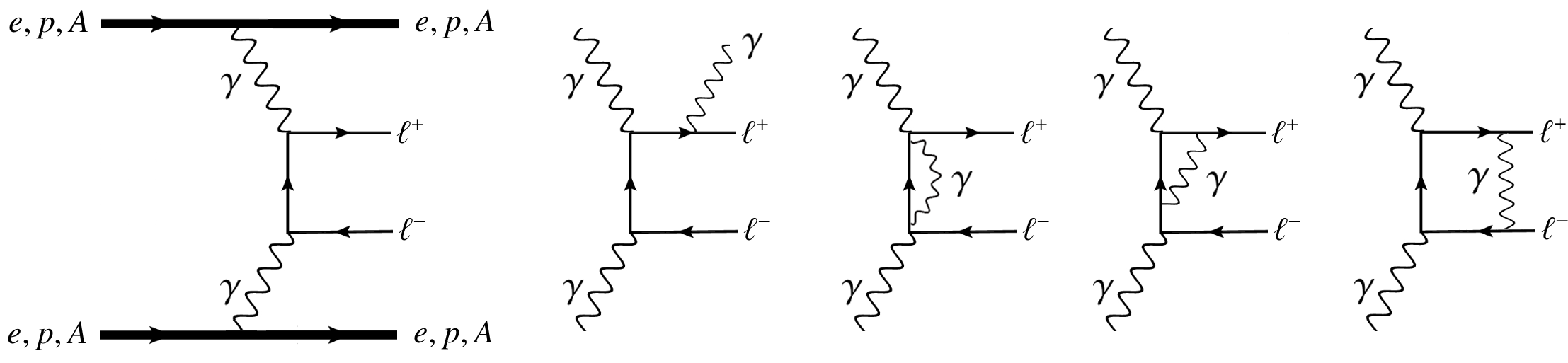}
\caption{Selection of diagrams for the photon-fusion production of a pair of charged leptons, $\gaga\to\ellell$ with $\ell^\pm = e^\pm,\mu^\pm,\tau^\pm$, in collisions of electrons, protons, and/or ions: Born (leftmost), and real (second) and virtual (rest of the diagrams)  NLO corrections.
\label{fig:nlo_diags}}
\end{figure}

Arguably, the simplest photon-photon collision process in the standard model (SM) is the $t$-channel production of a pair of charged leptons, $\gaga\to\ellell$ with $\ell^\pm = e^\pm,\mu^\pm,\tau^\pm$, whose elementary cross section is proportional to the square of the electromagnetic coupling, $\alpha = 1/137.036$ (Fig.~\ref{fig:nlo_diags}, left). Among those, the Breit--Wheeler (B--W) process $\gaga\to\epem$ by which pure light is transformed into matter, was the first one studied in quantum electrodynamics (QED)~\cite{Breit:1934zz}. %The B--W cross section in two-photon collisions of charges $Z_1$ and $Z_2$ rises with the beam energy following a $\ln\left(\gamma_\mathrm{L,1}\gamma_\mathrm{L,2}\right)$ dependence~\cite{LandauLifschit}, %the nucleon-nucleon (NN) center-of-mass (\cm) energy, $\ln^3(\sqrtsnn)$. 
The simplicity and relatively large cross section of the B--W process have facilitated its measurement in $\epem$ collisions~\cite{JADE:1986axr, L3:1997acq} as well as in UPCs at fixed-target and collider energies (by the WA93~\cite{Vane:1992ms}, CERES/NA45~\cite{CERESNA45:1994cpb}, STAR~\cite{STAR:2004bzo,STAR:2019wlg}, PHENIX~\cite{PHENIX:2009xtn}, CDF~\cite{CDF:2006apx,CDF:2009xwv}, ALICE~\cite{ALICE:2013wjo}, CMS~\cite{CMS:2012cve,CMS:2018erd,CMS:2018uvs,CMS:2024tfd}, and ATLAS~\cite{ATLAS:2015wnx,ATLAS:2017fur,ATLAS:2020mve,ATLAS:2022srr} experiments), 
%and has become a clean final state to test the theoretical ingredients of UPC cross section calculations.
%The two-photon production of dielectron, dimuon, and/or ditau pairs in $\gaga$ collisions has been observed since many years in $\epem$~\cite{X}, e-p~\cite{X}, and hadronic ultraperipheral~\cite{X} collisions, 
and found in overall agreement with the theoretical EPA predictions at leading-order (LO) accuracy in QED. However, incorporation of theoretical contributions from final-state $\gamma$ radiation (FSR) (second diagram of Fig.~\ref{fig:nlo_diags})~\cite{Klein:2020jom,CMS:2024tfd} ---as well as properly accounting for the polarization of the incoming photons in order to reproduce particular dielectron azimuthal distributions~\cite{Brandenburg:2022tna}--- are needed to reach accord with the latest more precise data. A good control of the B--W process where one or both $\rm e^\pm$ radiate a hard bremsstrahlung photon, is also of relevance as a background in studies of light-by-light (LbL) scattering $\gaga\to\gaga$~\cite{dEnterria:2013zqi}, as confirmed by the experimental measurements~\cite{ATLAS:2017fur,CMS:2018erd,ATLAS:2019azn,CMS:2024tfd}.\\

Exclusive dimuon production, $\gaga\to\mumu$, is even better controlled than the B--W process for experimental and theoretical reasons. Experimentally, high-energy muons can be often identified and reconstructed more accurately than electrons. Theoretically, muons are 200 times more massive than electrons, and corrections due to collinear and/or soft photon emissions, including lepton ``dressing'' to account for photon FSR, are under a better control. For these two reasons, the $\gaga\to\mumu$ process has been often considered as a useful ``standard candle'' %for $\gaga$ fusion processes in order 
to validate the experimental reconstruction, including its use as a ``luminometer''~\cite{Khoze:2000db,Shamov:2002yi}, and the theoretical ingredients ($\gamma$ fluxes, hadron survival probabilities, %implemented in EPA calculations, and 
radiative corrections, etc.)~\cite{Vermaseren:1982cz,Berends:1986if} in two-photon physics. A large number of experimental measurements of $\gaga\to\mumu$ exist in $\epem$~\cite{PLUTO:1984gmq, TOPAZ:1992bfv, OPAL:1993law, L3:1997acq,L3:2004ful}, e-p~\cite{H1:2003nlr}, \ppbar~\cite{CDF:2009xey,CDF:2009xwv}, \pp~\cite{CMS:2011vma,ATLAS:2015wnx,ATLAS:2017sfe,CMS:2018uvs,ATLAS:2020mve}, \pPb~\cite{ALICE:2023mfc} and \PbPb~\cite{CMS:2020skx,ATLAS:2020epq} collisions, with the latest, more precise, \PbPb\ UPC results showing increasing deviations in some differential distributions with respect to the predictions of LO QED Monte Carlo (MC) event generators~\cite{Klein:2016yzr,Harland-Lang:2020veo}.\\

The third dilepton process of interest is $\gaga\to\tautau$, which directly probes the photon-tau vertex and is thus sensitive to the anomalous electromagnetic couplings of the $\tau$ lepton. Given the short $\tau$ lifetime, which precludes measuring its properties at storage rings, the photon-fusion channel has been long proposed to probe the tau lepton anomalous magnetic, $a_\tau\equiv (g-2)_\tau/2$, and electric dipole, $d_\tau$, moments at $\epem$~\cite{Cornet:1995pw} and hadron~\cite{delAguila:1991rm,Atag:2010ja,Beresford:2019gww,Dyndal:2020yen,Shao:2023bga} colliders. Two-photon production of tau pairs has been measured in $\epem$~\cite{OPAL:1993law,L3:1997acq,DELPHI:2003nah} and in UPCs at the LHC~\cite{CMS:2022arf,ATLAS:2022ryk,CMS:2024skm}. The currently best experimental measurement of $ a_\tau = 0.0009_{-0.0031}^{+0.0032}$
%limit of $-0.0042 < a_\tau < 0.0062$ at 95\% confidence level (CL) 
has been achieved via $\gaga\to\tautau$ in p-p UPCs at 13~TeV~\cite{CMS:2024skm}, superseding the previous $-0.052 < a_\tau < 0.013$ limit from LEP~\cite{DELPHI:2003nah}. This latest LHC measurement has a precision and accuracy approaching those needed to probe the first ``Schwinger term'' quantum correction to $(g-2)_\tau$, which is common to all leptons and amounts to $a_\tau^\text{NLO} = \alpha/(2\pi) \approx 0.00116$~\cite{Schwinger:1948iu}. Future measurements will be sensitive to higher-order SM terms, which result in $a_\tau = 0.00117721(5)$~\cite{Eidelman:2007sb}, and to any potential beyond SM (BSM) modification. An improved measurement of $a_\tau$ has become an experimental priority given the long-standing data-theory discrepancies observed for its muon counterpart $a_\mu$~\cite{Aoyama:2020ynm}, and the fact that, by standard lepton mass-scaling arguments, the $\tau$ particle should be more sensitive than the muon to potential BSM physics by a factor of $m_\tau^2/m_\mu^2 \approx 280$.\\

The availability of larger $\gaga\to\ellell$ data samples measured at the LHC, with experimental cross section uncertainties reduced to the few percent level, calls for an improved theoretical description to match the experimental accuracy and precision. In this context, full next-to-leading-order (NLO) corrections accounting for virtual and real QED emissions, such as those shown in four rightmost diagrams of Fig.~\ref{fig:nlo_diags}, are now a requirement to extract precise physics results from photon-photon collisions. This is particularly true when aiming at comparing data with theory to extract precision SM parameters, such as $a_{\tau}$, and/or to search for deviations in the absolute or differential cross section from the SM predictions due to elusive new physics contributions. To our knowledge, the first theoretical calculations that included NLO (QCD and/or QED) corrections for photon-photon processes in UPCs were obtained for the $\gaga\to\ttbar$ and $\gaga\to\gaga$ processes employing the \gammaUPC\ code in Refs.~\cite{Shao:2022cly} and~\cite{AH:2023kor,AH:2023ewe}, respectively. Both studies showed that the NLO corrections augment the predicted cross sections (by about 20\% for $\ttbar$ and 5\% for LbL scattering) and reduce their theoretical QCD scale uncertainties, emphasizing the need to include them for the accurate and precise calculation of $\gaga$ processes. The purpose of this work is to extend the number of physics processes available with improved theoretical accuracy and precision, by calculating the NLO QED corrections for the photon-fusion production of pairs of muon and tau leptons.\\

The paper is organized as follows. Section~\ref{sec:th} describes the theoretical setup used to compute the LO and NLO predictions for two-photon production of dimuon and ditau final states. Results for exclusive $\gaga\to\mumu,\tautau$ processes in UPCs at hadron colliders ---for the ditau case, a short discussion of the photon-fusion production in $\epem$ collisions at LEP is also provided--- are presented in Sections~\ref{sec:dimuon} and~\ref{sec:ditau}, respectively, including total cross sections as well as differential distributions. The paper is closed with a summary in Section~\ref{sec:summ}. Appendix~\ref{sec:kTsmear} provides technical details on the implementation of $k_\perp$ smearing, needed to account for the small initial-state photon virtualities in UPCs, and its approximate matching to NLO distributions for lepton pairs.

%%%%%%%%%%%%%%%%%%%%%%%%%%%%%%%%%%%%%%%%%%%%%%%%%%%%%%%%%%%%%%%%%%%%%%%%%%%%%%%%%%%%
\section{Theoretical setup}
\label{sec:th}

The matrix elements calculations reported in this work have been performed within the \madgraph\ framework (\mgshort\ for short)~\cite{Alwall:2014hca} extended with the automated computations of NLO electroweak corrections discussed in Ref.~\cite{Frederix:2018nkq}. For the UPC results with protons and ions, the initial photon fluxes as well as the hadron survival probabilities for exclusive final states are obtained with the \gammaUPC\ MC code~\cite{Shao:2022cly}. %interfaced with \mgshort. 
The \gammaUPC\ program provides two types of elastic photon fluxes as a function of impact parameter $b$, $N_{\gamma/Z}(E_\gamma,b)$, based on the electric-dipole (EDFF) and charge (ChFF) form factors for proton and nuclei, respectively, as well as associated survival probabilities of the photon-emitting hadrons. At the LHC, the $\gaga$ cross sections at a given $\gaga$ \cm\ energy computed with the ChFF flux are larger than those obtained with the EDFF flux by about 10--20\% (for p-p and p-Pb UPCs) and 20--40\% (for Pb-Pb UPCs), with the difference rising for increasingly heavier final states~\cite{Shao:2022cly}. In general, the \gammaUPC\ cross sections computed with the ChFF flux match those derived with the \superchic\ MC~\cite{Harland-Lang:2020veo}, whereas those computed using the EDFF photon fluxes agree with the ones obtained with the \starlight\ code~\cite{Klein:2016yzr}. The EDFF photon number density is divergent for small impact parameters, $b\to 0$, and an arbitrary infrared (IR) cutoff needs to be imposed in the calculations. Usually, the cutoff is chosen as $b>R_\mathrm{A}$, where $R_\mathrm{A}$ is the radius parameter of the hadron transverse profile. Such a limitation prevents from using the EDFF fluxes in nonexclusive $\gaga$ collisions with hadron overlap. Since the ChFF photon flux is well-behaved for all hadron-hadron impact parameters, the results obtained with it are considered more realistic, are found to agree better with the most precise B--W results to date~\cite{CMS:2024tfd}, and will be our default choice hereafter. Results obtained with the EDFF flux will be shown below in some cases for comparison purposes given that such a flux has been often used in the past.\\

The ChFF flux contains an explicit dependence on the photon transverse momentum $\kt$ through its conjugate dependence on the impact parameter $b$, which is related to the photon virtuality via $Q^2 =  \kt^2 + E_{\gamma}^2/\gamma_\mathrm{L}^2$. 
%Since the impact parameter $b$ is conjugate to the $\kt$ variable, we integrate over $b$ in order to get the $\kt$ dependence. 
The unintegrated photon number density at a given photon energy $E_\gamma$ and transverse momentum $\kt$ (see its $b$-dependent counterpart in Eq.~(13) of Ref.~\cite{Shao:2022cly}) reads
\begin{eqnarray}
%\begin{equation}
%\resizebox{.95\hsize}{!}{
N_{\gamma/Z}^\mathrm{\tiny ChFF}(E_\gamma,\kt) &=&
2\pi \kt \frac{Z^2\alpha}{\pi^2}\left|\int_0^{\infty}\frac{\mathrm{d}b \kt^2}{\kt^2+E_\gamma^2/\gamma_\mathrm{L}^2}F_{\mathrm{ch},A}\left(\sqrt{\smash[b]\kt^2 + E_{\gamma}^2/\gamma_\mathrm{L}^2}\right)J_1(b\kt)\;\right|^2 \nonumber\\ 
&=& \frac{2Z^2\alpha}{\pi}\frac{\kt^3}{\left(\kt^2 + E_{\gamma}^2/\gamma_\mathrm{L}^2\right)^2}\left[F_{\mathrm{ch},A}\left(\sqrt{\kt^2 + E_{\gamma}^2/\gamma_\mathrm{L}^2}\right)\right]^2.
\label{eq:photonnumdenkT}
%}
%\end{equation}
\end{eqnarray}
where $F_{\mathrm{ch},A}(Q^2)$ is the form factor of the ion A emitting the photon (the full analytical expressions for protons and ions can be found in Ref.~\cite{Shao:2022cly}), and $J_1$ is the Bessel function of the first kind. However, the matrix element calculations in \mgshort\ expect collinear (\ie\ $\kt$-independent) input photon distributions, and therefore the photon fluxes need to be fully integrated over $Q$ and $\kt$ (cf.\ Eq.~(13) in Ref.~\cite{Shao:2022cly}). Thus, when computing a given photon-photon process at LO accuracy with the \gammaUPC\ plus \mgshort\ (or \helaconia\footnote{Quarkonium final states in $\gaga$ fusion can also be simulated at LO accuracy with \gammaUPC\ and the \helaconia\ event generator~\cite{Shao:2012iz,Shao:2015vga}.}) codes  out of the box, the central system  will be produced exactly at rest in the transverse plane. For the $\gaga\to \ellell$ case, this means $\pT^{\lele}=0$ with both leptons emitted back-to-back in azimuth, \ie\ $A_\phi = 0$. Since the photon density follows a $1/\kt^2$ dependence and its $\kt$ values are many orders-of-magnitude smaller than their longitudinal energy $E_\gamma$, the EPA assumption that both photons are real ($Q\approx 0$) is well fulfilled, and has no actual numerical impact on the computed fully integrated cross sections. However, %the assumption that the colliding photons have zero $\kt$, \ie\ that the central system is produced exactly at rest, has no real experimental implication either because the detector resolution smears out the energies of the decay products of the central system leading to $\pT^{X}$ values that, though nonzero, are still well below the experimental analysis $\pT^{X}\approx 1$~GeV (the usual upper limit imposed in the experimental analyses to remove nonexclusive backgrounds). Nonetheless, as discussed in the introduction, 
in reality the coherently emitted photons can have very small but nonzero virtualities up to about $Q_\mathrm{max}^{2} \approx 0.08$~GeV$^2$ for protons and $Q^2_\text{max}\approx 8\cdot10^{-4}$~GeV$^2$ for Pb nuclei, and as a result the lepton pairs feature a steeply falling spectrum reaching up to $\pT^{\lele}\approx 0.4$ and 0.2~GeV for p-p, and Pb-Pb UPCs, respectively, with very small (but nonzero) acoplanarity tails. In general, the experimental selection criteria applied to select (semi)exclusive photon-photon events require $\pT^{\lele}$ and $A_\phi$ very close to zero, though not exactly null because of detector smearing effects. It is thus more realistic to generate MC events that also account for the small virtuality-induced variations of the $\pT^{\lele}$ and $A_\phi$ distributions. To do so, we explicitly include in our \gammaUPC\ setup  the initial $\gamma$ virtualities through a small extra $\kt$ implemented by directly modifying the 4-momenta of the incoming photons and outgoing particles in the Les~Houches (\lhe) file output of the generated MC events. Such a ``$\kt$ smearing'' of the initial and final states is technically implemented as explained in detail in App.~\ref{sec:kTsmear}. For each independent \lhe\ event, we sample the virtualities of the transverse momenta of the two initial photons according to Eq.~\eqref{eq:photonnumdenkT}, and the final four-momenta of the produced particles are reshuffled to guarantee the onshell and momentum conservation conditions event-by-event, as done similarly in Ref.~\cite{Frixione:2019fxg}. 

We note that, in general, such a $\kt$-smearing procedure can only be applied directly onto the \lhe\ dilepton events output when computed at LO accuracy. This is so because at NLO, the different Born, virtual, real, and infrared-divergence-subtraction counterterms are strongly correlated, and one cannot just smear the initial photon $\kt$ of each contribution independently. However, when a particular photon-photon final state does not have IR-divergence cancellations between the NLO real and virtual contributions, such as in the LbL $\gaga\to\gaga$ case~\cite{AH:2023kor,AH:2023ewe}, the NLO \lhe\ files can indeed be smeared out following the method outlined above. A procedure to combine, in an ad-hoc manner, the impact of the initial photon $k_\perp$ smearing and the NLO corrections in the generated $\pT^{\lele}$ and $A_\phi$ distributions is also presented in App.~\ref{sec:kTsmear}. This simplistic $k_\perp\,+\,$NLO combination procedure, with uncertainties of $\mathcal{O}(25\%)$ in the overlapping region, %is certainty ad-hoc and simplistic and 
is however a temporary solution to be improved with a proper matching scheme between the NLO calculation and the QED parton shower in the future, but such developments are beyond the scope of this work.

For the determination of NLO QED corrections, we proceed as follows. Since the initial photons are almost onshell, we work in the so-called $\alpha(0)$ renormalization scheme, found suitable for final states with similar topologies~\cite{Pagani:2021iwa}. Using the alternative $G_\mu$ renormalization scheme with $\alpha_{G_\mu}=1/132.183$ would slightly increase the size of the NLO contributions from emitted real and exchanged virtual photons by a factor of $\alpha_{G_\mu}/\alpha(0) -1 = 3.7\%$ (the incoming photons are always basically onshell, and the tree-level calculations should always use $\alpha(0)$). Since the NLO contributions are themselves a small correction (${\approx}5\%$ at most) to the total yields, the uncertainty due to the choice of the scheme-dependent QED coupling propagates into negligible (few permille) uncertainties into the total theoretical cross sections.

In order to avoid the complications of ``dressed'' massless leptons (affected by collinear $\gamma$ emission)~\cite{Frederix:2016ost}, the masses of the charged leptons are kept nonzero (since the electron mass is very small, this prevents us from considering the NLO corrections to the B--W process, $\gaga\to\epem$, in this work). Because of the large hierarchies between the hard scale given by the invariant mass of the lepton pairs and the low scale of the lepton masses, the code has been optimized to avoid numerical instabilities. The relevant ultraviolet and rational $R_2$ counterterms of the NLO QED amplitudes calculation have been coded in the Universal Feynman Output %(\ufo) 
format~\cite{Degrande:2011ua,Darme:2023jdn} generated with an ad-hoc {\sc\small Mathematica} program. In the $\alpha(0)$ scheme, thanks to the Ward identity, only external lepton masses are relevant, while all other charged fermion masses that enter into the renormalization constant of $\alpha$ cancel out exactly with the corresponding terms in the wavefunction renormalization constants of the external photons. The numerical results have been obtained with $\alpha = 1/137.036$, and  $m_\mu= 0.10566$~GeV and $m_\tau = 1.77686$~GeV masses~\cite{ParticleDataGroup:2022pth}.\\

The exclusive dilepton cross sections in an UPC of hadrons A and B (\ABgagaX) are computed with the latest version (v1.6) of the \gammaUPC\ code (some of whose improvements are described more in detail in~\cite{Crepet:2024dvv,gammaUPCv2}), using the EPA-based factorized product of the elementary cross section at a given $\gaga$ \cm\ energy, $\sigmagagaX(W_{\gaga})$, times the two-photon differential distribution of the colliding beams, integrated over the photon energies, 
\begin{equation}
\sigma(\mathrm{A}\; \mathrm{B}\,\xrightarrow{\gaga} \mathrm{A} \; X \; \mathrm{B})=
\int \frac{\mathrm{d}E_{\gamma_{1}}}{E_{\gamma_{1}}} \frac{\mathrm{d}E_{\gamma_{2}}}{E_{\gamma_{2}}} \, \frac{\mathrm{d}^2N^{(\mathrm{AB})}_{\gamma_1/\mathrm{Z}_1,\gamma_2/\mathrm{Z}_2}}{\mathrm{d}E_{\gamma_{1}}\mathrm{d}E_{\gamma_{2}}} \sigmagagaX(W_{\gaga})\,,
\label{eq:two-photon}
\end{equation}
where 
\begin{equation}
\frac{\mathrm{d}^2N^{(\mathrm{AB})}_{\gamma_1/\mathrm{Z}_1,\gamma_2/\mathrm{Z}_2}}{\mathrm{d}E_{\gamma_1}\mathrm{d}E_{\gamma_2}} =  \int{\mathrm{d}^2\pmb{b}_1\mathrm{d}^2\pmb{b}_2\,P_\text{no\,inel}(\pmb{b}_1,\pmb{b}_2)\,N_{\gamma_1/\mathrm{Z}_1}(E_{\gamma_1},\pmb{b}_1)N_{\gamma_2/\mathrm{Z}_2}(E_{\gamma_2},\pmb{b}_2)},
\label{eq:2photonintegral}
\end{equation}
is derived from the convolution of the two photon number densities $N_{\gamma_i/\mathrm{Z}_i}(E_{\gamma_i},\pmb{b}_i)$ with energies $E_{\gamma_{1,2}}$ at impact parameters $\pmb{b}_{1,2}$ from hadrons A and B, respectively (the vectors $\pmb{b}_{1}$ and $\pmb{b}_{2}$ have their origins at the center of each hadron, and, therefore, $|\,\pmb{b}_{1}-\pmb{b}_{2}|$ is the impact parameter between them); and $P_\text{no\,inel}(\pmb{b}_1,\pmb{b}_2)$ encodes the probability of hadrons A and B to remain intact after their interaction, which depends on their relative impact parameter. 
The probability $P_\text{no\,inel}(b)$ to have no inelastic hadronic interaction at impact parameter $b$ is obtained from the standard opacity (optical density), or eikonal Glauber, expressions:
\begin{eqnarray}
P_\text{no\,inel}\left(b\right)&=&\left\{\begin{array}{ll} 
e^{-\,\sigma^{\mathrm{NN}}_{\text{inel}}\cdot T_{\mathrm{AB}}(b)}, & \text{for nucleus-nucleus UPCs,}\\
%e^{-\,\sigma^{\mathrm{NN}}_{\text{inel}}\cdot T_\mathrm{A}(b)}, & \text{for proton-nucleus UPCs}\\
\left|1-\Gamma(s_{_\text{NN}},b)\right|^2,\; \mbox{ with }\;\Upgamma(s_{_\mathrm{NN}},b)\propto e^{-b^2/(2b_0)} & \text{for \pp\ UPCs.}\\
\end{array}\right.
\label{eq:Psurv}
\end{eqnarray}
Here, $T_{\mathrm{AB}}(b)$ is the nuclear overlap function derived from the transverse density profile parametrized as a Woods--Saxon distribution with radius $R_A$ and diffusivity $a_A$, $\sigma^{\mathrm{NN}}_{\text{inel}} \equiv \sigma^{\mathrm{NN}}_{\text{inel}}(\!\sqrtsnn)$ is the inelastic NN scattering cross section parametrized as a function of $\sqrtsnn$, and $\Upgamma(s_{_\mathrm{NN}},b)$ is the Fourier transform of the \pp\ elastic scattering amplitude modelled by an exponential function~\cite{Frankfurt:2006jp} with inverse slope $b_0 \equiv b_0(\!\sqrtsnn)$ dependent on the NN \cm\ energy. The $\sqrtsnn$ dependencies of both quantities, $\sigma^{\mathrm{NN}}_{\text{inel}}$ and $b_0$, are those of the fits provided in Ref.~\cite{dEnterria:2020dwq}.

\begin{figure}[htp!]
\centering
\includegraphics[width=0.49\textwidth]{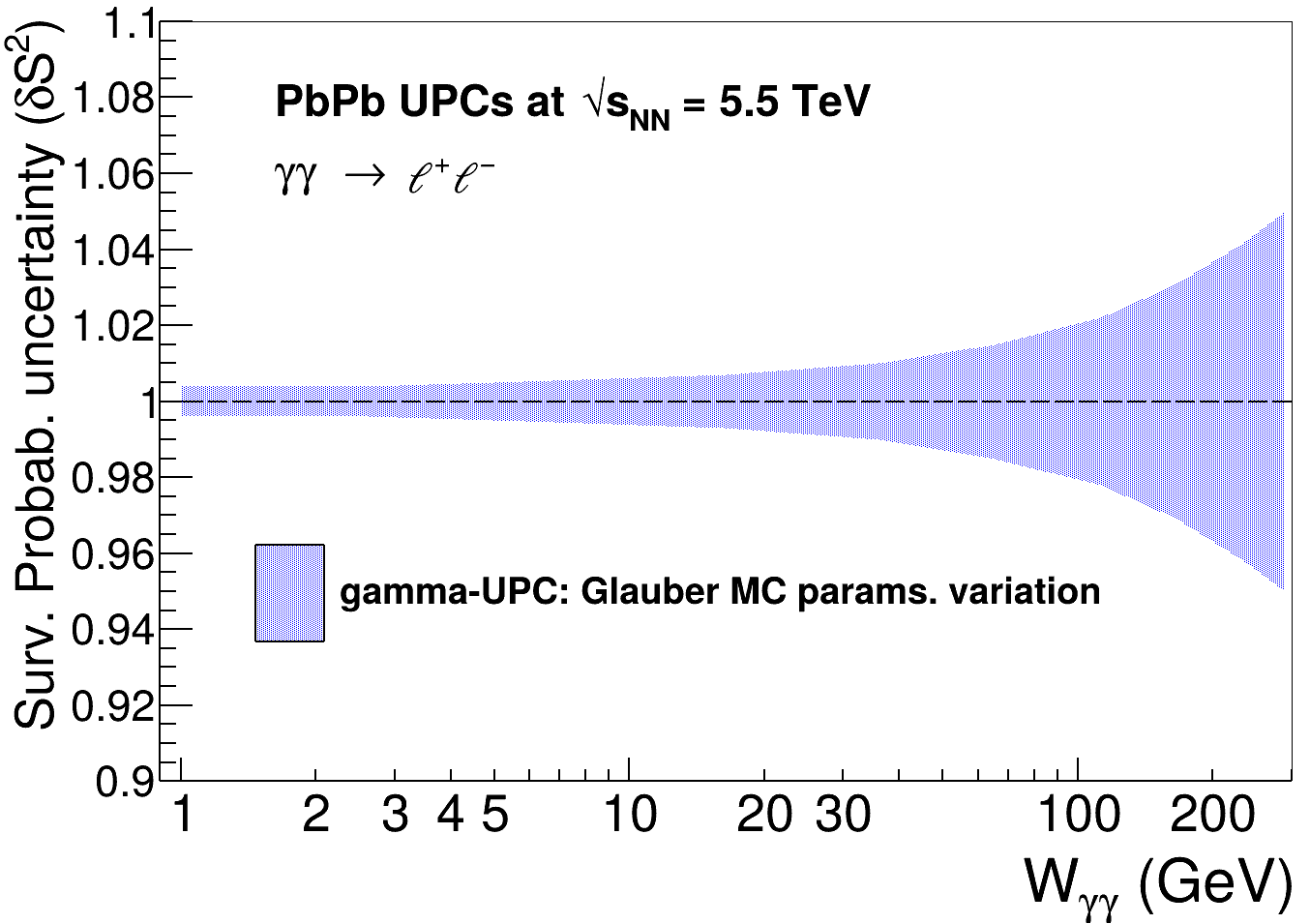}
\includegraphics[width=0.49\textwidth]{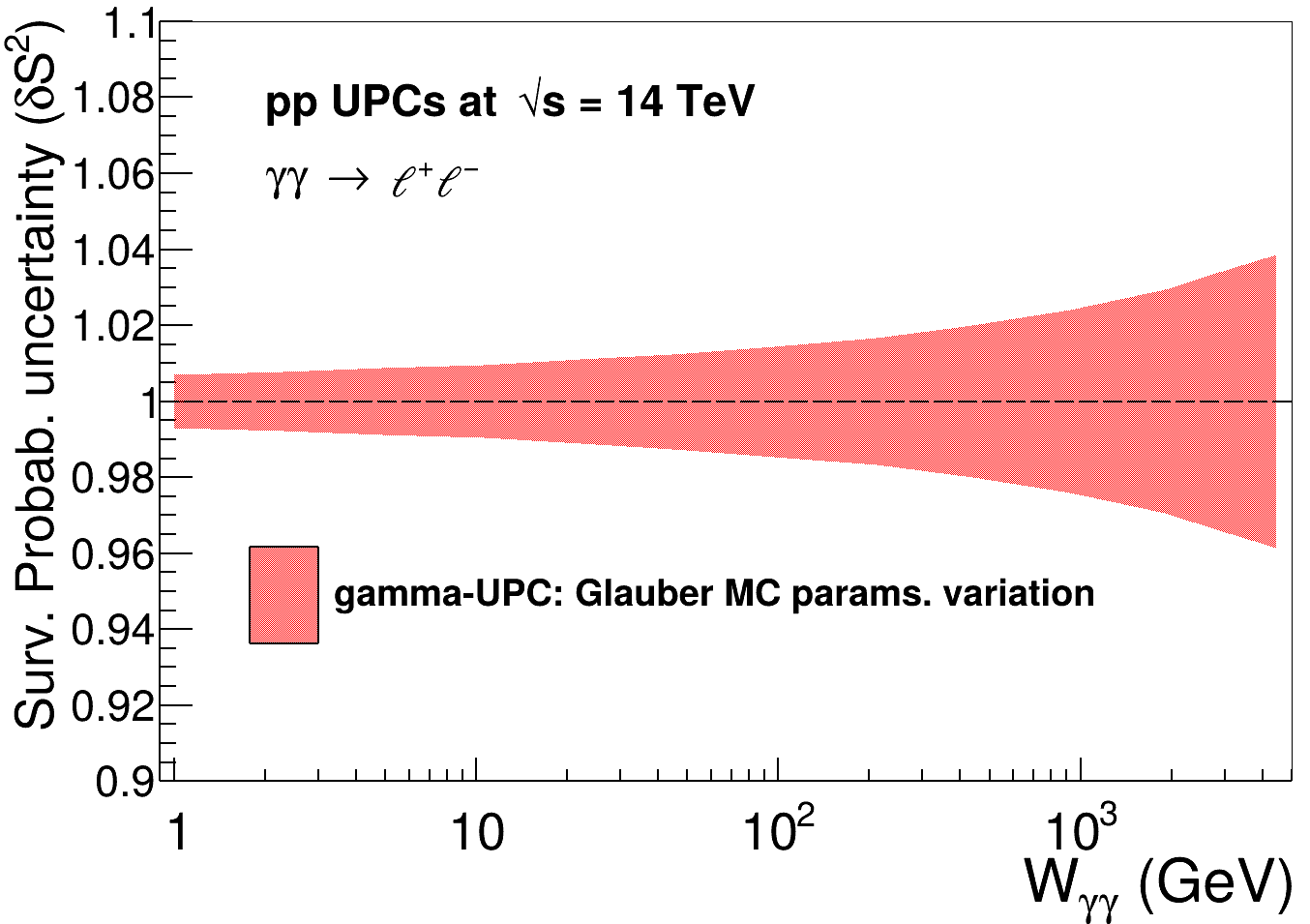}
\caption{Parametric uncertainties in the calculation of exclusive $\gaga\to\ellell$ production cross sections as a function of the $\gaga$ \cm\ energy, due to the hadronic survival probabilities in PbPb (left) and p-p (right) UPCs at the LHC.
\label{fig:gaga_surv_uncert}}
\end{figure}

Compared to the v1.0 version of \gammaUPC~\cite{Shao:2022cly}, our latest code includes several improvements, such as updated photon flux of the proton to account for corrections beyond the dipole elastic form factor, as well as recalculated ion survival probabilities based on fits of the $T_{\mathrm{AB}}(b)$ factors obtained with a MC Glauber code~\cite{Loizides:2017ack}, rather than based on analytical ``optical Glauber'' expressions. The leading parametric uncertainty in the $\gaga\to\ellell$ UPC cross section arises from the modeling of the $P_\text{no\,inel}$ probability or, equivalently, the survival probability of the hadrons,
\begin{equation}
S^2_{\gaga} = \frac{\int{\mathrm{d}^2\pmb{b}_1\mathrm{d}^2\pmb{b}_2\,P_\text{no\,inel}(\pmb{b}_1,\pmb{b}_2)\,N_{\gamma_1/\mathrm{Z}_1}(E_{\gamma_1},\pmb{b}_1)N_{\gamma_2/\mathrm{Z}_2}(E_{\gamma_2},\pmb{b}_2)}
}
{\int{\mathrm{d}^2\pmb{b}_1\mathrm{d}^2\pmb{b}_2\,N_{\gamma_1/\mathrm{Z}_1}(E_{\gamma_1},\pmb{b}_1)N_{\gamma_2/\mathrm{Z}_2}(E_{\gamma_2},\pmb{b}_2)}
},
\label{eq:S2}
\end{equation}
where the numerator is the two-photon density requiring the hadron survival in the photon-photon interaction, Eq.~(\ref{eq:2photonintegral}), and the denominator represents the integral of the two photon fluxes over all impact parameters without any constraint on the hadronic overlap. The parametric uncertainties of the modeling of the proton and ion survival probabilities have been estimated by varying the values of $R_A$, $a_A$, $b_0$, and $\sigma^{\mathrm{NN}}_{\text{inel}}$ in Eq.~(\ref{eq:Psurv}) within their individual uncertainties, and propagating them to the final $\gaga\to\ellell$ cross section. Figure~\ref{fig:gaga_surv_uncert} shows the corresponding uncertainties as a function of photon-photon c.m.\ energy. 
%two-photon luminosities in PbPb and p-p UPCs as a function of $W_{\gaga}$, determined from the following expression
%\begin{equation}
%\frac{\mathrm{d} \mathcal{L}_{\gaga}^{(\mathrm{AB})}}{\mathrm{d}W_{\gaga}}=\frac{2 W_{\gaga}}{s_{\mathrm{NN}}} \int \frac{\mathrm{d} E_{\gamma_1}}{E_{\gamma_1}} \frac{\mathrm{d} E_{\gamma_2}}{E_{\gamma_2}} \delta\left(\frac{W_{\gaga}}{s_{\mathrm{NN}}}-\frac{4 E_{\gamma_1} E_{\gamma_2}}{s_{\mathrm{NN}}}\right) \frac{\mathrm{d}^2 N_{\gamma_1 / \mathrm{Z}_1, \gamma_2 / \mathrm{Z}_2}^{(\mathrm{AB})}}{\mathrm{d} E_{\gamma_1} \mathrm{~d} E_{\gamma_2}},
\label{eq:gagalumi}
%\end{equation}
The parametric uncertainties for PbPb and p-p UPCs amount to about 0.5--2\% for the range of dilepton masses measured in UPCs at the LHC, $W_{\gaga} = m_{\lele}\approx 5$--100~GeV (although they can reach up to $\pm5\%$ at the very high-mass tails of the dilepton distributions: $m_{\lele}\approx 0.2,\,3$~TeV for PbPb and p-p, respectively), and are smaller than the LO versus NLO differences discussed here. 

%%%%%%%%%%%%%%%%%%%%%%%%%%%%%%%%%%%%%%%%%%%%%%%%%%%%%%%%%%%%%%%%%%%%%%%%%%%%%%%%%%%%
\section{Two-photon dimuon production at NLO in QED}
\label{sec:dimuon}

This section studies the case of exclusive dimuon production, $\gaga\to\mumu$, in UPCs with protons and Pb ion beams. Such a final state is considered as a clean standard-candle process that can be used to calibrate our theoretical understanding of the photon fluxes and hadron survival probabilities. At the LHC, the process has been precisely measured in \pp~\cite{CMS:2011vma,ATLAS:2015wnx,ATLAS:2017sfe,CMS:2018uvs,ATLAS:2020mve} and \PbPb~\cite{ATLAS:2020epq,CMS:2020skx} collisions. In Ref.~\cite{Shao:2022cly}, we compared our \gammaUPC\ LO results with the ATLAS \PbPb\ UPC measurement at $\sqrtsnn=5.02$~TeV~\cite{ATLAS:2020epq}, finding that the ChFF predictions normalized to the total inclusive cross section reproduced better the shape of the kinematic distributions compared to the EDFF case. There exist also detailed comparisons of the LO calculations from the \superchic~\cite{Harland-Lang:2020veo} and \starlight~\cite{Klein:2016yzr} programs with these ATLAS data in Ref.~\cite{Harland-Lang:2021ysd}. Here, we confront our LO and NLO QED predictions with the four existing measurements of $\gaga\to\mumu$ cross sections in \PbPb\ and \pp\ UPCs at the LHC performed within the fiducial phase spaces listed in Table~\ref{tab:dimuoncuts}.
%Table~\ref{tab:dimuoncuts} lists several $\gaga\to\mumu$ measurements at the LHC, along with their fiducial regions. 

\begin{table}[htbp!]
\tabcolsep=3.mm
\caption{Exclusive dimuon measurements at the LHC (compared with the LO and NLO predictions computed in this work) with their kinematic phase space indicated.}\label{tab:dimuoncuts}
\vspace{0.2cm}
\centering
\begin{tabular}{lc}\hline
System, experiment & $\gaga\to\mumu$ fiducial phase space\\\hline
\pp\ at 7 TeV, CMS~\cite{CMS:2011vma} & $\pT^{\,\mu}>4$ GeV, $|\eta^{\,\mu}|<2.1$, $m_{\mm}>11.5$ GeV\\ 
\pp\ at 7 TeV, ATLAS~\cite{ATLAS:2015wnx} & $\pT^{\,\mu}>10$ GeV, $|\eta^{\,\mu}|<2.4$, $m_{\mm}>20$ GeV\\ %\hline%\hline
\multirow{2}{*}{\pp\ at 13 TeV, ATLAS~\cite{ATLAS:2017sfe} } & $\pT^{\,\mu}>6$ GeV, $|\eta^{\,\mu}|<2.4$, $m_{\mm}= 12\mbox{--}30$~GeV\\
 & $\pT^{\,\mu}>10$ GeV, $|\eta^{\,\mu}|<2.4$, $m_{\mm}= 30\mbox{--}70$~GeV\\ %\hline
\PbPb\ at 5.02 TeV, ATLAS~\cite{ATLAS:2020epq} & $\pT^{\,\mu}>4$ GeV, $|\eta^{\,\mu}|<2.4$, $m_{\mm}>10$ GeV \\\hline %, $\pT^{\mm}<2$ GeV\\\hline
\end{tabular}
\end{table}

\subsection*{Exclusive two-photon dimuon production in \pp\ collisions at the LHC}

For the \pp\ UPC case, measurements have been carried out so far by CMS at $\sqrtsnn=7$ TeV~\cite{CMS:2011vma} and by ATLAS at $\sqrtsnn=7,\,13$~TeV~\cite{ATLAS:2015wnx,ATLAS:2017sfe}. The muon pseudorapidity $\eta^{\,\mu}$ selections are relatively similar in both experiments, but the CMS measurement is carried out at somehow lower $\pT^\mu$ and dimuon mass values. Table~\ref{tab:xsecsdimuonpp} presents our LO and NLO fiducial cross sections ($\sigma^\mathrm{LO}$ and $\sigma^\mathrm{NLO}$, respectively) compared with the four experimental results. Our default results are those obtained with ChFF fluxes, but we provide for comparison in parenthesis the EDFF values, which are about 12--20\% smaller than those derived with the former. The first observation is that the NLO corrections decrease the theoretical cross section by 3.5--5.0\% with respect to the LO results  (independently of the photon flux used). The impact of the NLO corrections increases with $m_{\mm}$, due to the mass-dependence of the $\alpha\ln{(m_\mu^2/m_{\mm}^2)}$ logarithm %$\alpha\ln{\frac{m_\mu^2}{\mu_{\mm}^2}}$ 
that controls the quasi-collinear emission of FSR photons and ``moves'' the cross section of dimuon pairs from higher to lower $m_{\mm}$ values. If the momentum of any collinear FSR photon emitted by the muons was experimentally reconstructed together with the parent lepton four-momentum (``dressed'' lepton), then the impact of such corrections would be reduced.  The experimental uncertainties of the cross sections are relatively small, of the order of 3--15\% depending on the bin, so the precision of the measurement is commensurate with the size of the NLO corrections. In general, the NLO ChFF predictions agree with the data (as indicated by the $\sigma^\mathrm{data}/\sigma^\mathrm{NLO}\approx 1$ ratio in the last column of Table~\ref{tab:xsecsdimuonpp}), whereas the LO ChFF results tend to overestimate the data by 4--10\%, and the NLO EDFF calculations to undershoot the measurements by about 10\%.
%At LO, EDFF results are 5-8\% below the data, while ChFF are 4-9\% above the data. The NLO QED corrections reduce the LO cross sections by around 4-5\%, largely independent of the photon flux. 
%The NLO QED corrections improve the agreement between ChFF results and the data, while the comparisons between the EDFF prediction and the measured data are worse than LO. The NLO ChFF calculations lie in the one sigma standard deviation in the three quoted experimental errors, which is clearly not the case for either LO or EDFF. 
These results further confirm that the ChFF photon flux is more realistic than the EDFF one, which is already theoretically justified since the latter has an arbitrary infrared cutoff on the UPC impact parameter.

\begin{table}[htpb!]
\centering
\tabcolsep=2.75mm
\caption{Fiducial exclusive dimuon cross sections measured in UPCs at the LHC (within the phase space defined in Table~\ref{tab:dimuoncuts}), compared with the theoretical LO and NLO QED results obtained with \gammaUPC\ using the ChFF and EDFF $\gamma$ fluxes. The last column lists the corresponding $\sigma^\mathrm{data}/\sigma^\mathrm{NLO}$ ratios.
%The columns denoted with ChFF$_\mathrm{corr}$ means the ChFF with the proton form factor corrected beyond the dipole form.
\label{tab:xsecsdimuonpp}}
\vspace{0.2cm}
%\begin{tabular}{l|c|ccc|ccc} \hline%\hline
%$\gaga\to\mumu$ & Measured $\sigma$ & \multicolumn{3}{c|}{\gammaUPC~$\sigma^\mathrm{LO}$} & \multicolumn{3}{c}{\gammaUPC~$\sigma^\mathrm{NLO}$} \\%\hline
%Experiment, system & & EDFF & ChFF  & ChFF$_\mathrm{corr}$ & EDFF & ChFF & ChFF$_\mathrm{corr}$\\\hline
 %CMS, \pp\ at 7 TeV~\cite{CMS:2011vma} & $3.38^{+0.62}_{-0.59}$~pb & $3.20$~pb & $3.63$~pb & $3.62$~pb & $3.10$~pb & $3.51$~pb & $3.50$~pb \\
 %ATLAS, \pp\ at 7 TeV~\cite{ATLAS:2015wnx} & $0.628\pm 0.038$~pb & $0.590$~pb & $0.688$~pb & $0.687$~pb & $0.560$~pb & $0.654$~pb & $0.653$~pb \\
 %ATLAS, \pp\ at 13 TeV~\cite{ATLAS:2017sfe} & $3.12\pm 0.16$~pb & $2.88$~pb & $3.23$~pb & $3.23$~pb & $2.76$~pb & $3.10$~pb & $3.09$~pb \\
 %ATLAS, \PbPb\ at 5.02 TeV~\cite{ATLAS:2020epq} & $34.1\pm 0.8$~$\mu$b & $32.1$~$\mu$b & $40.4$~$\mu$b & -- & $30.5$~$\mu$b & $38.5$~$\mu$b & --\\\hline
\begin{tabular}{l|c|c|c|c} \hline%\hline
$\gaga\to\mumu$ & measured $\sigma^\mathrm{data}$ & {\gammaUPC~$\sigma^\mathrm{LO}$} & {\gammaUPC~$\sigma^\mathrm{NLO}$} & ratio $\sigma^\mathrm{data}/\sigma^\mathrm{NLO}$ \\%\hline
System, experiment &  &  ChFF (EDFF) & ChFF (EDFF) & ChFF (EDFF) \\\hline
\pp\ at 7 TeV, CMS~\cite{CMS:2011vma} & $3.38^{+0.62}_{-0.59}$~pb & $3.62~(3.20)$~pb & $3.50~(3.10)$~pb & $0.97^{+0.18}_{-0.17}$ ($1.09^{+0.20}_{-0.19}$) \\
\pp\ at 7 TeV, ATLAS~\cite{ATLAS:2015wnx} & $0.628\pm 0.038$~pb & $0.687~(0.59)$~pb & $0.653~(0.56)$~pb & $0.96\pm0.06$ ($1.12\pm 0.07$)\\
\pp\ at 13 TeV, ATLAS~\cite{ATLAS:2017sfe} & $3.12\pm 0.16$~pb & $3.23~(2.88)$~pb & $3.09~(2.76)$~pb & $1.00\pm0.05$ ($1.13\pm 0.06$)\\
%\PbPb\ at 5.02 TeV, ATLAS~\cite{ATLAS:2020epq} & $34.1\pm 0.8$~$\mu$b & $40.4~(32.1)$~$\mu$b & $38.5~(30.5)$~$\mu$b & $0.89\pm0.02$ ($1.12\pm 0.03$) \\\hline
\PbPb\ at 5.02 TeV, ATLAS~\cite{ATLAS:2020epq} & $34.1\pm 0.8$~$\mu$b & $39.4~(31.5)$~$\mu$b & $37.5~(30.0)$~$\mu$b & $0.91\pm0.02$ ($1.14\pm 0.03$) \\\hline
\end{tabular}
\end{table}

\begin{figure}[htp!]
\centering
\includegraphics[width=0.49\textwidth]{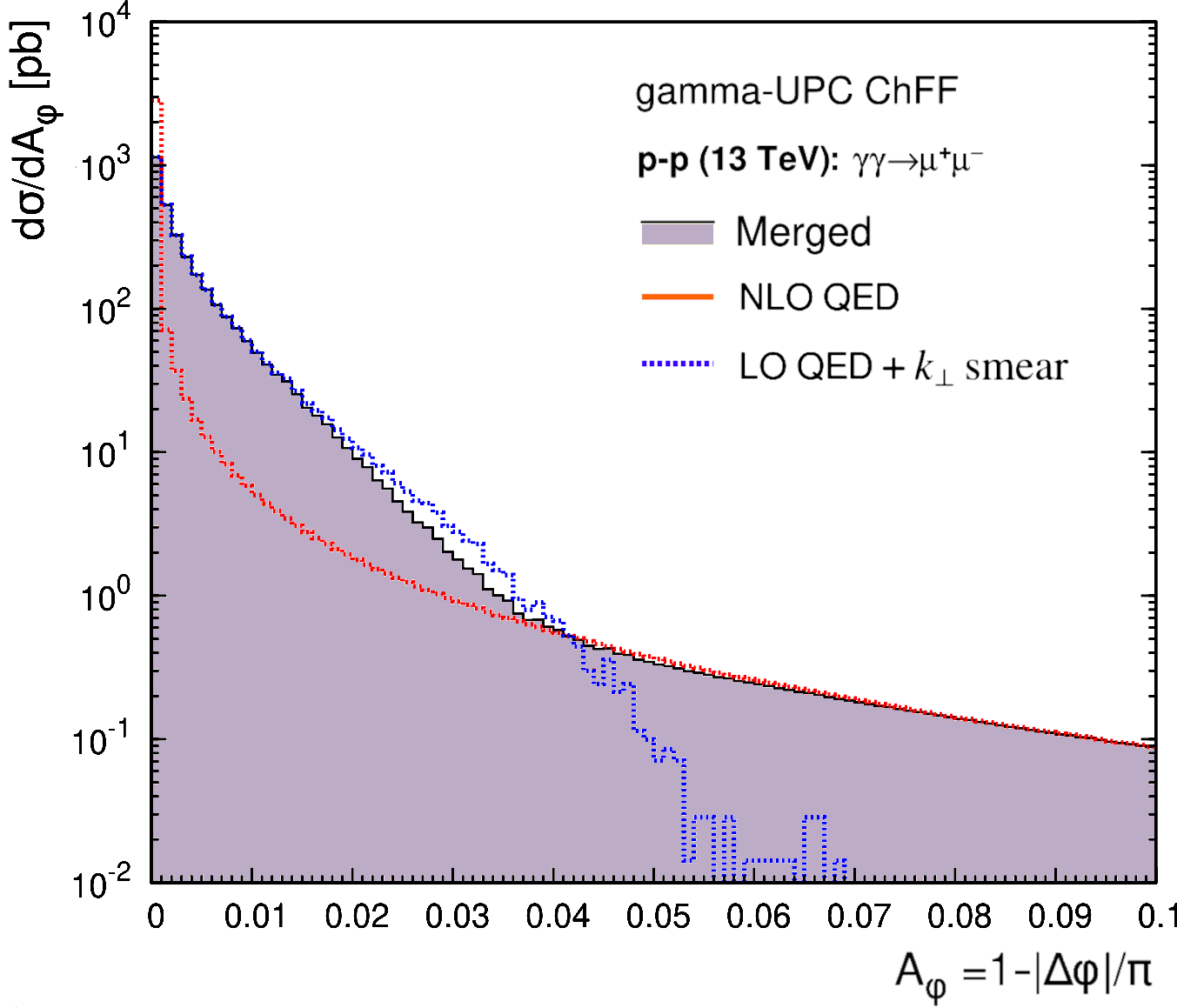}
\includegraphics[width=0.49\textwidth]{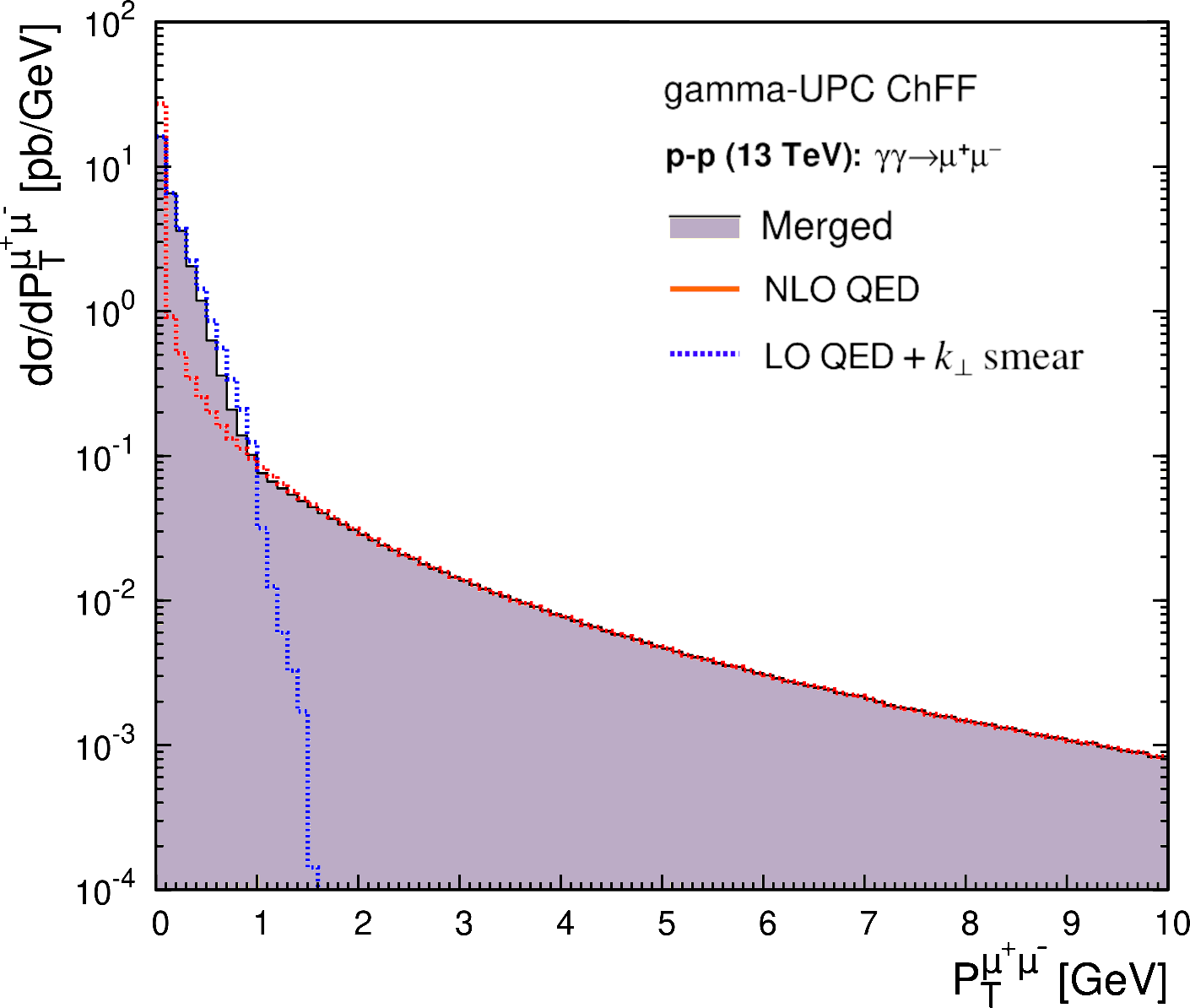}
\caption{Distributions of dimuon acoplanarity $A_{\phi}^{\mm}$ (left) and $\pT^{\mm}$ (right) in \pp\ UPCs at $\sqrts=13$ TeV (within the phase space defined in Table~\ref{tab:dimuoncuts}, third row) computed with \gammaUPC\ (ChFF photon flux) at LO$\,+\,\kt$-smearing (blue dashed histograms) and NLO QED (red histograms) accuracy, and their corresponding combined distributions (filled grey histograms).
\label{fig:dimuonPTAco_pp13TeV}}
\end{figure}
It is important to notice that the exclusive $\gaga\to\mumu$ cross sections in \pp\ UPCs are usually obtained from a fit of the acoplanarity distribution measured in data over the $A_\phi \approx 0$--0.1 range in order to evaluate the contributions from semiexclusive (single and double proton dissociation) processes constrained by templates from MC simulations. It is, thus, important to have a good theoretical control of the spectral shape of the signal acoplanarity. Although the large majority of the signal is clearly located at very small acoplanarities ($A_\phi \lesssim 0.02$), Fig.~\ref{fig:dimuonPTAco_pp13TeV} (left) shows that the NLO QED corrections are needed to properly describe the dimuon acoplanarity tail beyond $A_\phi \approx 0.04$. Similarly, the NLO contributions enhance also the tail of the dimuon transverse momentum, beyond $\pTmumu\gtrsim 1$~GeV (Fig.~\ref{fig:dimuonPTAco_pp13TeV}, right). Any experimental selection applied on the dimuon $\pT^{\mm}$ distribution has an impact on the determination of (a fraction of) the full exclusive cross section and, thus, requires the use of NLO predictions.\\

\begin{figure}[htbp!]
\centering
\includegraphics[width=0.49\textwidth]{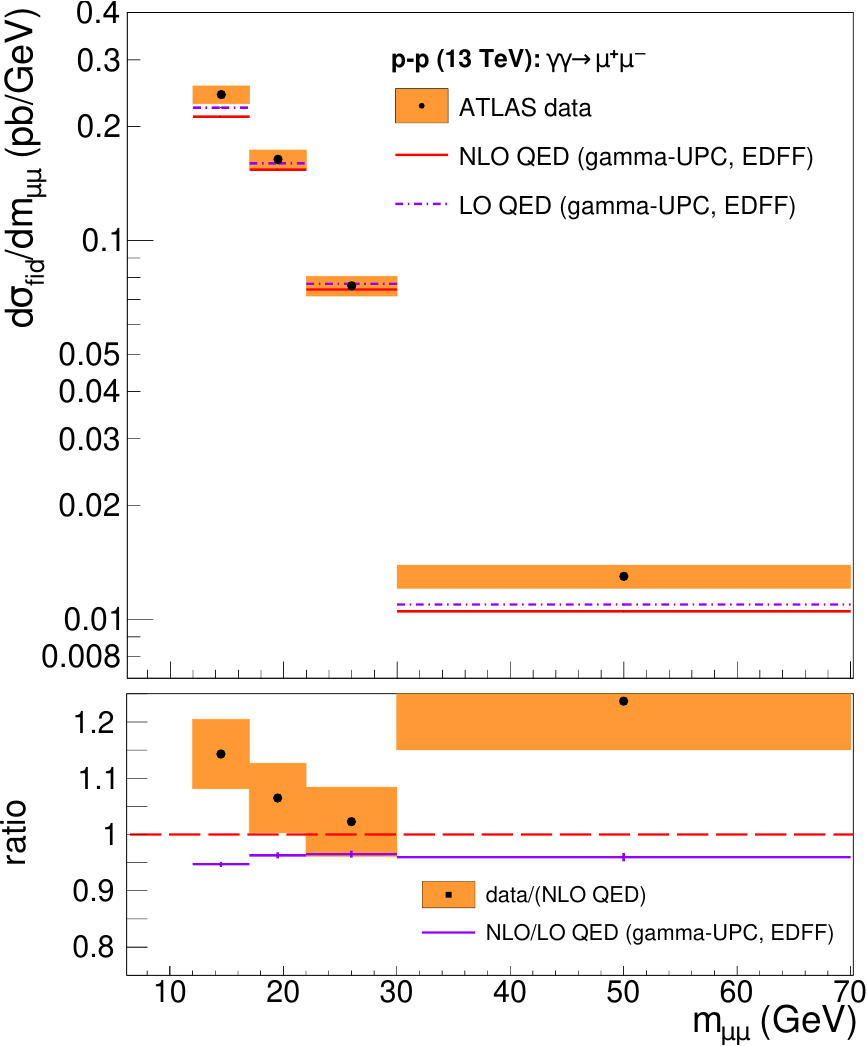}
\includegraphics[width=0.49\textwidth]{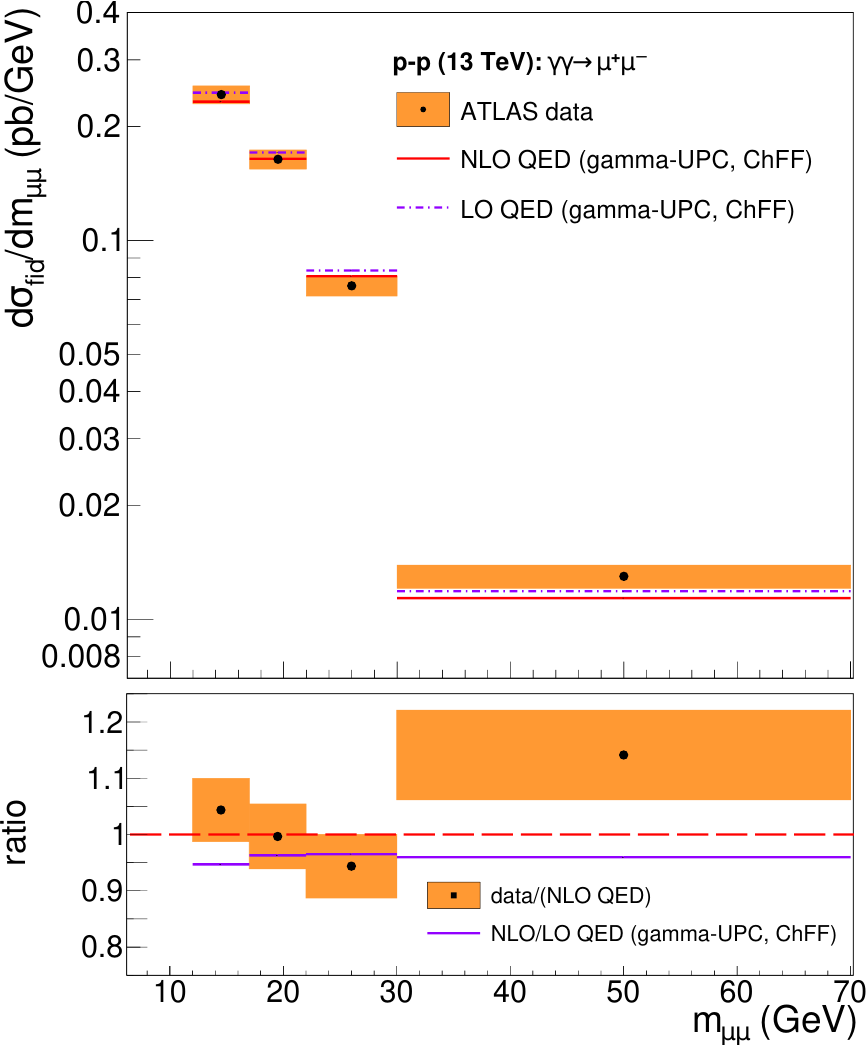}
\caption{Differential cross section for exclusive dimuon production as a function of invariant mass, $\mathrm{d}\sigma/\mathrm{d}m_{\mu^+\mu^-}$, in \pp\ UPCs at $\sqrts=13$~TeV within the fiducial cuts of Table~\ref{tab:dimuoncuts} (third row). The data (dots with orange boxes)~\cite{ATLAS:2017sfe} are compared with LO (blue dashed histogram) and NLO  (red histogram) QED predictions with EDFF (left) and ChFF (right) photon fluxes. The lower panels of each plot show the corresponding data/NLO ratios (dots with orange uncertainties boxes) and NLO/LO $K$ factors (violet line).
\label{fig:dimuondMpp13TeV}}
\end{figure}

Besides the fiducial integrated cross section, the ATLAS analysis~\cite{ATLAS:2017sfe} provided measurements with small uncertainties of the differential dimuon cross section as a function of pair invariant mass over $m_{\mm}=12$--70~GeV. Figure~\ref{fig:dimuondMpp13TeV} shows the data (black dots with orange uncertainties boxes) compared to our LO (blue dashed histogram) and NLO (red histogram) predictions with the EDFF (left) and ChFF (right) photon fluxes. The top panels show the data--theory comparison of the distributions, and the bottom panels present the data/NLO ratios (dots with orange uncertainties boxes) and the $K=$~NLO/LO factors (violet line). The $K\approx 0.945$--0.965 factors are relatively constant as a function of $m_{\mm}$ and independent of the photon flux type. Both, the NLO corrections and ChFF fluxes improve the theoretical reproduction of the data. A data--theory goodness-of-fit yields $\chi^2/N_\mathrm{dof}\approx 1.2,\,1.8$ (for $N_\mathrm{dof} =4$ degrees of freedom) for the NLO predictions obtained with the ChFF and EDFF fluxes, respectively. This result further confirms the optimized choice of ChFF photon fluxes plus NLO corrections for accurate calculations of dilepton production in $\gaga$ processes.

\subsection*{Exclusive two-photon dimuon production in \PbPb\ collisions at the LHC}

\begin{figure}[htbp!]
\centering
\includegraphics[width=0.44\textwidth]{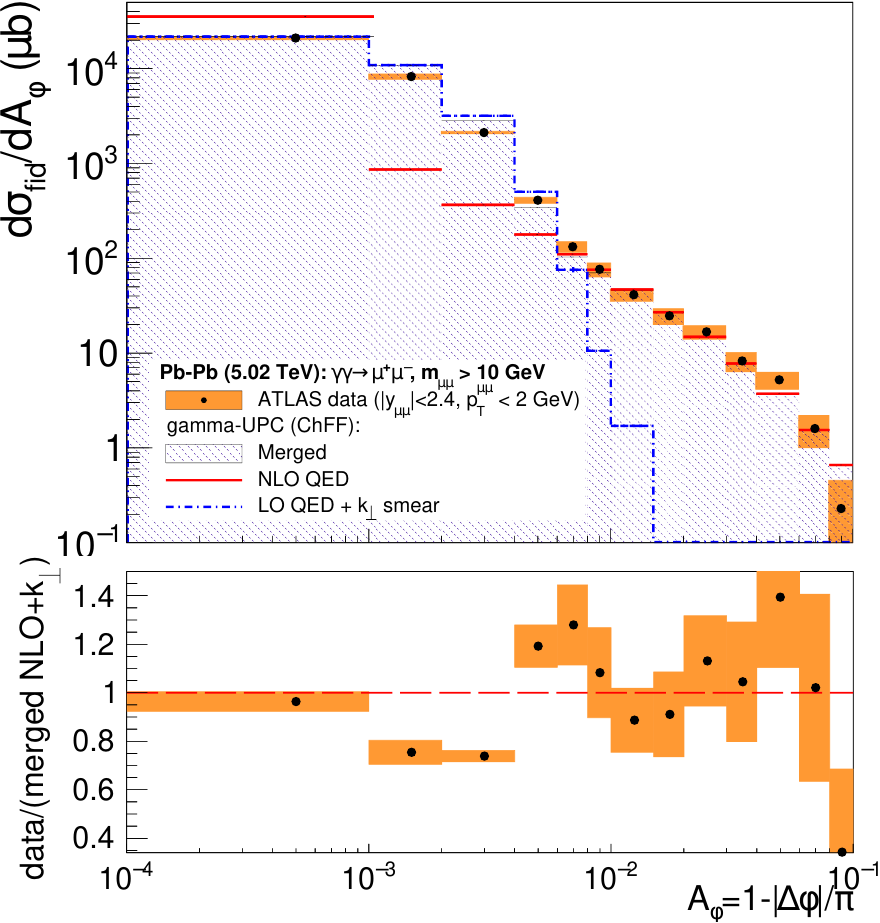}
\includegraphics[width=0.55\textwidth]{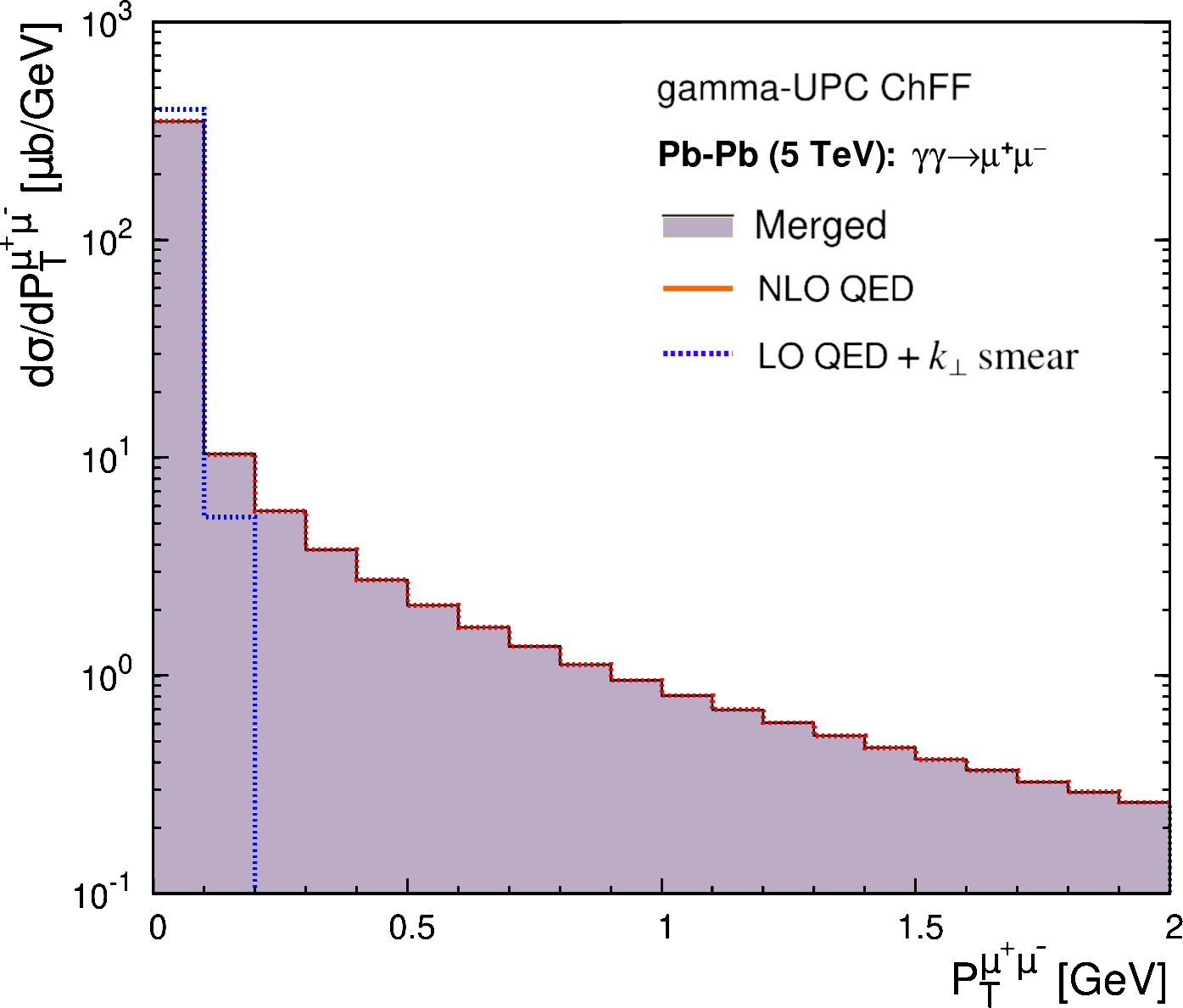}
\caption{Distributions for dimuon acoplanarity $A_{\phi}^{\mm}$ (left) and $\pT^{\mm}$ (right) in \PbPb\ UPCs at $\sqrtsnn=5.02$ TeV computed with \gammaUPC\ (ChFF photon flux) at LO$\,+\,\kt$-smearing (blue dashed histograms) and NLO (red histogram) QED accuracy, and their corresponding combined distributions (filled grey histograms). In the left panel, the acoplanarity distributions are compared to the $\mathrm{d}\sigma/\mathrm{d}A_{\phi}$ measurement of Ref.~\cite{ATLAS:2020epq} (black dots with orange uncertainties), and the ratio of data over merged NLO$\,+\,\kt$-smearing prediction is shown in the bottom left panel.
\label{fig:dimuonPTAco_PbPb5TeV}}
\end{figure}

Exclusive dimuon production has been also measured in \PbPb\ UPCs, whose cross section are enhanced by a $Z^4\approx 4.5\cdot 10^7$ factor with respect to their \pp\ counterpart. The ATLAS experiment has measured the fiducial cross section of $\sigma(\gaga\to\mumu) = 34.1\pm0.3\,(\mathrm{stat}.)\pm 0.7\,(\mathrm{syst}.)~\mu$b in \PbPb\ at 5.02~TeV, reaching a 2.3\% precision~\cite{ATLAS:2020epq}. As shown in the last row of Table~\ref{tab:xsecsdimuonpp}, the LO ChFF cross section overshoots the data by 13\%, whereas the LO EDFF cross section undershoots them by 8\%. The NLO QED corrections reduce the LO cross sections by about 5\% and, as a conclusion, the NLO ChFF (EDFF) deviates from the ATLAS data by $+9\%$ ($-13\%)$. All in all, the NLO$\,+\,$ChFF result features the closest agreement with the experimental fiducial cross section. %As similarly observed in Ref.~\cite{Shao:2022cly} for the comparison between \starlight\ and LO EDFF, NLO EDFF gives us a close value $30.5~\mu$b to the quoted \starlight+\pythia8 value $30.8~\mu$b by ATLAS~\cite{ATLAS:2020epq}. The latter has additionally taken into account the final-state radiation by using QED shower. 
Figure~\ref{fig:dimuonPTAco_PbPb5TeV} shows the impact of the NLO corrections on the dimuon acoplanarity $A_{\phi}^{\mm}$ (left plot) and pair $\pT^{\mm}$ (right plot). As found in the \pp\ case, the higher-order corrections extend the tails of both distributions to values much above those obtained with LO$\,+\,\kt$-smearing calculations. In the left plot of Fig.~\ref{fig:dimuonPTAco_PbPb5TeV}, the predicted acoplanarity distributions are compared to the $\mathrm{d}\sigma/\mathrm{d}A_{\phi}^{\mm}$ measurement of Ref.~\cite{ATLAS:2020epq} (black dots with orange uncertainties), and the ratio of data over merged NLO$\,+\,\kt$-smearing prediction is shown on the bottom panel. As expected, the  LO$\,+\,\kt$-smearing result (blue dashed histogram) describes well the back-to-back region ($A_{\phi}^{\mm}\lesssim 5\cdot 10^{-3}$), but clearly misses the experimental data in the tail above $A_{\phi}^{\mm}\approx 10^{-2}$. On the other hand, the NLO-alone calculation (red histogram) exhibits the opposite behaviour. A proper combination of both regimes (as discussed in App.~\ref{sec:kTsmear}) can reproduce the data within $\pm20\%$ over the full acoplanarity range, as displayed in the bottom left ratio. Those results emphasize once more the need to properly account for NLO corrections in the extraction of the exclusive dimuon signal from the experimental \PbPb\ UPC data.\\

\begin{figure}[htbp!]
\centering
\includegraphics[width=0.33\textwidth]{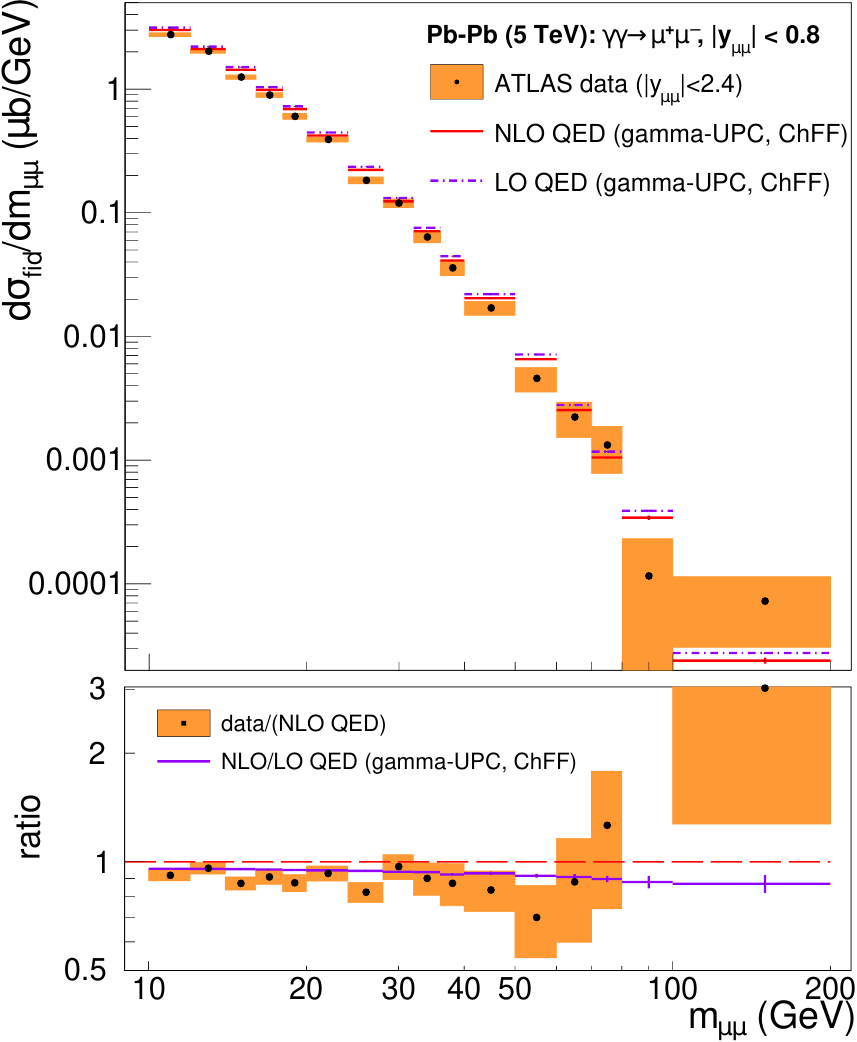}
\includegraphics[width=0.33\textwidth]{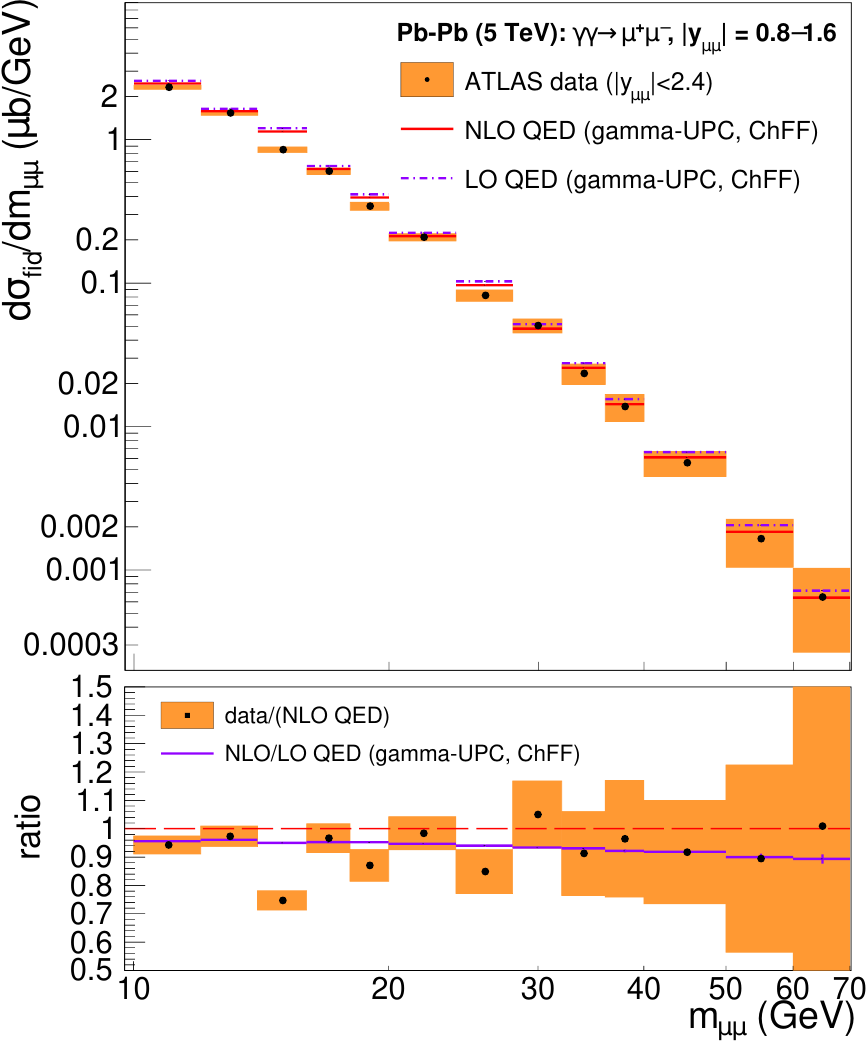}
\includegraphics[width=0.33\textwidth]{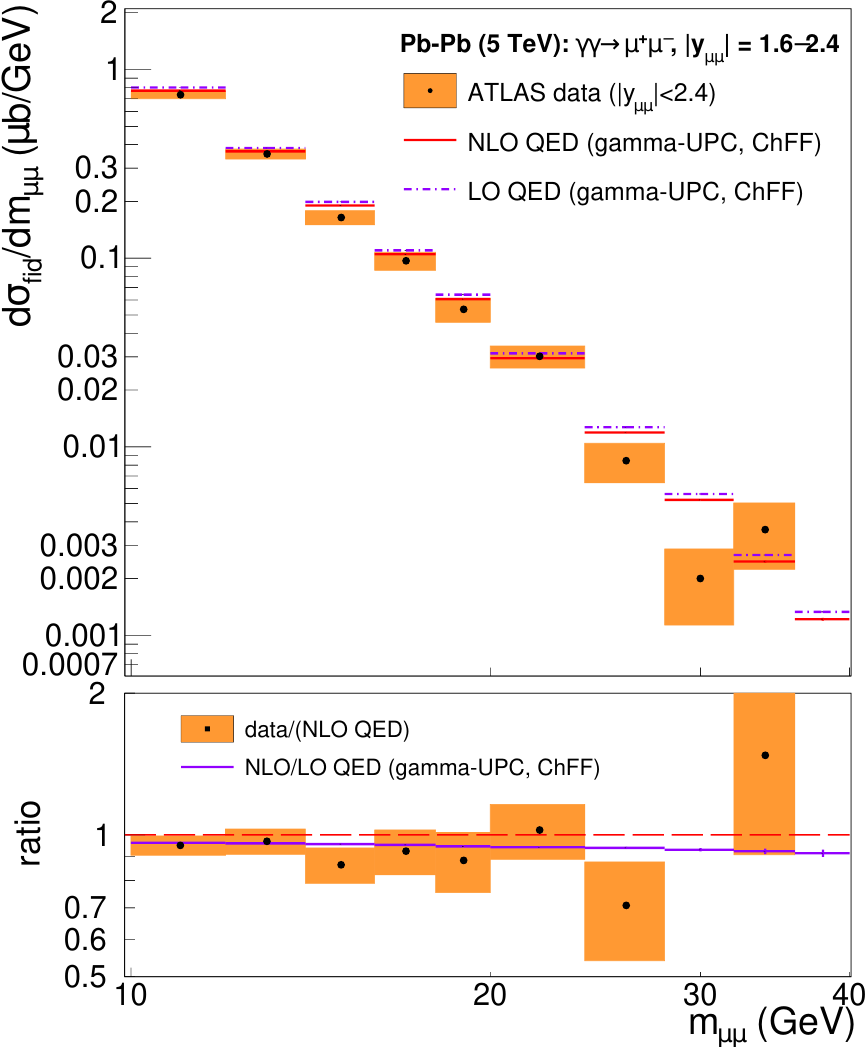}
\includegraphics[width=0.33\textwidth]{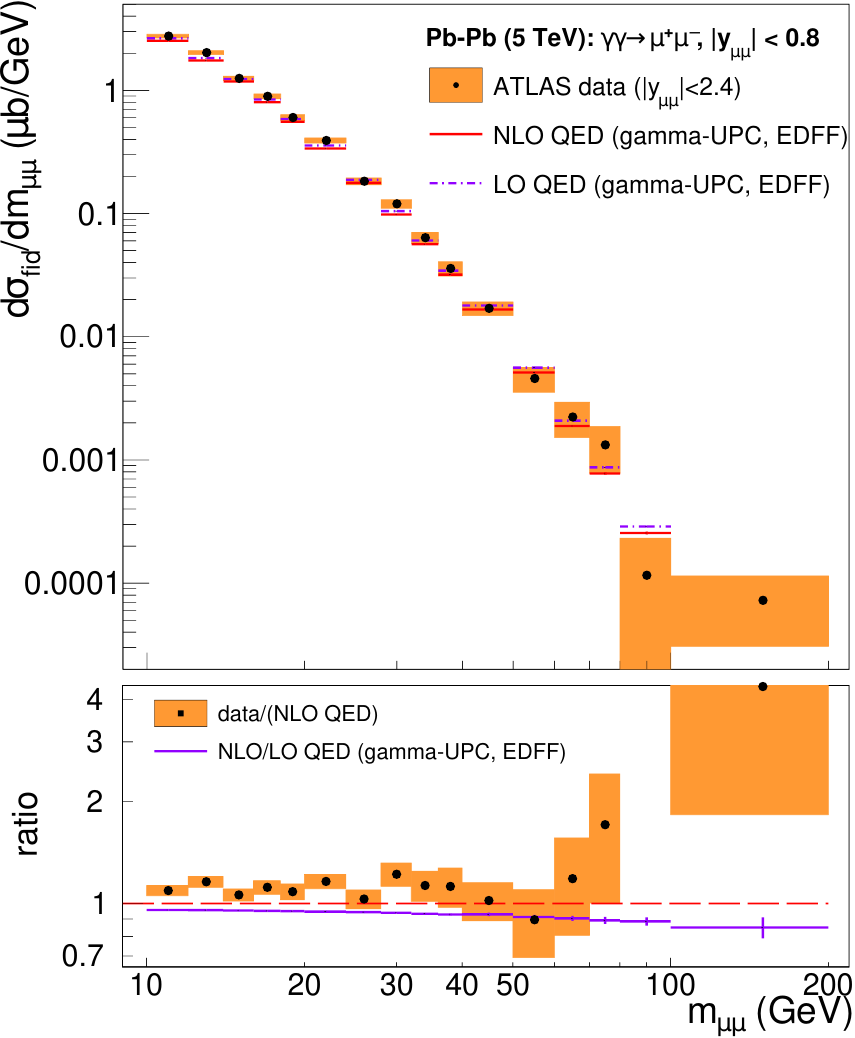}
\includegraphics[width=0.33\textwidth]{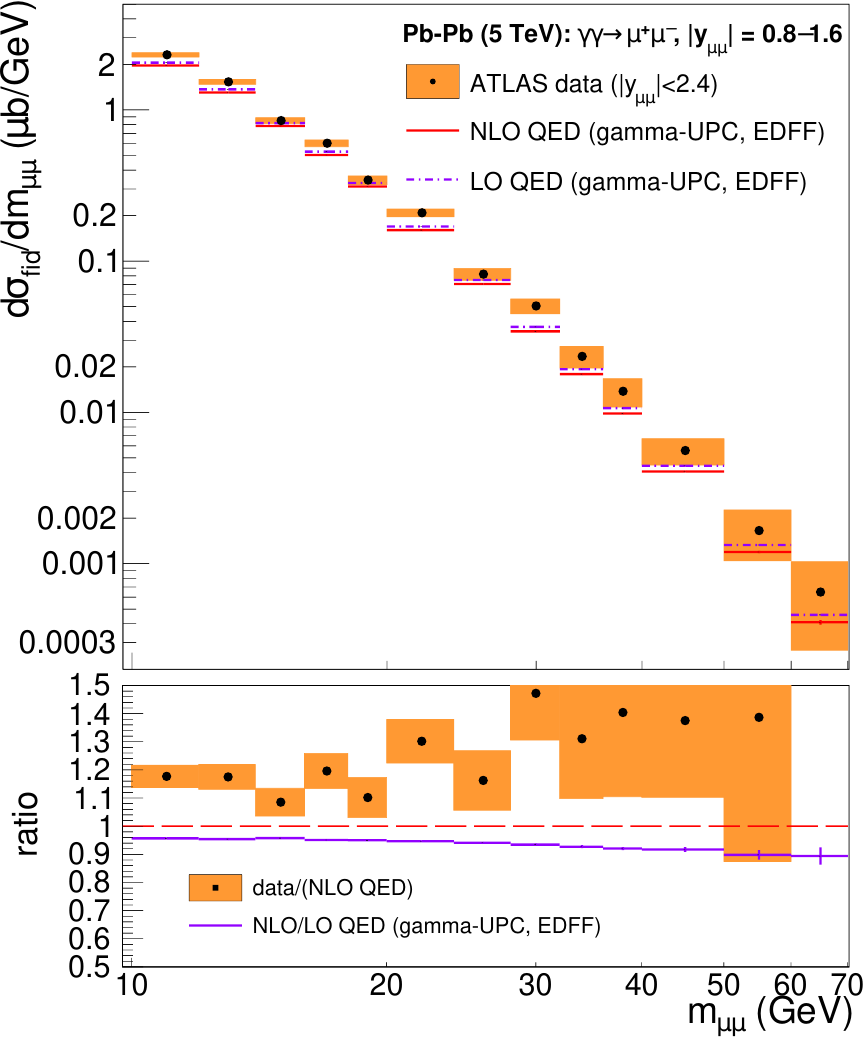}
\includegraphics[width=0.33\textwidth]{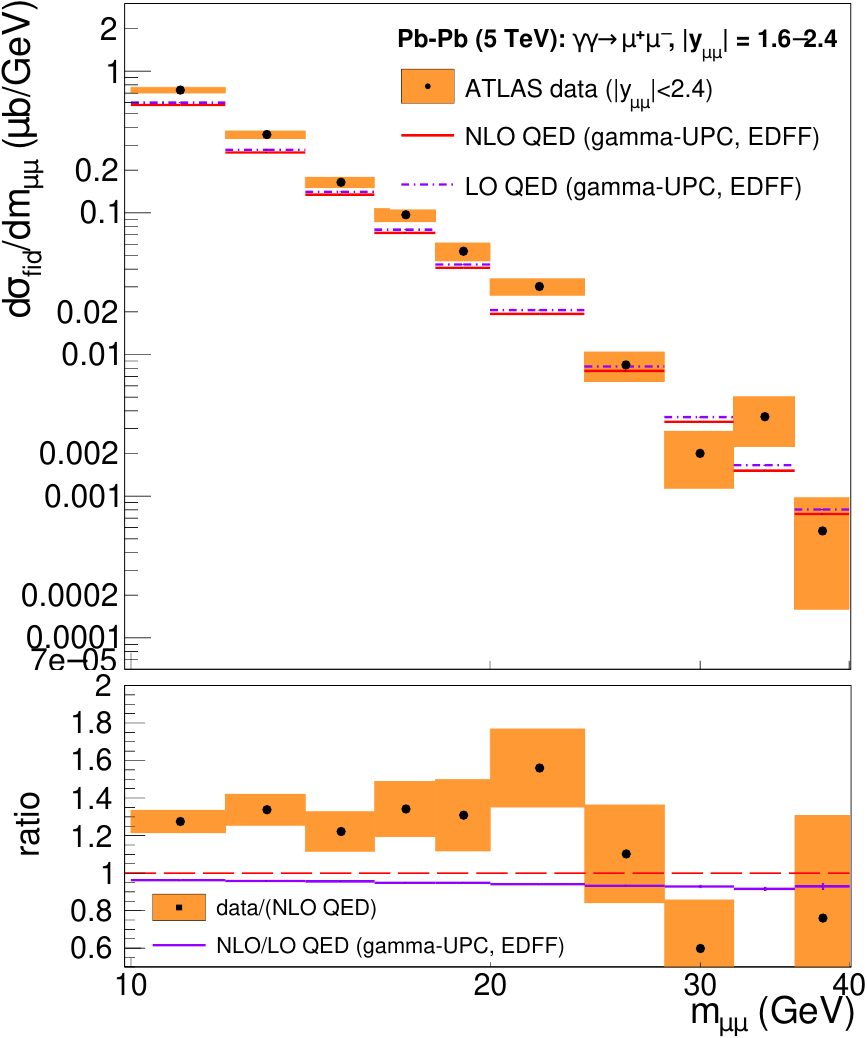}
\caption{Differential cross section as a function of invariant mass, $\mathrm{d}\sigma/\mathrm{d}m_{\mu^+\mu^-}$, for exclusive dimuon production in \PbPb\ UPCs at $\sqrtsnn=5.02$~TeV in different ranges of dimuon rapidity. The data (black dots with orange uncertainties)~\cite{ATLAS:2020epq} are compared with LO (blue dashed histogram) and NLO (red histogram) QED \gammaUPC\ predictions with ChFF (top) and EDFF (bottom) $\gamma$ fluxes. The lower panels of each plot show the corresponding data/NLO ratios (black dots with orange uncertainties) and NLO/LO $K$ factors (violet histogram).
%Data-theory $\chi^2$ values are quoted for nEDFF and nChFF fluxes. In the middle inset, we have the normal $K$ factors $\frac{\mathrm{d}\sigma^\mathrm{NLO}}{\mathrm{d}\sigma^\mathrm{LO}}$, while the lower inset shows the ratios of the NLO results over the ATLAS data.
\label{fig:dimuondM}}
\end{figure}

In the data-theory comparisons of differential distributions in \PbPb\ UPCs carried out in Ref.~\cite{Shao:2022cly}, we concluded that the normalized shapes of the measured spectra favoured the ChFF photon flux over the EDFF one, albeit the overall normalization of the LO ChFF prediction appeared somewhat worse than the LO EDFF one. We examine now how such a conclusion is altered by including NLO QED corrections. The first differential cross section considered is that as a function of the invariant mass of the muon pair shown in Fig.~\ref{fig:dimuondM}. From left to right, the three panels present the data-theory comparisons for three different pair rapidity intervals ($|y_{\mm}|<0.8$, $0.8<|y_{\mm}|<1.6$, and $1.6<|y_{\mm}|<2.4$, respectively), with the upper (lower) plots showing the ChFF (EDFF) predictions. The top panels of each plot show the differential cross sections in data and at LO and NLO accuracies, and the bottom panels display the data/NLO ratios (black dots with orange uncertainties) and NLO/LO $K$ factors (violet histogram). The first observation is that the NLO ChFF results are in better agreement with data than the EDFF ones for all kinematic ranges: the data/NLO(ChFF) ratios are close to unity for the three measured distributions.
%In order to quantify the (dis)agreements between theory and experiment, we evaluate the $\chi^2$ values in each panel. The $\chi^2$ values are computed by considering only the experimental errors. The overall $\chi^2$ values for the invariant mass distribution are $47.4$ and $58.0$ for NLO ChFF and EDFF respectively. These numbers slightly decrease with respect to the LO values $(\chi^2_\mathrm{LO~ChFF},\chi^2_\mathrm{LO~EDFF})=(53.0,60.2)$. The agreements are in general good, because the $\chi^2$ per degree-of-freedom ($N_\mathrm{dof}=N_\mathrm{data}-1=38$) is 1.24 and 1.53 for NLO ChFF and EDFF. Once more, we observe from the middle insets that the $K$ factors are independent of the photon flux. 
The second observation is that the NLO QED corrections are small in the low $m_{\mm}$ range, but become increasingly relevant and reduce the differential cross section by up to 15\% at the highest dimuon invariant mass. This can be understood from the aforementioned $m_{\mm}$ dependence of the logarithmic FSR correction. This result shows that it would be incorrect to apply a global constant $K$ factor to account for NLO corrections in the exclusive dimuon mass distributions.\\

\begin{figure}[htbp!]
\centering
\includegraphics[width=0.33\textwidth]{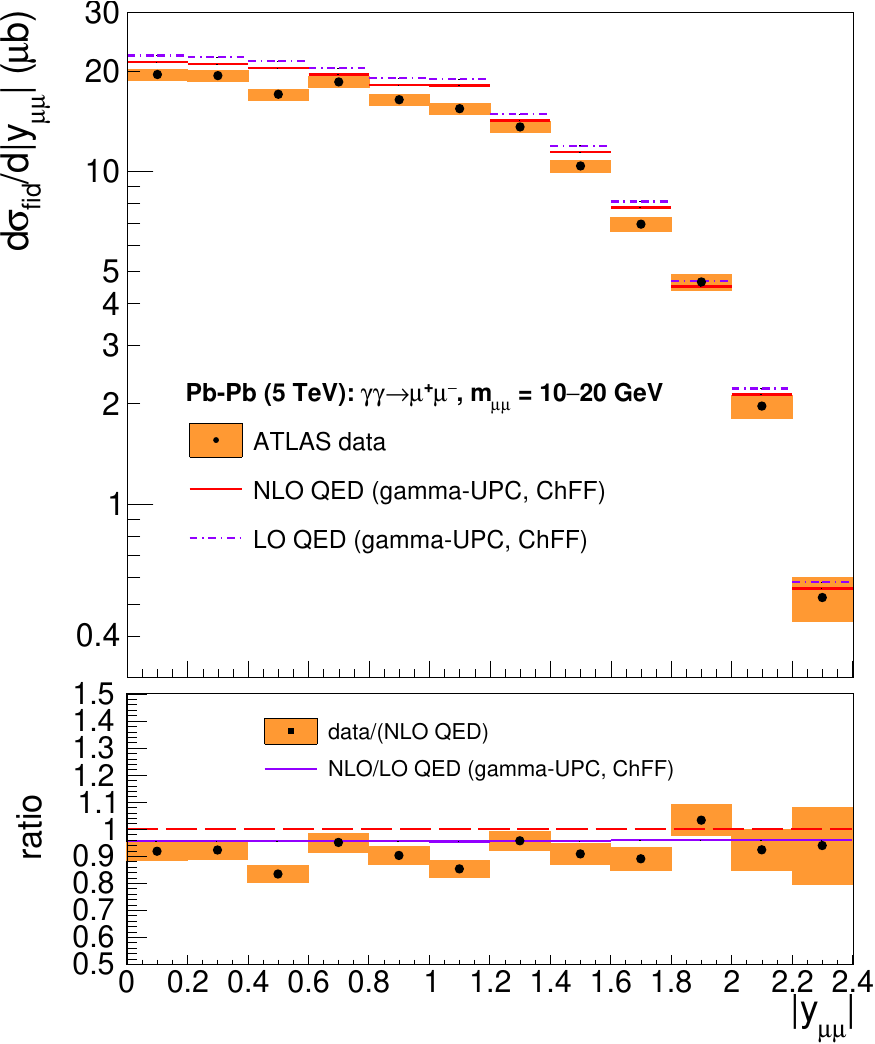}
\includegraphics[width=0.33\textwidth]{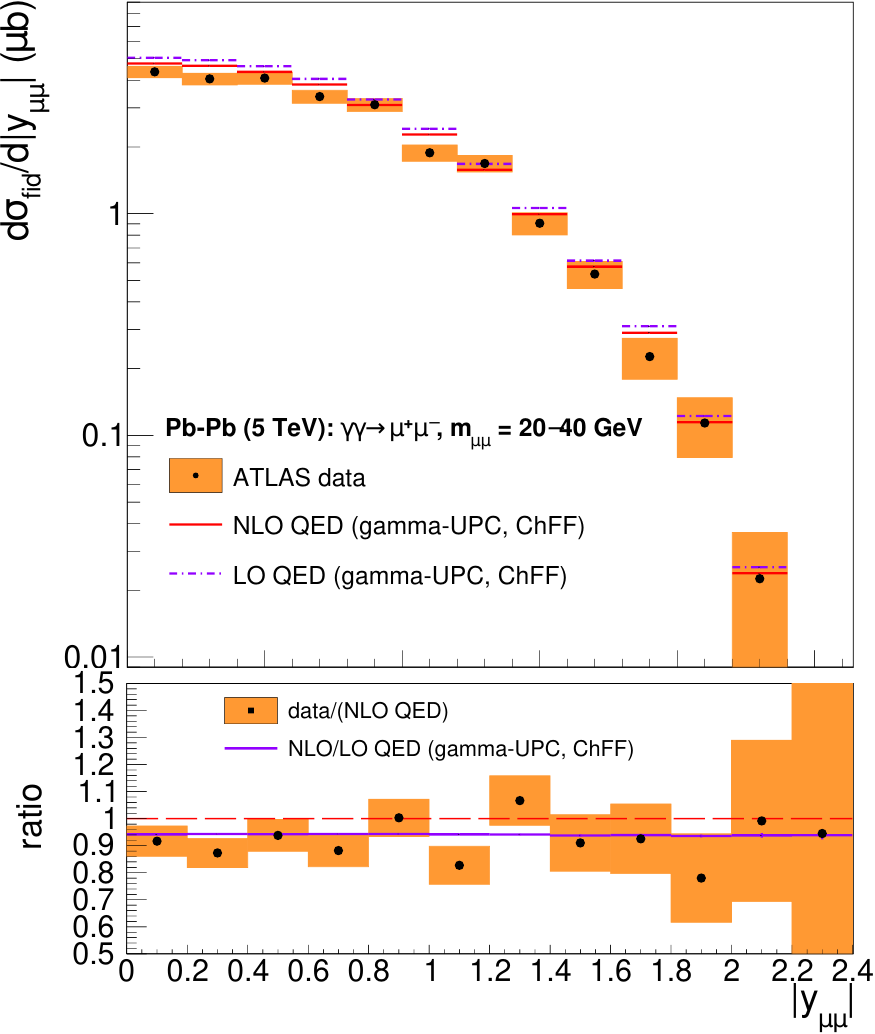}
\includegraphics[width=0.33\textwidth]{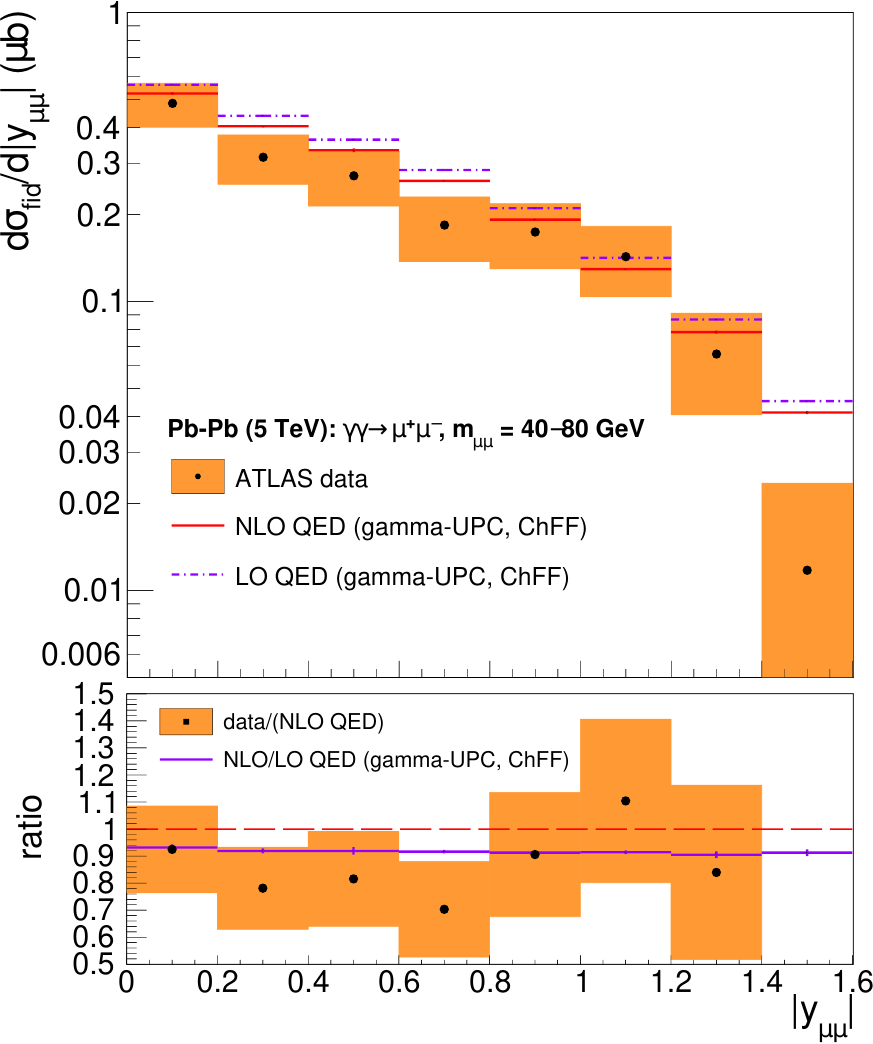}
\includegraphics[width=0.33\textwidth]{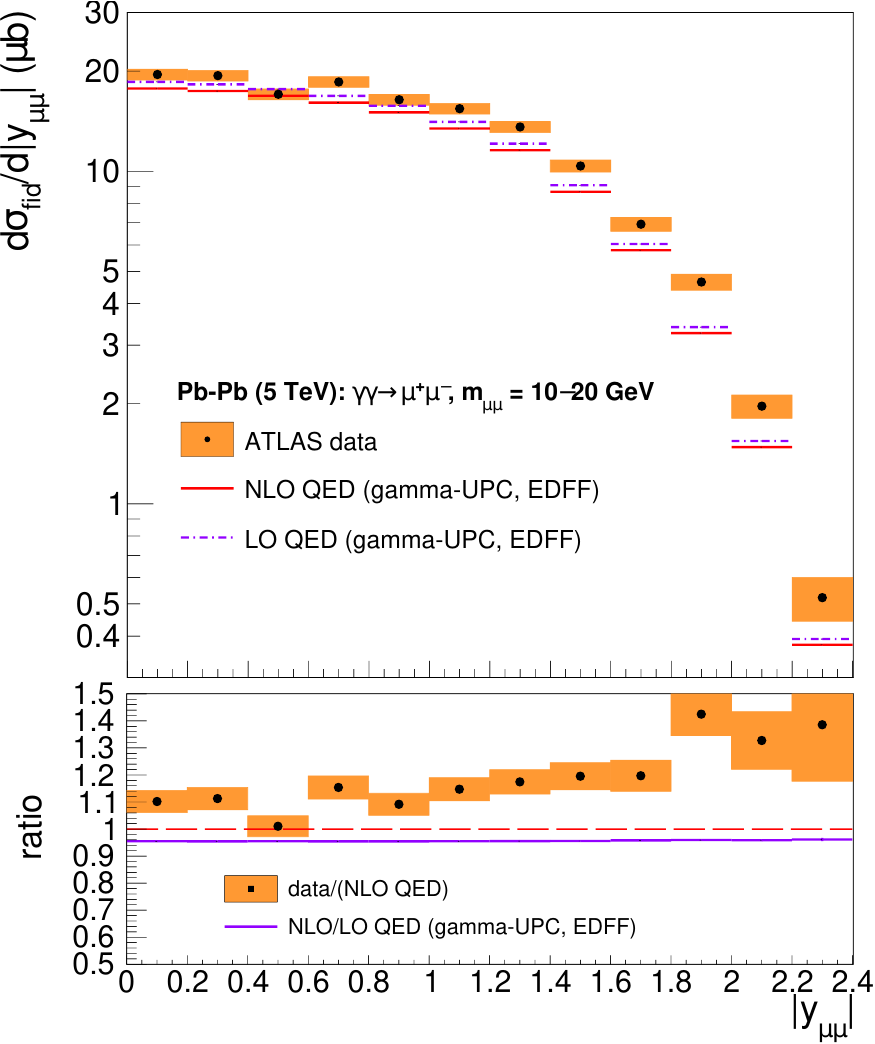}
\includegraphics[width=0.33\textwidth]{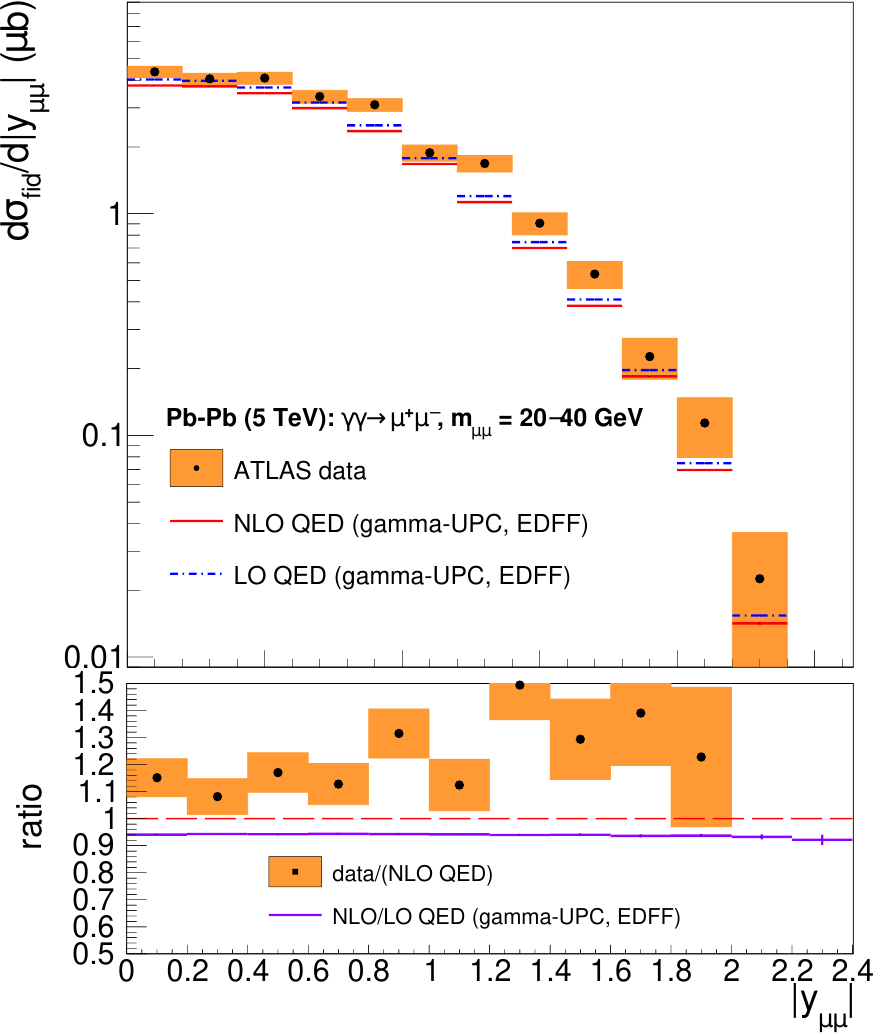}
\includegraphics[width=0.33\textwidth]{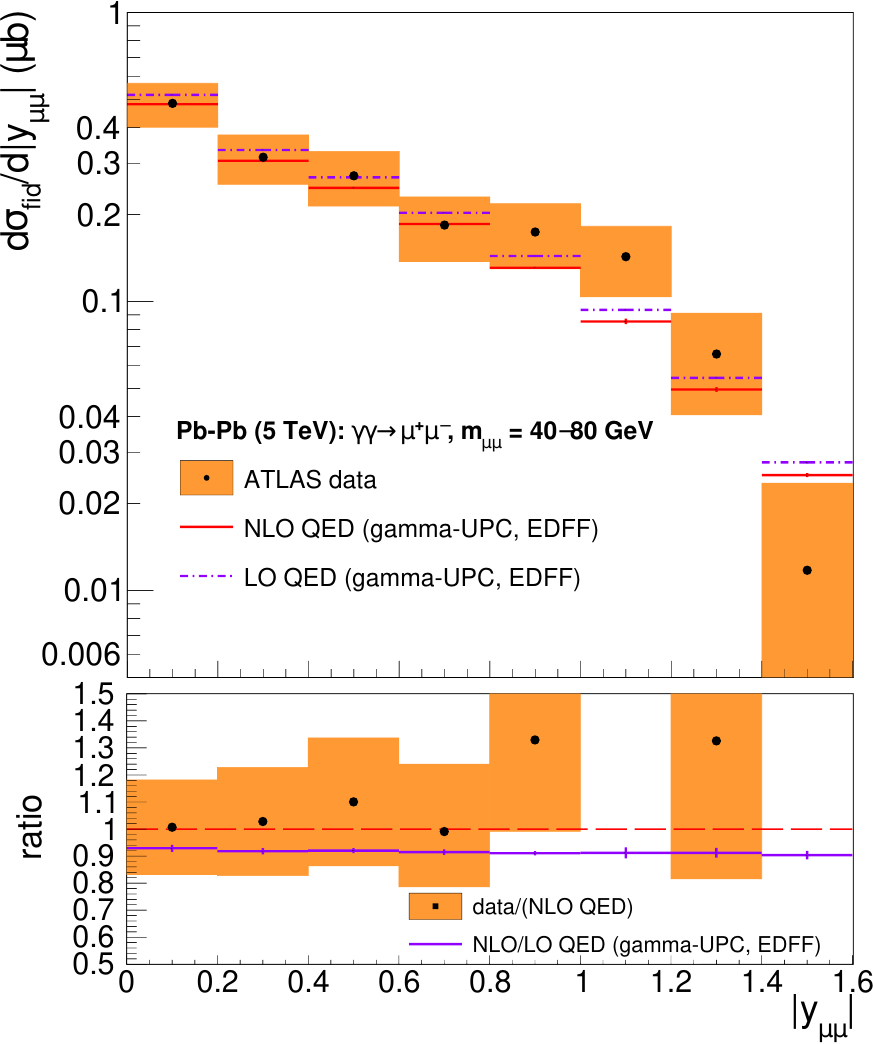}
\caption{Differential cross section as a function of dimuon absolute rapidity, $\mathrm{d}\sigma/\mathrm{d}|y_{\mu^+\mu^-}|$, for exclusive dimuon production in \PbPb\ UPCs at $\sqrtsnn=5.02$~TeV in different ranges of invariant mass. The data (black dots with orange uncertainties)~\cite{ATLAS:2020epq} are compared with LO (blue dashed histogram) and NLO (red histogram) QED \gammaUPC\ predictions with ChFF (top) and EDFF (bottom) $\gamma$ fluxes. The lower panels of each plot show the corresponding data/NLO ratios (black dots with orange uncertainties) and NLO/LO $K$ factors (violet histogram).
%Data-theory $\chi^2$ values are quoted for nEDFF and nChFF fluxes. In the middle inset, we have the normal $K$ factors $\frac{\mathrm{d}\sigma^\mathrm{NLO}}{\mathrm{d}\sigma^\mathrm{LO}}$, while the lower inset shows the ratios of the NLO results over the ATLAS data.
\label{fig:dimuondy}}
\end{figure}

The second differential cross section of interest is that as a function of absolute pair rapidity, $\mathrm{d}\sigma/\mathrm{d}|y_{\mm}|$. The ATLAS study~\cite{ATLAS:2020epq} revealed systematic deviations between the measured distribution, in particular in the tail, and the LO \starlight\ predictions. Our comparison of the data to the LO and NLO distributions with ChFF and EDFF fluxes is shown in Fig.~\ref{fig:dimuondy} with the same layout as in Fig.~\ref{fig:dimuondM}. The three panels correspond to three invariant mass intervals: $m_{\mm}\in [10,20], [20,40],[40,80]$~GeV, from left to right. Again, the NLO ChFF predictions agree nicely with the data for all kinematic ranges within the experimental uncertainties, whereas the NLO EDFF calculations underestimate them, especially for increasing $|y_{\mm}|$.
%In the lowest mass window $10<m_{\mm}<20$ GeV, NLO EDFF starts to systematically undershoot the data, yielding $\chi^2/N_\mathrm{dof}=3.3$. In contrast, NLO ChFF agrees well with the data with $\chi^2/N_\mathrm{dof}=1.2$ in the same $m_{\mm}$ bin. Overall, after summing all three mass windows, $\chi^2/N_\mathrm{dof}$ is 1.0 (1.9) for NLO ChFF (EDFF). NLO ChFF shows a slightly better agreement than LO ChFF ($\chi^2/N_\mathrm{dof}=1.2$), while EDFF has the similar $\chi^2$ value from LO to NLO. 
However, at variance with the behaviour seen in the invariant mass distribution, the $K$ factors are quite flat in $|y_{\mm}|$.\\

\begin{figure}[htbp!]
\centering
\includegraphics[width=0.33\textwidth]{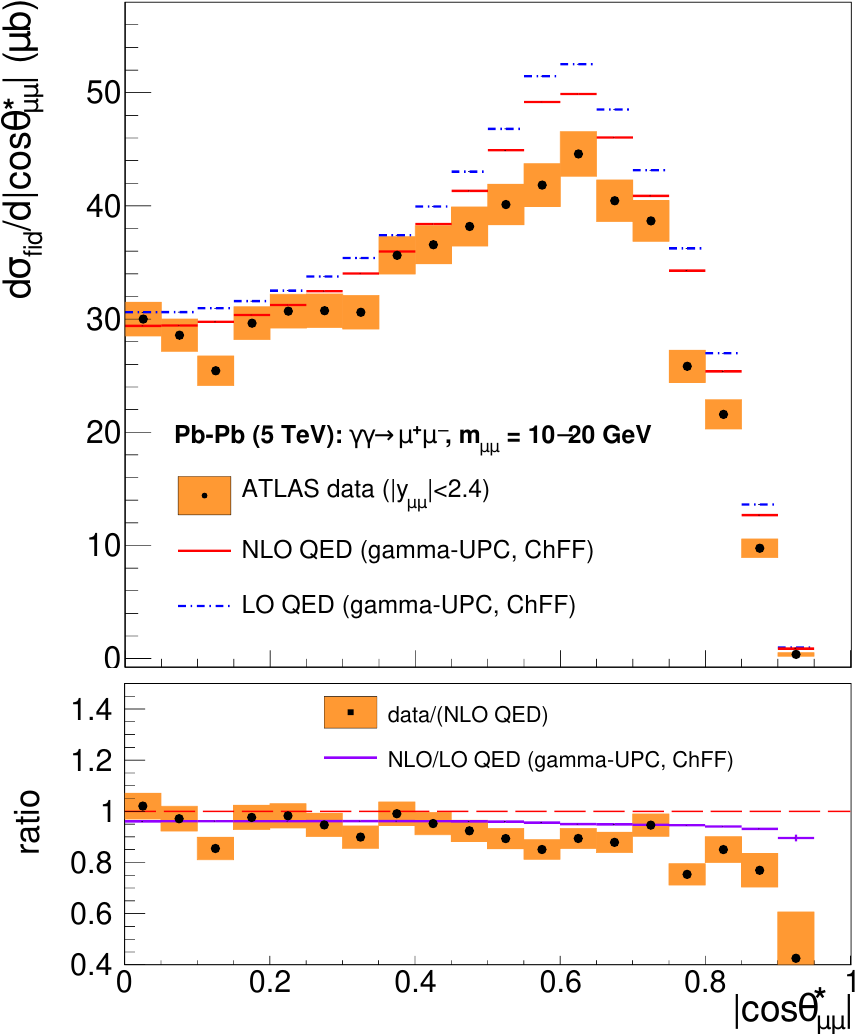}
\includegraphics[width=0.33\textwidth]{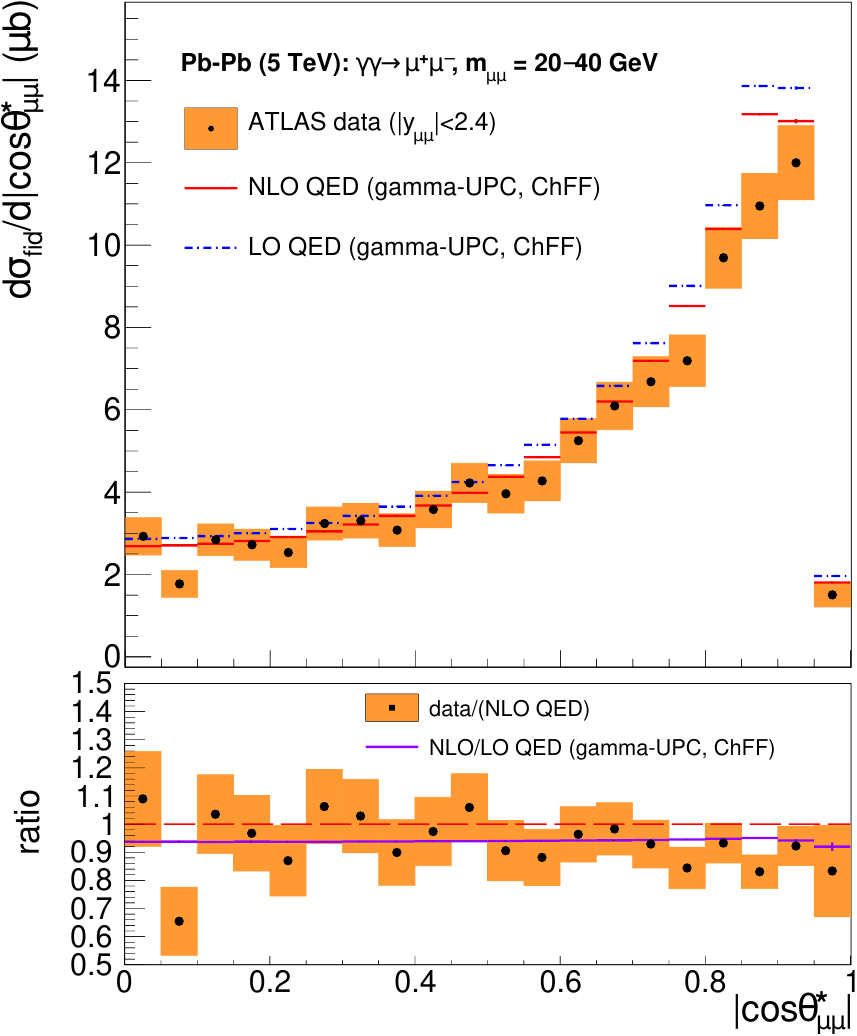}
\includegraphics[width=0.33\textwidth]{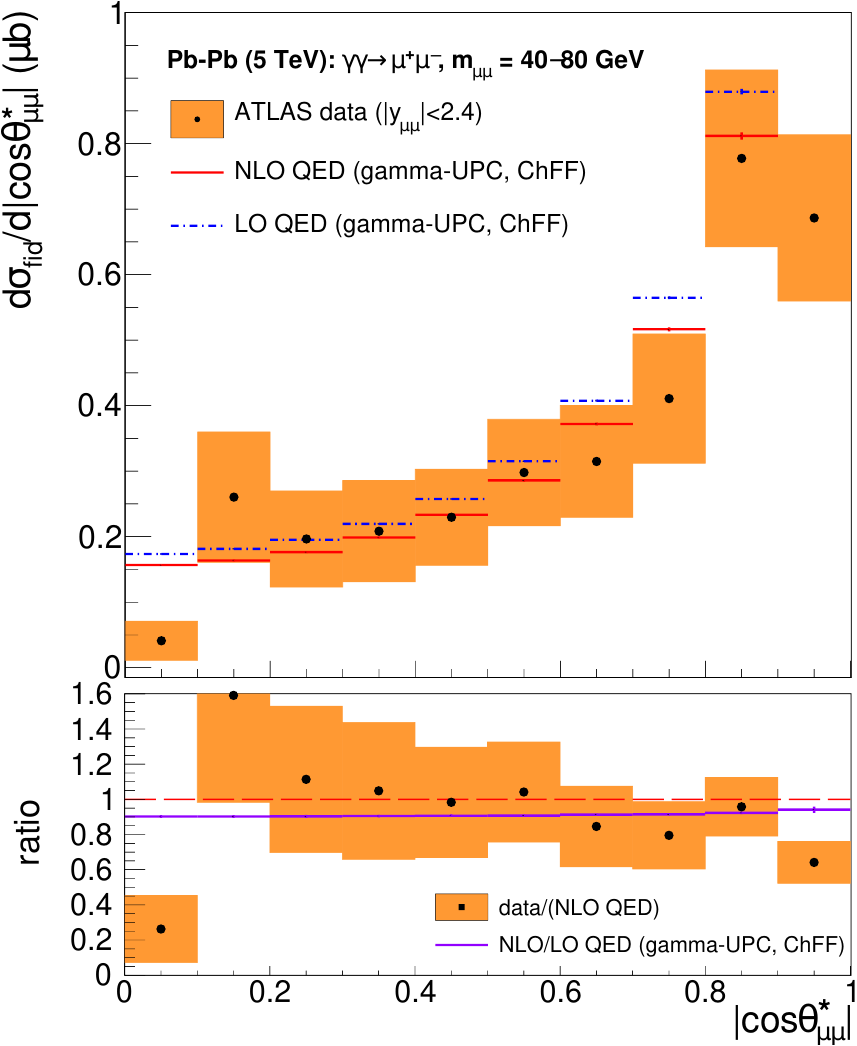}
\includegraphics[width=0.33\textwidth]{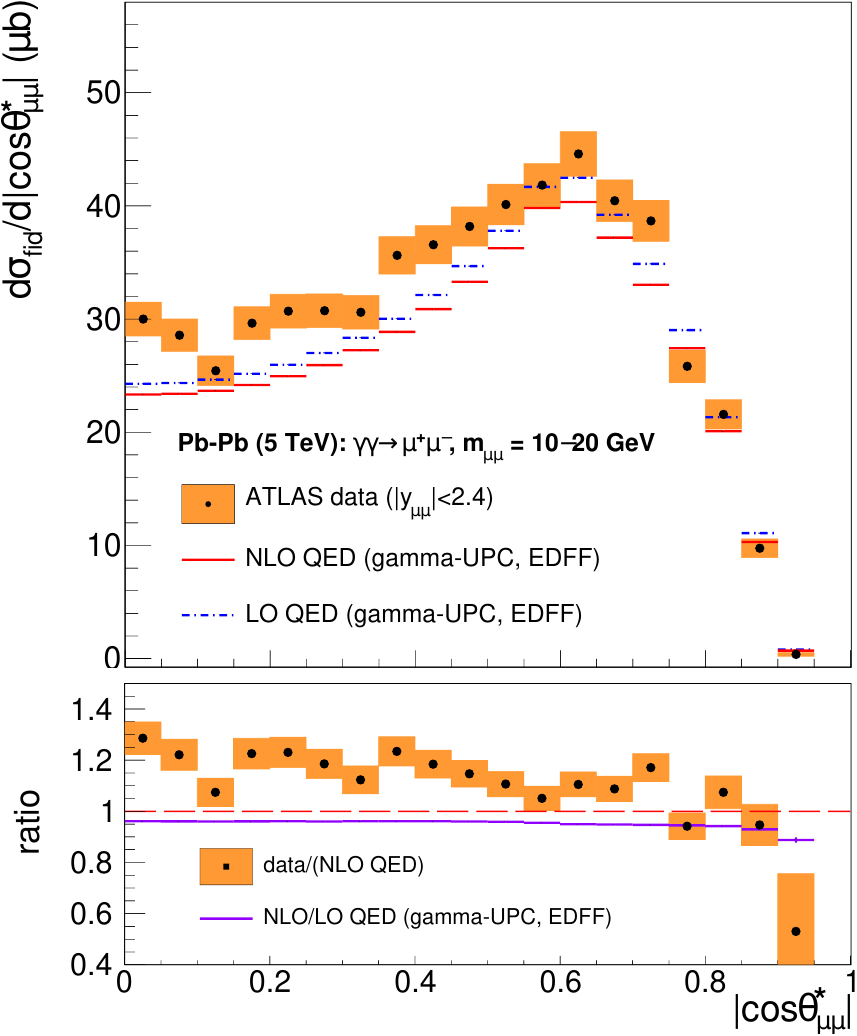}
\includegraphics[width=0.33\textwidth]{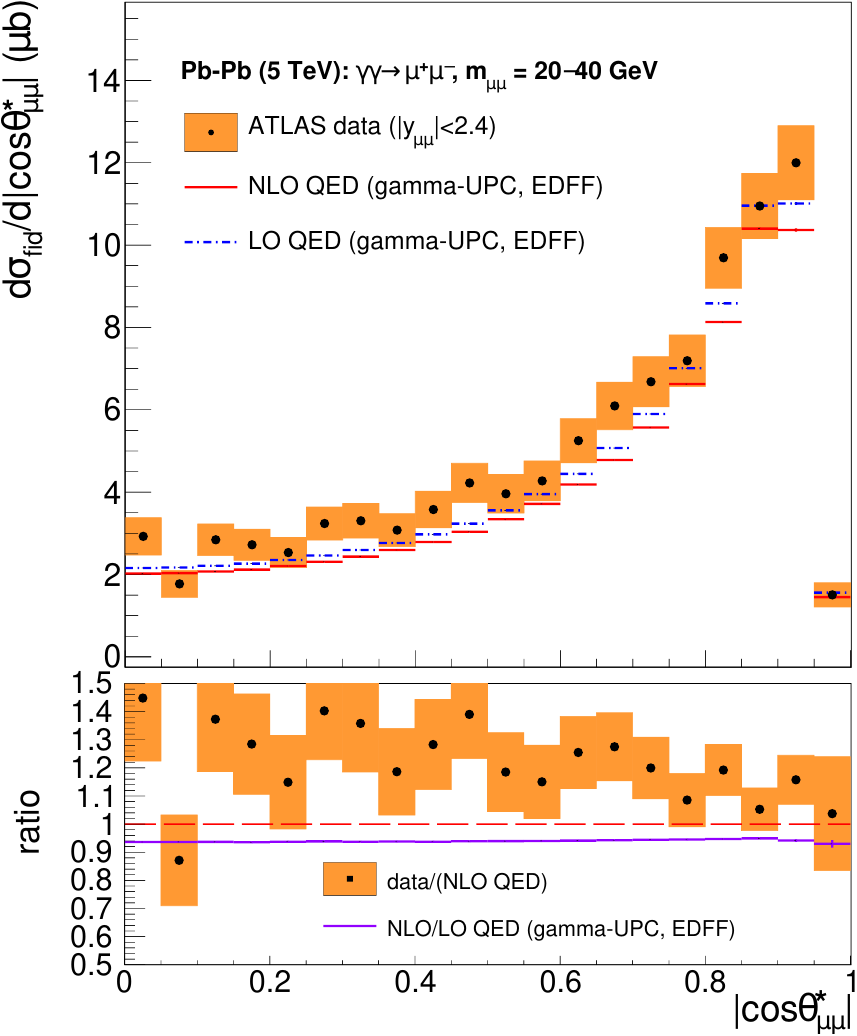}
\includegraphics[width=0.33\textwidth]{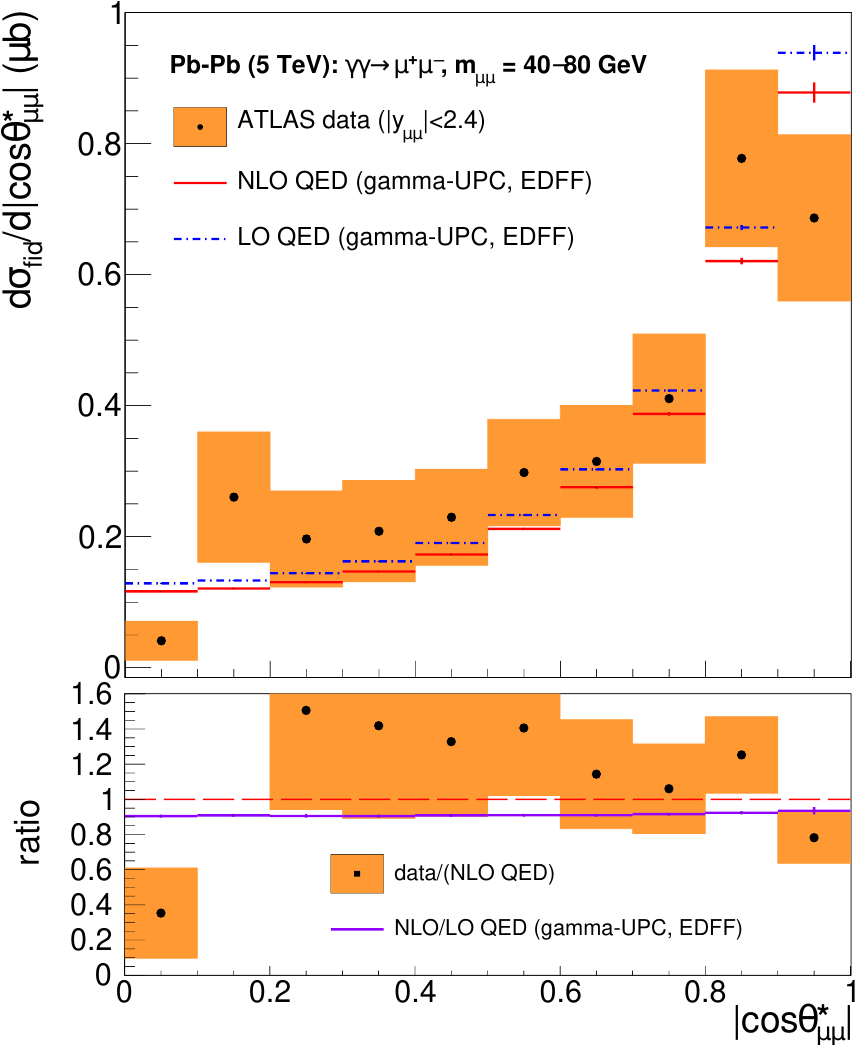}
\caption{Differential cross section as a function of the cosine of the scattering angle in the rest frame of the dimuon system, $\mathrm{d}\sigma/\mathrm{d}|\cos{\theta^\star_{\mm}}|$, for exclusive dimuon production in \PbPb\ UPCs at $\sqrtsnn=5.02$~TeV (over $|y_{\mu\mu}|<2.4$). The data (black dots with orange uncertainties)~\cite{ATLAS:2020epq} are compared with LO (blue dashed histogram) and NLO (red histogram) QED \gammaUPC\ predictions with ChFF (top) and EDFF (bottom) $\gamma$ fluxes. The lower panels of each plot show the corresponding data/NLO ratios (black dots with orange uncertainties) and NLO/LO $K$ factors (violet histogram).
\label{fig:dimuondcosth}}
\end{figure}

The last differential cross section studied is as a function of the scattering angle $\theta^\star_{\mm}$ of a muon in the rest frame of the dimuon system. The absolute value of the cosine of $\theta^\star_{\mm}$ can be obtained from the rapidity separation of the two muons via the relationship $|\cos{\theta^\star_{\mm}}|=\tanh{\left(\Delta y/2\right)}$, where $\Delta y=y^{\,\mu^+}-y^{\,\mu^-}$. Figure~\ref{fig:dimuondcosth} shows this angular distribution over the full pair rapidity coverage ($|y_{\mm}|<2.4$) in three dimuon invariant mass windows: $10<m_{\mm}<20$~GeV, $20<m_{\mm}<40$~GeV, and $40<m_{\mm}<80$~GeV, from left to right. The rapid depletion of the distribution approaching $|\cos{\theta^\star_{\mm}}|=1$ is due to the experimental fiducial cuts, in particular on $\eta^{\,\mu}$, that remove events with the muons flying along the beam direction. As found in previous differential cross sections, the NLO ChFF predictions feature a much better agreement with the data than the NLO EDFF ones, as indicated by the data/NLO ratios for the former centered at unity in all bottom panels.
%NLO ChFF exhibits $\chi^2/N_\mathrm{dof}=1.68$ and $1.27$ in the full rapidity and central rapidity regions respectively, while the corresponding two numbers for NLO EDFF are $1.78$ and $1.33$. The inclusion of the NLO QED corrections has also lowered $\chi^2/N_\mathrm{dof}$ a bit. 
The $K$ factor has nontrivial dependencies on both $m_{\mm}$ and $|\cos{\theta^\star_{\mm}}|$. In the lowest mass window, the $K$ factor decreases from 0.96 at $|\cos{\theta^\star_{\mm}}|=0$ to 0.88 at $|\cos{\theta^\star_{\mm}}|\approx 0.92$. The experimental fiducial criteria remove the population of events in the region of $|\cos{\theta^\star_{\mm}}|\gtrsim 0.92$. In the $20<m_{\mm}<40$~GeV range, the $K$ factor is quite flat as a function of $|\cos{\theta^\star_{\mm}}|$ up to $|\cos{\theta^\star_{\mm}}|\gtrsim 0.8$. Finally, in the highest mass regime, the $K$ factor increases from 0.9 at $|\cos{\theta^\star_{\mm}}|=0$ to 0.93 for $|\cos{\theta^\star_{\mm}}|>0.95$. Such a nontrivial kinematic dependence calls for a rigorous NLO QED computation as presented in this paper for any more precise future studies of the $\theta^\star_{\mm}$ scattering angle in data.\\

\begin{figure}[htbp!]
\centering
\includegraphics[width=0.45\textwidth]{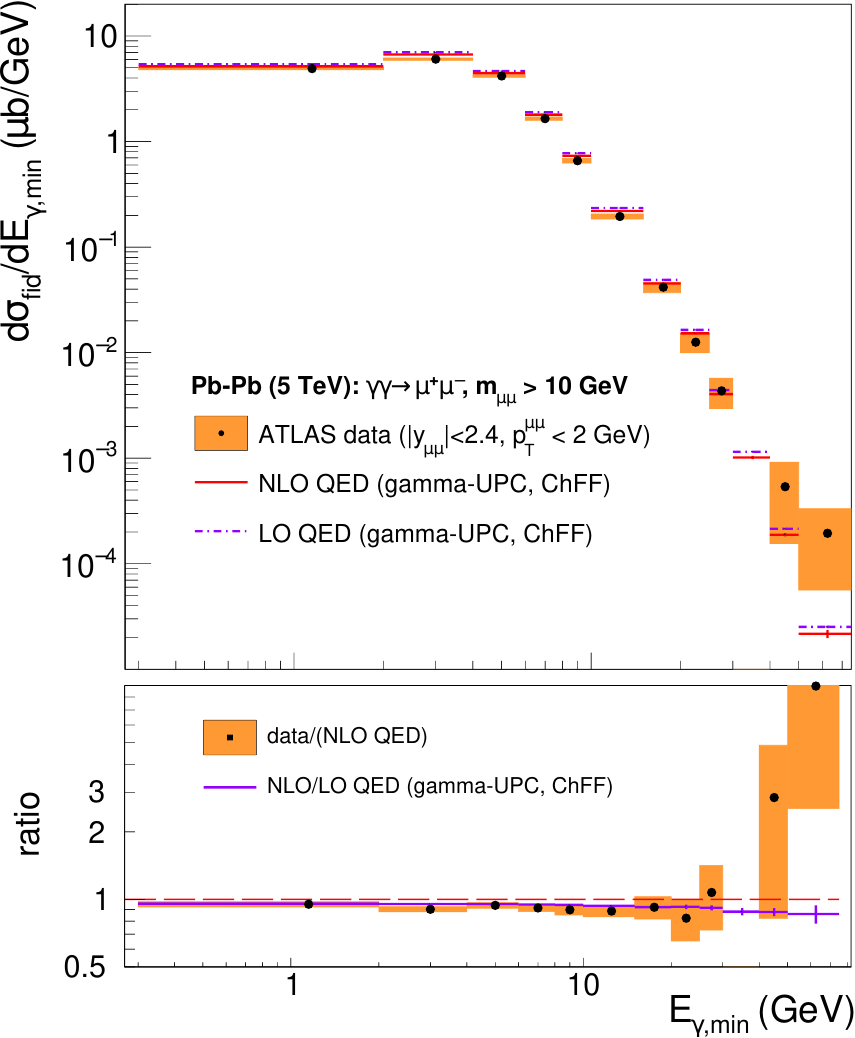}\hspace{0.4cm}
\includegraphics[width=0.45\textwidth]{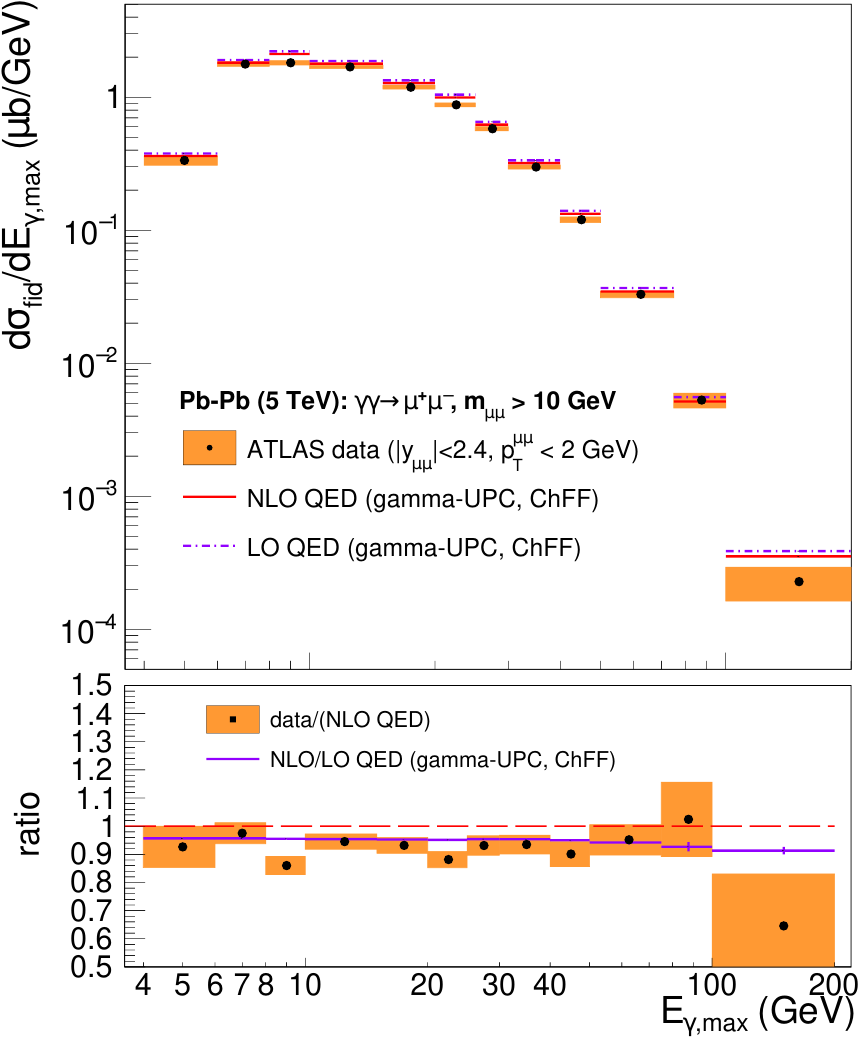}
\includegraphics[width=0.45\textwidth]{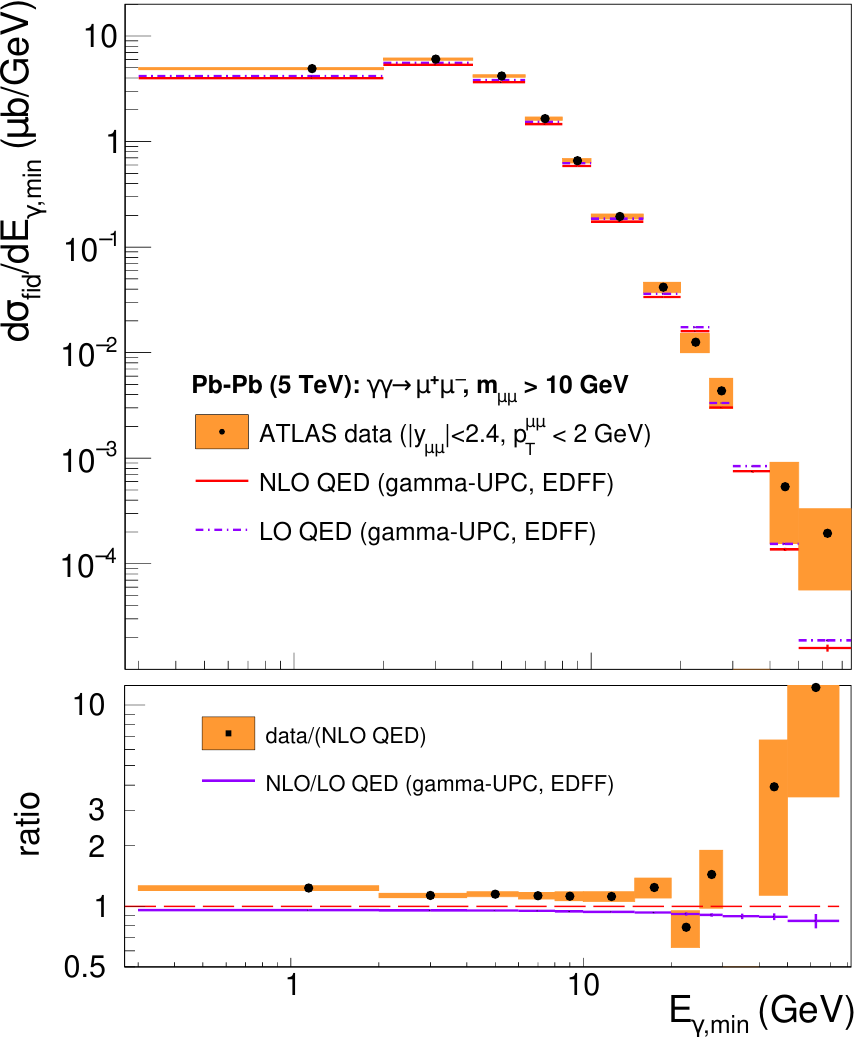}\hspace{0.4cm}
\includegraphics[width=0.45\textwidth]{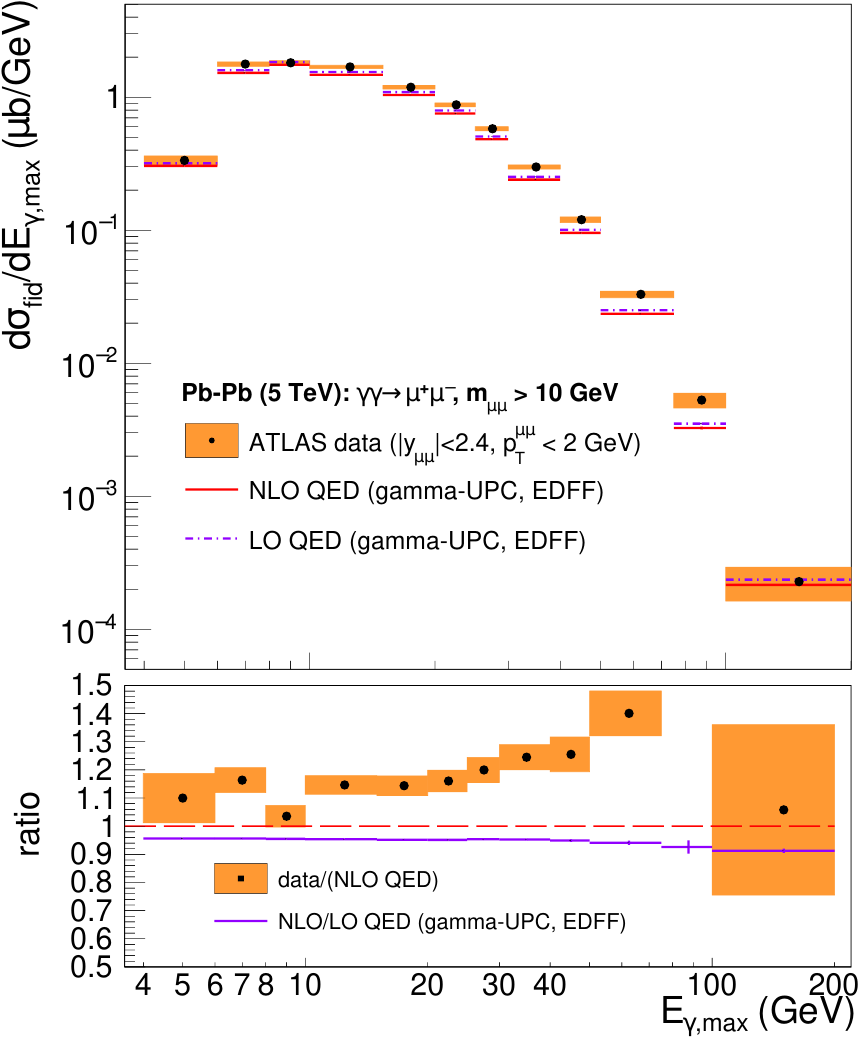}
\caption{Differential cross section as a function of the minimum ($\mathrm{d}\sigma/dE_{\gamma,\mathrm{min}}$, left) and maximum ($\mathrm{d}\sigma/dE_{\gamma,\mathrm{max}}$, right) energy of the initial photon in exclusive dimuon production in \PbPb\ UPCs at $\sqrtsnn=5.02$~TeV. The data (black dots with orange uncertainties)~\cite{ATLAS:2020epq} are compared with LO (blue dashed histogram) and NLO (red histogram) QED \gammaUPC\ predictions with ChFF (top) and EDFF (bottom) $\gamma$ fluxes. The lower panels of each plot show the corresponding data/NLO ratios (black dots with orange uncertainties) and NLO/LO $K$ factors (violet histogram).
\label{fig:dimuondE}}
\end{figure}

A more direct study of the EPA photon flux can be performed by inferring the energies $E_{\gamma,1}$ and $E_{\gamma,2}$ of the two initial photons in the process $\gaga\to\mumu$ from the four-momenta of the final-state muons and energy conservation. At LO accuracy, they can be obtained from the dimuon mass and rapidity as $E_{\gamma,1}=(m_{\mm}/2)\exp{(+y_{\mm})}$ and $E_{\gamma,2}=(m_{\mm}/2)\exp{(-y_{\mm})}$, respectively. At NLO accuracy, both expressions are not fully exact but remain useful for most of the cases. %As been done in the ATLAS analysis, we use the above two equations for our definitions of $E_{\gamma,1}$ and $E_{\gamma,2}$ even beyond LO. 
Following the ATLAS analysis, we define the maximum and minimum photon energies in an event as $E_{\gamma,\mathrm{max}}=\mathrm{max}(E_{\gamma,1},E_{\gamma,2})$ and  $E_{\gamma,\mathrm{min}}=\mathrm{min}(E_{\gamma,1},E_{\gamma,2})$, respectively. The distributions of $E_{\gamma,\mathrm{min}}$ and $E_{\gamma,\mathrm{max}}$ measured in data are shown in Fig.~\ref{fig:dimuondE} left and right, respectively, compared to our theoretical LO and NLO predictions obtained with the ChFF (top) and EDFF (bottom) photon fluxes. The measured $E_{\gamma,\mathrm{min,max}}$ distributions are well reproduced by the NLO ChFF calculations except in the last bin\footnote{Note that the similar data-theory comparison shown at LO accuracy in our previous work~\cite{Shao:2022cly} (Fig.~10 left of the paper) is incorrect in the first $E_{\gamma,\mathrm{min}}$ bin, as we had wrongly normalized the distribution by a 2-GeV (instead of 1.7~GeV) bin size.} %This is why the obtained $\chi^2$ values in Ref.~\cite{Shao:2022cly} are much larger than here.} 
where the theory underestimates the data although the latter suffers from large statistical uncertainties. %With regards to the $E_{\gamma,\mathrm{max}}$ distribution  (Fig.~\ref{fig:dimuondE}, right), the NLO ChFF result has very good 
%The level of agreement with the data is quantified by a goodness-of-fit $\chi^2/N_\mathrm{dof}=0.9$.
On the other hand, the calculations with the EDFF $\gamma$ flux undershoot both initial $E_{\gamma,\mathrm{min}}$ and $E_{\gamma,\mathrm{max}}$ photon energy distributions over all their measured ranges, as observed already at LO accuracy in the comparison of the ATLAS data to the \starlight\ MC predictions~\cite{ATLAS:2020epq}, which are based on this same photon flux.  %On the other hand, a significant deviation between the NLO EDFF and the data can be observed in the tail of the spectrum, which results into a large $\chi^2/N_\mathrm{dof}=3.49$ value. 
The $K$ factors for both $E_{\gamma,\mathrm{min}}$ and $E_{\gamma,\mathrm{max}}$ spectra, shown in the lower panels of both plots, feature a monotonic decrease with energy, from $K\approx 0.95$ down to 0.85 and 0.91 respectively, due again to the effects of the quasi-collinear FSR emission.\\

In summary, the detailed data-theory comparisons presented here for four differential observables measured in $\gaga\to\mumu$ processes in \pp\ and \PbPb\ UPCs at the LHC, confirm that the NLO QED calculations with the ChFF photon flux can accurately describe them %the $\gaga\to\mumu$ measurements 
in all kinematic regions of phase space probed. The measured dimuon invariant masses (Figs.~\ref{fig:dimuondMpp13TeV} and~\ref{fig:dimuondM}), absolute pair rapidity (Fig.~\ref{fig:dimuondy}), $|\cos{\theta^\star_{\mm}}|$ (Fig.~\ref{fig:dimuondcosth}) and the $E_{\gamma,\mathrm{min,max}}$ (Fig.~\ref{fig:dimuondE}) distributions agree nicely with the NLO predictions obtained with the ChFF $\gamma$ flux, whereas the calculations with the EDFF flux systematically undershoot the experimental results in all regions of phase space.

%\clearpage

%%%%%%%%%%%%%%%%%%%%%%%%%%%%%%%%%%%%%%%%%%%%%%%%%%%%%%%%%%%%%%%%%%%%%%%%%%%%%%%%%%%%
\section{Two-photon ditau production at NLO in QED}
\label{sec:ditau}

Let us start by considering $\gaga\to\tautau$ collisions at $\epem$ colliders where the process was first studied. More concretely, we consider the measurement of exclusive ditau in photon-photon collisions performed by the DELPHI experiment from which the previous best limits on $(g-2)_{\tau}$ had been derived~\cite{DELPHI:2003nah}. In this study, the extraction of $(g-2)_{\tau}$ from the measured $\gaga\to\tautau$ cross section %carried out in the original DELPHI analysis 
was performed with a LO EPA calculation, with freely floating $(g-2)_{\tau}$ values fitted to the data, where the upper limit of the integration over 4-momenta of the emitted photons ($Q^2_\mathrm{max}$) was chosen such that the fiducial cross section predicted by the EPA (for the SM value of the anomalous magnetic moment) agreed with the more theoretically accurate \textsc{radcor} prediction~\cite{Berends:1986if}. This latter code calculated the process at LO QED accuracy including FSR corrections on the electron and positron lines (second diagram in Fig.~\ref{fig:nlo_diags}). We compute here the cross sections for $\epem \xrightarrow[]{\gaga} \epem\tautau$ at LO and NLO accuracy within our setup using the default Weizs\"acker-Williams EPA photon flux of \mgshort. Our cross section calculations implement the original fiducial selection criteria for the signal, mostly applied to remove highly virtual $\gaga$ processes: $m_{\tata}< 40$~GeV, $\theta^{\rm e+}<10^\circ$, $\theta^{\rm e-} < 2^\circ$ (where $\theta$ is the angle between incoming and outgoing $\epem$, ranging from 0 to 180$^\circ$) for the various \cm\ energies considered in the DELPHI measurement: $\sqrts = 182.7,\, 188.7,\, 197.6,\,$ and 206.3~GeV. Within this phase space, and limiting the maximum virtuality of the colliding photons to values $Q^2_\mathrm{max}\approx (10$--15~GeV)$^2$ as in the original EPA approach employed by the DELPHI analysis, we obtain cross sections consistent with the experimental collision-energy/luminosity-weighted value of $\sigma(\gaga\to\tautau) = 429 \pm 17$~pb. Interestingly, we find that the NLO corrections are positive for this process and augment the LO cross sections by a factor of about 0.9\%. Although the DELPHI experimental uncertainty of 4\% is larger than the size of the NLO corrections, and though probably a leading fraction of the latter were already accounted for in the original LO$\,+\,$FSR \textsc{radcor} calculations, it would have been more theoretically accurate to go beyond the real radiative corrections and take also into account any potential cancellations between NLO real and virtual terms as considered here. This result underscores the importance of accounting for higher-order QED terms for future precise extractions of $(g-2)_{\tau}$ via $\gaga\to\tautau$ measurements at $\epem$ colliders such as at Belle-II~\cite{Belle-II:2018jsg} and, in the longer term, FCC-ee~\cite{FCC:2018evy,Dam:2018rfz}, with expected subpercent or even subpermil experimental uncertainties.\\
%In addition what DELPHI did to compute (g-2)tau was to take LO QED modifying $Q^2$ until matching the RADCOR prediction for sigma it is not one-to-one-correspondence: $Q^2 = 4*E*(1-x) Sin[theta/2]^2$ It means that a cut on theta will yield on cut on $Q^2$ but depending on x if I apply theta cut, we need to change Qmax2 according to x then in iWW at x=0.01, $Q2max=251 GeV^2$. but at 2 degree, it goes to $10 GeV^2$. Now, I get a large cross section: 5.308e+02 pb at 182.7 GeV with Q2max=251 GeV$^2$ %without any cuts at all (Qmax = 200^2 GeV^2)
%for theta(e+)<10 deg, theta(e-)< 2 deg, I get 4.709e+02 pb
%for theta(e-)<10 deg, theta(e+)< 2 deg, you get the same
%If both are theta(e+,e-)<2 degree, I got 4.112e+02 pb
%those are the RADCOR values for the sigma(gamma-gamma->tautau) fiducial cross section %(m(tautau)< 40 GeV, theta_e+<10deg, theta_e- < 2 deg) at LEP:
%sqrt(s) sigma:
%182.7 GeV 428.2 ± 0.5 pb
%188.7 GeV 436.7 ± 0.5 pb
%197.6 GeV 448.5 ± 0.5 pb
%206.3 GeV 459.4 ± 0.5 pb 
%In fact, the NLO correction is positive for this: LO: 4.0754765e+02, NLO: 4.1115395e+02

To date, there exist two measurements of $\gaga\to\tautau$ production cross sections in UPCs at the LHC. First, at low ditau invariant mass ($m_{\tata} \gtrsim 5$~GeV) using PbPb data at 5.02~TeV~\cite{CMS:2022arf} and, secondly, at higher masses ($m_{\tata}>50$~GeV) in p-p collisions at 13~TeV~\cite{CMS:2024skm}.
%within the multiple fiducial criteria applied to the leptonic and hadronic decay products of both tau leptons 
The measured fiducial cross sections have, however, still large  uncertainties in the 16\%--30\% range. We have transformed the experimental fiducial cross sections (including convoluted $\eta,\pT$ acceptance criteria applied at the detector-level to the different leptonic and hadronic decay products of both tau leptons) into more inclusive cross sections that are simpler to compare to the corresponding \gammaUPC\ predictions at LO and NLO QED accuracy with ChFF and EDFF photon fluxes. The experimental and theoretical results are listed in Table~\ref{tab:xsecsditau}. Based on the data--theory comparisons of Ref.~\cite{CMS:2022arf}, we find that the extrapolation of the measured \PbPb\ fiducial result, $\sigma^\mathrm{CMS}_\mathrm{fid}(\gaga\to\tautau) = 4.8 \pm 0.8$~fb, to a fully inclusive cross section depends on whether one uses the calculations of Ref.~\cite{Beresford:2019gww} or~\cite{Dyndal:2020yen}. In the former case, one obtains $570 \pm 100$~$\mu$b, whereas in the latter case the result is $850 \pm 150$~$\mu$b. Given the relatively poor experimental precision and the additional apparent extrapolation uncertainty, we quote the range span by both values. %and anticipate that upcoming more accurate data will solve this apparent discrepancy.
In addition, we provide predictions for $\gaga\to\tautau$ in \pp\ collisions at $\sqrts = 13.6$~TeV in the phase space region of very high masses ($m_{\tata}>300$~GeV) where both protons can be tagged by the forward near-beam spectrometers~\cite{CMS:2018uvs}.

\begin{table}[htpb!]
\centering
\tabcolsep=2.5mm
\caption{Exclusive ditau cross sections in \PbPb\ UPCs at $\sqrtsnn=5.02$~TeV and in p-p collisions at $\sqrts = 13$~TeV ($m_{\tata}>50$~GeV) and~13.6~TeV ($m_{\tata}>500$~GeV) experimentally measured and obtained with \gammaUPC\ at LO and NLO QED accuracy using the EDFF and ChFF $\gamma$ fluxes. The first column lists the experimental measurements of Refs.~\cite{CMS:2022hly,CMS:2024skm} extrapolated to the broader phase space considered in the calculations.
The last column lists the corresponding NLO-over-LO $K$ factors.
\label{tab:xsecsditau}}
\vspace{0.2cm}
\begin{tabular}{l|c|c|c|c} \hline%\hline
%$\gaga\to\tautau$ & Measured $\sigma^\mathrm{data}$ & {\gammaUPC~$\sigma^\mathrm{LO}$} & {\gammaUPC~$\sigma^\mathrm{NLO}$} & K factor \\%\hline
%Colliding system (approx. fiducial cuts)  & (exact fiducial cuts) & ChFF (EDFF) & ChFF (EDFF) &  $\sigma^\mathrm{NLO}/\sigma^\mathrm{LO}$  \\\hline
%\PbPb\ at 5.02 TeV (inclusive) & $4.8 \pm 0.8$~$\mu$b~\cite{CMS:2022hly} & $1062$ $(859.4)$~$\mu$b & $1072$ $(867.6)$~$\mu$b & $1.010$ \\
%\pp\ at 13 TeV ($m_{\tata}=50$--300~GeV, $|\eta^\tau|< 2.1$, $\pT^\tau>60$~GeV)  & $4.6_{-1.1}^{+1.4}$~fb~\cite{CMS:2024skm} & $6.84$ $(5.50)$~fb & $6.57$ $(5.28)$~fb  & $0.960$ \\
%\pp\ at 13.6 TeV ($m_{\tata}>300$~GeV, $|\eta^\tau|< 2.5$)    & -- & $1.350$ $(1.015)$~fb & $1.305$ $(0.981)$~fb  &  $0.966$ \\\hline
$\gaga\to\tautau$ & Measured $\sigma^\mathrm{data}$ & {\gammaUPC~$\sigma^\mathrm{LO}$} & {\gammaUPC~$\sigma^\mathrm{NLO}$} & $K$ factor \\%\hline
Colliding system (kinematic cuts)  & (extrapolated) & ChFF (EDFF) & ChFF (EDFF) &  $\sigma^\mathrm{NLO}/\sigma^\mathrm{LO}$  \\\hline
\PbPb\ at 5.02 TeV (inclusive)     & 580--850~$\mu$b & $1060$ $(860)$~$\mu$b & $1070$ $(870)$~$\mu$b & $1.01$ \\
\pp\ at 13 TeV ($m_{\tata}>50$~GeV)  & $855_{-215}^{+260}$~fb & $900$ $(730)$~fb & $895$ $(725)$~fb  & $0.995$ \\
\pp\ at 13.6 TeV ($m_{\tata}>300$~GeV, $|y^\tau|< 2.5$)    & -- & $1.35$ $(1.02)$~fb & $1.31$ $(0.98)$~fb  &  $0.965$ \\\hline
\end{tabular}
\end{table}
%$12.4_{-3.1}^{+3.8}$~fb~\cite{CMS:2024skm}

On the one hand, as similarly found for the $\gaga\to\mumu$ case (Table~\ref{tab:xsecsdimuonpp}), the theoretical cross sections derived with the ChFF fluxes are about 20\% (30\%) larger than those obtained with the EDFF ones in \PbPb\ (\pp) UPCs. The difference between the cross sections obtained with both photon fluxes rises with ditau mass because the ChFF spectrum becomes harder than the EDFF one for increasing $E_\gamma$ energies~\cite{Shao:2022cly}. On the other hand, the impact of the NLO corrections on the theoretical cross sections is only of about $+1\%$ at the low $m_{\tata}$ masses probed in PbPb UPCs, about $-0.5\%$ at the intermediate $m_{\tata}$ probed in \pp\ UPCs, whereas at high masses it is more significant and leads to a reduction by about $-3.5\%$ of the LO $\sigma(\gaga\to\tautau)$ predictions. We note that for very large ditau masses, weak boson corrections~\cite{Godunov:2023myj,Demirci:2021zwz} should be of the same size (\ie\ percent level) as the QED corrections and, therefore, would need to be also taken into account, but we leave their calculation for future work. Unfortunately, the current accuracy and precision of the experimental fiducial cross sections precludes any detailed conclusion on the level of data--theory agreement, but one should expect that the outcome of the more precise total and differential $\gaga\to\mumu$ cross section studies discussed in the previous section, \ie\ that ChFF-plus-NLO constitutes the state-of-the-art prediction, should hold also for the $\gaga\to\tautau$ process.\\

\begin{figure}[htbp!]
\centering
\includegraphics[width=0.49\textwidth]{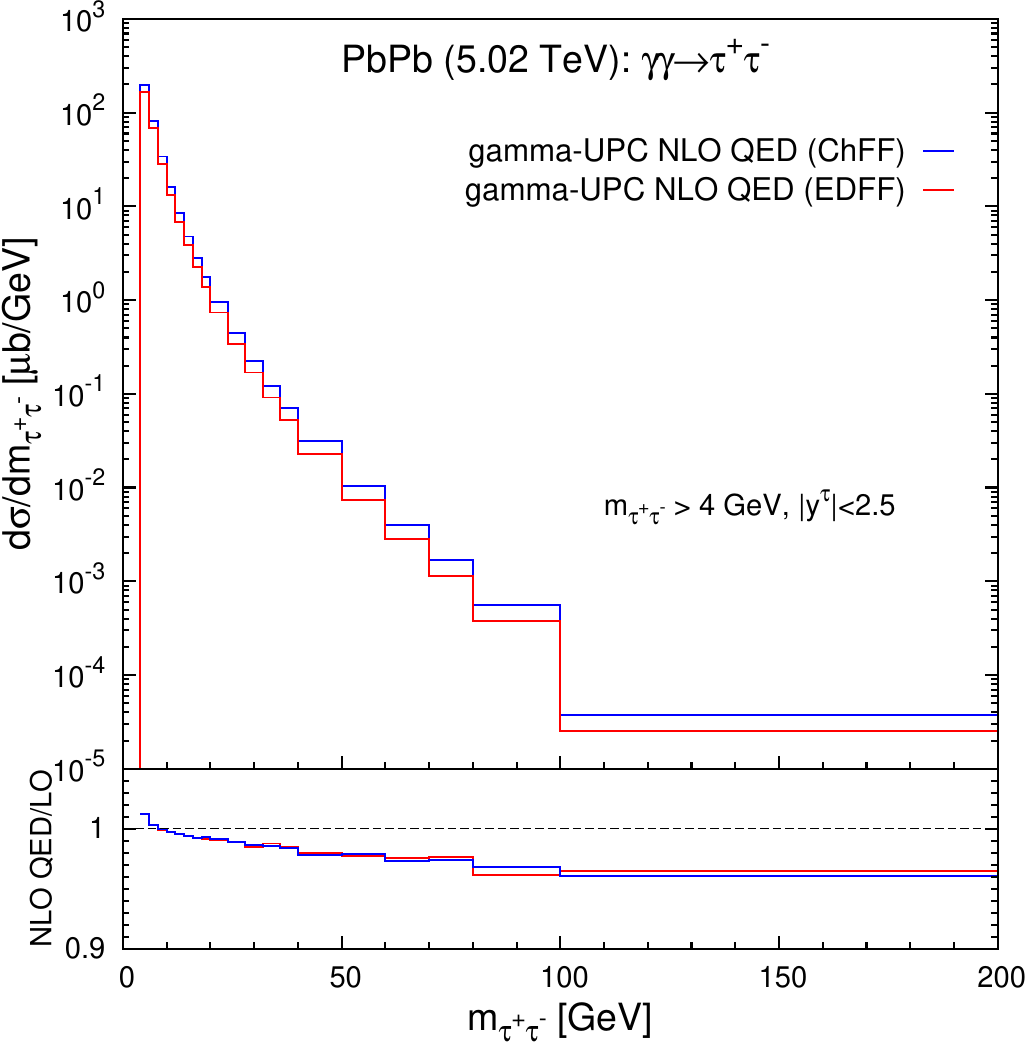}
\includegraphics[width=0.49\textwidth]{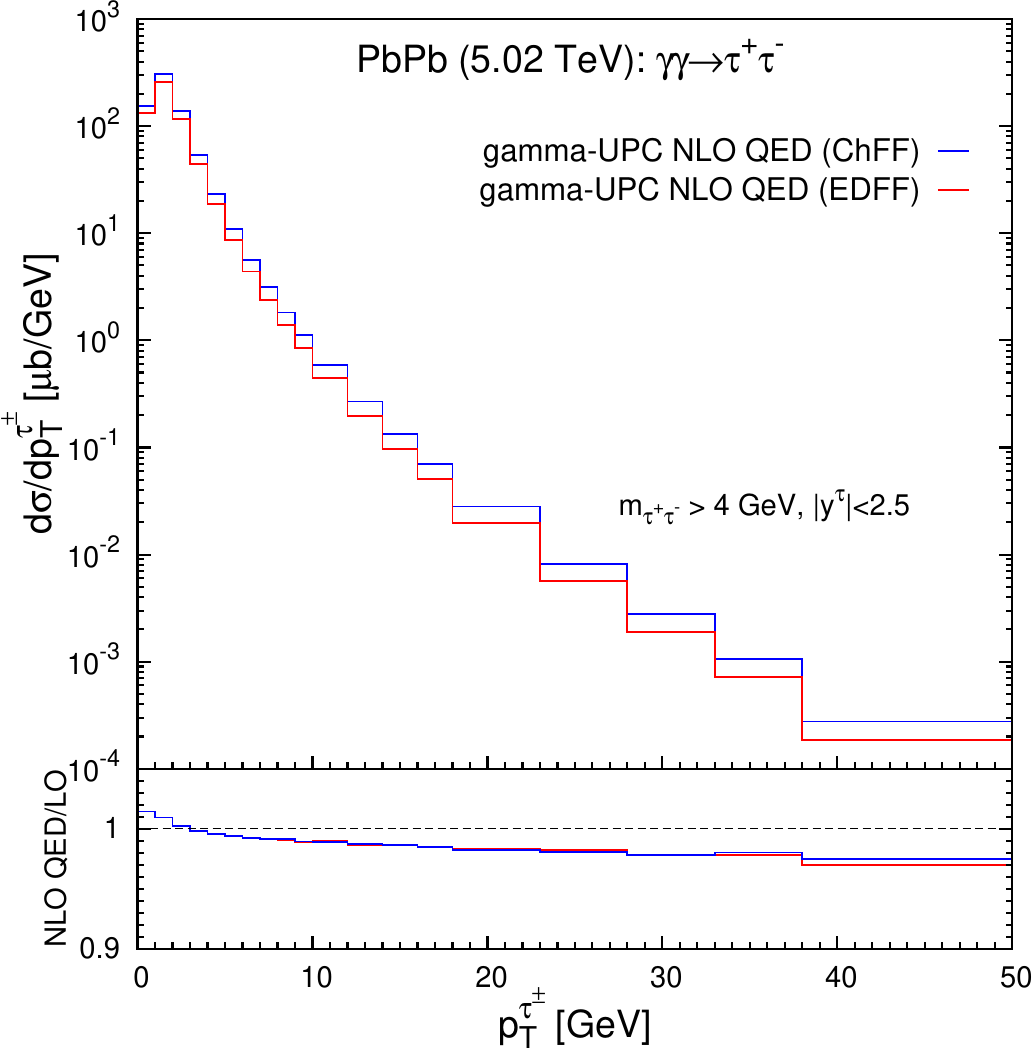}
\caption{Predicted $\gaga\to\tautau$ differential cross sections at NLO QED accuracy as a function of ditau mass $\mathrm{d}\sigma/\mathrm{d}m_{\tata}$ (left) and $\mathrm{d}\sigma/\mathrm{d}\pT^{\tau}$ (right) in \PbPb\ UPCs at $\sqrtsnn=5.02$~TeV  (for $m_{\tata} > 4$~GeV, $|y^\tau|< 2.5$) using the ChFF (red) and EDFF (blue) photon fluxes. The lower plots show the corresponding $\mathrm{d}\sigma^\mathrm{NLO}/\mathrm{d}\sigma^\mathrm{LO}$ $K$ factors.
\label{fig:ditauPbPb}}
\end{figure}

\begin{figure}[htbp!]
\centering
\includegraphics[width=0.49\textwidth]{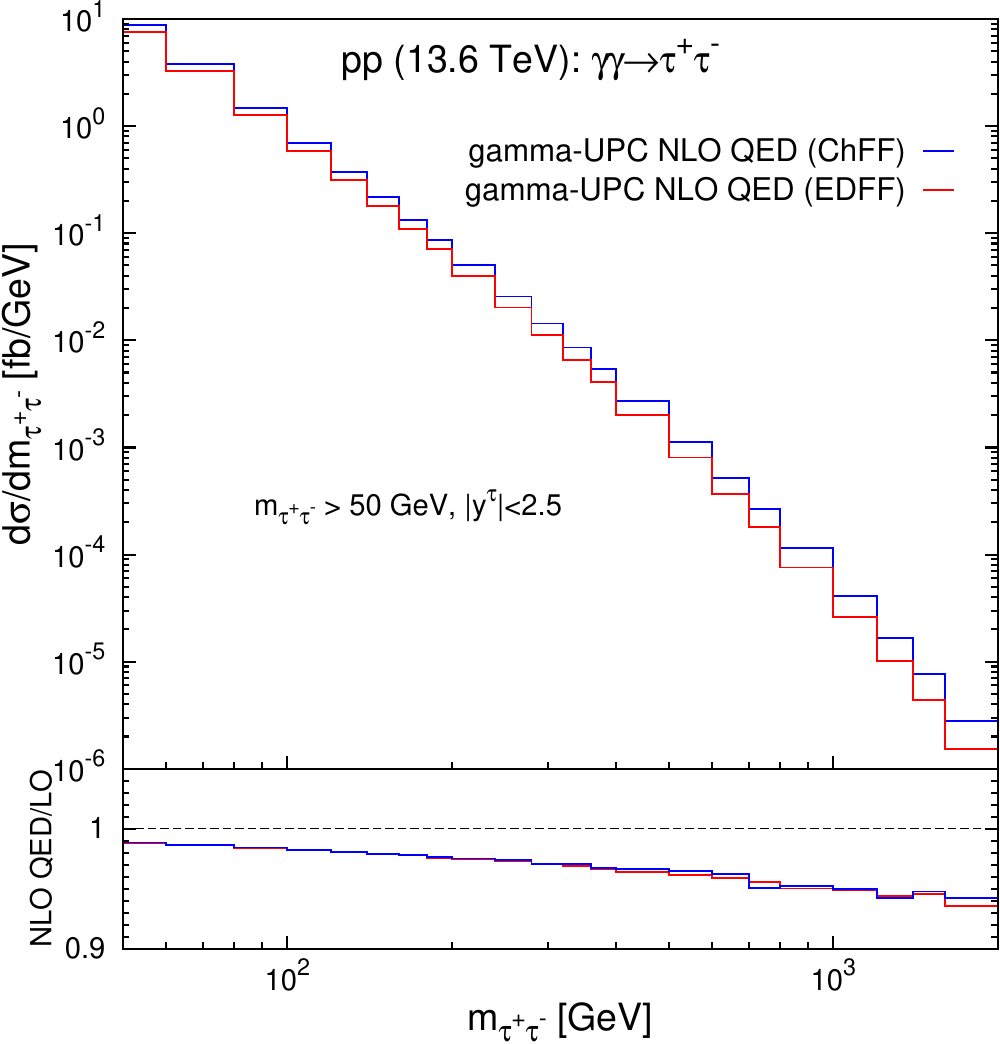}
\includegraphics[width=0.50\textwidth]{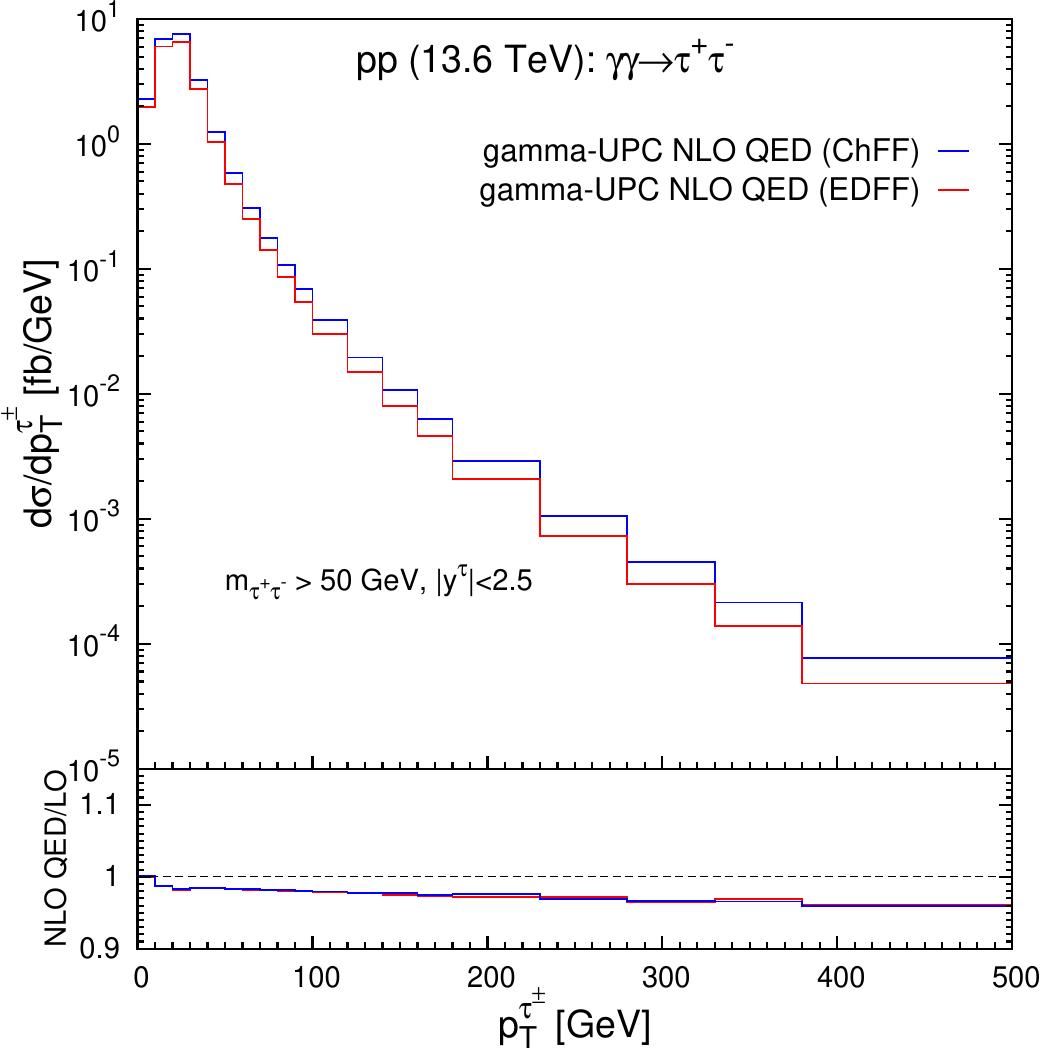}
\caption{Predicted $\gaga\to\tautau$ differential cross sections at NLO QED accuracy as a function of ditau mass $\mathrm{d}\sigma/\mathrm{d}m_{\tata}$ (left) and $\mathrm{d}\sigma/\mathrm{d}\pT^{\tau}$ (right) in \pp\ UPCs at $\sqrts=13.6$~TeV (for $m_{\tata}>50$~GeV, $|y^\tau|< 2.5$) using the ChFF (red) and EDFF (blue) photon fluxes. The lower plots show the corresponding $\mathrm{d}\sigma^\mathrm{NLO}/\mathrm{d}\sigma^\mathrm{LO}$ $K$ factors.
\label{fig:ditaupp}}
\end{figure}

The impact of NLO corrections is more visible in the $\gaga\to\tautau$ differential cross sections as a function of pair invariant mass $m_{\tata}$ and tau $\pT$ as shown for \PbPb\ UPCs at $\sqrtsnn = 5.02$~TeV in Fig.~\ref{fig:ditauPbPb}, and for high-mass \pp\ UPCs at $\sqrtsnn = 13.6$~TeV in Fig.~\ref{fig:ditaupp}. One can see that although the integrated cross sections change by a few percent going from LO to NLO, the differential distributions show in general small enhancements in the $K$ factors at low ditau masses and $\pT^\tau$ values that are then compensated by larger reductions of the $K$-factor for increasing $m_{\tata}$ and tau $\pT$, by up to a factor of $-6\%$ in the tail of the distributions. This latter decrease at high masses is due to the emission of quasi-collinear FSR photons that ``moves'' the tau pair yields to comparatively lower $m_{\mm}$ values. Since both kinematic distributions are the most sensitive ones for the extraction of $(g-2)_\tau$ through comparison of the experimental data to the predictions, it is important to have a good theoretical control of their shape accounting for the NLO corrections presented here.

%%%%%%%%%%%%%%%%%%%%%%%%%%%%%%%%%%%%%%%%%%%%%%%%%%%%%%%%%%%%%%%%%%%%%%%%%%%%%%%%%%%%
\section{Summary}
\label{sec:summ}

We have presented theoretical calculations for the production cross sections of muon and tau pairs in photon-photon collisions, $\gaga\to\mumu,\tautau$, at next-to-leading (NLO) accuracy in quantum electrodynamics (QED). The total and differential cross sections have been computed in the equivalent photon approximation (EPA) implemented in the \gammaUPC$\,+\,$\madgraph\ codes. The theoretical predictions have been confronted to the existing exclusive $\gaga\to\mumu,\tautau$ data from ultraperipheral collisions (UPCs) of proton (p) and lead (Pb) ions at the LHC, as well as to results from $\epem$ collisions at LEP for the $\gaga\to\tautau$ process. For UPCs at the LHC, two types of EPA photon fluxes, as well as associated survival probabilities of the photon-emitting hadrons, based on the charge (ChFF) and electric-dipole (EDFF) form factors have been tested. We have first determined that the parametric uncertainties associated with the modeling of the survival probabilities of Pb ions and protons amount to about 0.5--2\% in the range of dilepton masses measured in UPCs at the LHC, and are smaller than the differences between the leading-order (LO) and next-to-leading-order (NLO) cross-section corrections calculated here. Three other main findings are worth highlighting. 

First, the NLO terms are found to modify the total fiducial cross sections by up to 5\%, as well as to increase the tails of the dilepton acoplanarity $A_{\phi}^{\lele}$ and transverse momentum $\pT^{\lele}$ distributions, with respect to the LO predictions including the effects of the (very small) virtuality of the colliding photons. Since the experimental measurements of the $\gaga\to\ellell$ processes are based on applying selection criteria on the $A_{\phi}^{\lele}$ and $\pT^{\lele}$ distributions, and/or fitting them to Monte Carlo templates, the improved NLO calculations have quantitative implications for the cross section determinations, which are being measured with increasing experimental precision (few percent uncertainties) at the LHC. The NLO corrections also modify the differential cross sections as a function of dilepton invariant masses $m_{\lele}$, rapidity $y_{\lele}$, and $|\cos{\theta^\star_{\lele}}|$. In general, the corrections are small (and, in some cases, positive) at low dilepton masses and lepton $\pT^\ell$, but become increasingly large (and negative) for higher $m_{\lele}$ and $\pT^{\ell}$ values, with up to a 15\% depletion observed in the tails of some kinematic distributions compared to the LO predictions. This latter behaviour is due to the emission of final-state-radiation photons that reduces the invariant masses of the lepton pairs following a $\alpha\ln{(m_{\ell}^2/m_{\lele}^2)}$ dependence. Those results will have implications for upcoming, more precise, experimental measurements of the anomalous magnetic moment of the tau lepton, $(g-2)_{\tau}$, via the $\gaga\to\tautau$ process at different $m_{\tautau}$ regions in UPCs at the LHC. 

Second, the detailed UPC data-theory comparisons clearly favour the ChFF photon fluxes over the EDFF ones, and the former should be used in the calculations when confronting the data to the predictions for accurate phenomenological interpretation. With the ChFF flux, one can also compute nonexclusive photon-photon collisions with proton or nucleus overlap at NLO accuracy, since there is no kinematic restriction on the photon fluxes as a function of impact parameter (at variance with the EDFF case). For the $\gaga\to\mumu$ processes, the distributions of dimuon masses and rapidities, $|\cos{\theta^\star_{\mm}}|$, and $E_{\gamma,\mathrm{min,max}}$ agree nicely with the NLO prediction obtained with the ChFF $\gamma$ flux, whereas the calculations with the EDFF flux systematically underestimate the experimental results in all regions of phase space.

Third, for photon-photon collisions at LEP, the NLO corrections for the $\gaga\to\tautau$ final state measured within the DELPHI phase space are positive and augment the LO cross sections by a factor of about 1\%. Although the 4\% LEP experimental uncertainty is larger than the size of the NLO modifications, it would have been more theoretically accurate to consider them in order to extract  $(g-2)_{\tau}$ from the data. Such considerations are of particular relevance for future precise determinations of $(g-2)_{\tau}$ via $\gaga\to\tautau$ measurements at $\epem$ colliders, such as at Belle-II and FCC-ee, with expected subpercent or even subpermil experimental uncertainties.

In summary, the inclusion of NLO QED terms, and the use of ChFF photon fluxes for hadron beams, in the calculations of $\gaga\to\ellell$ processes improves the data--theory agreement, and proves an indispensable ingredient for the extraction of precision quantities, such as $(g-2)_{\tau}$. Accounting for NLO QED corrections appears as a requisite in searches for deviations from the SM predictions in the absolute or differential $\gaga$ cross sections due to elusive new physics contributions in future UPC measurements at the LHC and at $\epem$ colliders. With the developments presented in this work, we have extended the number of photon-fusion physics processes calculable with the $\gammaUPC$ code at NLO QED (or QCD) theoretical accuracy including, to date, the exclusive production of muon and tau leptons, light-by-light scattering, and top-quark pairs.\\

%%%%%%%%%%%%%%%%%%%%%%%%%%%%%%%%%%%%%%%%%%%%%%%%%%%%%%%%%%%%%%%%%%%%%%%%%%%%%%%%%%%
%\vspace{0.2cm}

\paragraph*{Acknowledgments.---} %Discussions with ... are gratefully acknowledged. 
Support from the European Union's Horizon 2020 research and innovation program (grant agreement No.\ 824093, STRONG-2020, EU Virtual Access ``NLOAccess''), the ERC (grant 101041109 `BOSON'), the French ANR (grant ANR-20-CE31-0015, ``PrecisOnium''), and the French LIA FCPPN, are acknowledged. Views and opinions expressed are however those of the authors only and do not necessarily reflect those of the European Union or the European Research Council Executive Agency. Neither the European Union nor the granting authority can be held responsible for them. For the purpose of Open Access, a CC-BY public copyright license has been applied by the authors to the present document and will be applied to all subsequent versions up to the Author Accepted Manuscript arising from this submission.

%%%%%%%%%%%%%%%%%%%%%%%%%%%%%%%%%%%%%%%%%%%%%%%%%%%%%%%%%%%%%%%%%%%%%%%%%%%%%%%%%%%%
%\clearpage

\bibliographystyle{myutphys}
\bibliography{references.bib}

%%%%%%%%%%%%%%%%%%%%%%%%%%%%%%%%%%%%%%%%%%%%%%%%%%%%%%%%%%%%%%%%%%%%%%%%%%%%%%%%%%%%
\clearpage

\appendix

\section{Initial photon virtualities in the \gammaUPC\ code}
\label{sec:kTsmear}

The virtuality dependence of the EPA photon fluxes emitted by a charge $Z$ is given by Eq.~\eqref{eq:photonnumdenkT}. Such an equation can be more simply written as a function of the photon virtuality $Q^2$ and longitudinal energy $E_\gamma$ (or, equivalently, fraction of the beam momentum $x_\gamma$ carried by the photon) alone, $N_{\gamma/Z}^{\mathrm{ChFF}}(x_\gamma, Q^2)$, using two similar expressions. The first expression is obtained by taking $Q^2\equiv k_\perp^2+\tfrac{E_\gamma^2}{\gamma_\mathrm{L}^2}=k_\perp^2+m_\mathrm{N}^2x_\gamma^2$ and $Q_\mathrm{min}^2=m_\mathrm{N}^2x_\gamma^2$, and reads
\begin{eqnarray}
\mathrm{d}N_{\gamma/Z}^{\mathrm{ChFF}}(x_\gamma, Q^2)&=&\frac{Z^2\alpha}{\pi}\frac{\mathrm{d}x_\gamma}{x_\gamma}\frac{\mathrm{d}Q^2}{Q^2} \left(1-\frac{Q^2_\mathrm{min}}{Q^2}\right)\left[F_{\mathrm{ch},A}(Q)\right]^2.\label{eq:dnChFFForm1}
\end{eqnarray}
The second form uses the exact expression for the virtuality, $Q^2=\tfrac{k_\perp^2+E_\gamma^2/\gamma_\mathrm{L}^2}{1-x_\gamma}=\tfrac{k_\perp^2+m_\mathrm{N}^2x_{\gamma}^2}{1-x_\gamma}$, and is obtained by replacing $\tfrac{E_\gamma^2}{\gamma_\mathrm{L}^2}=\left(x_\gamma m_\mathrm{N}\right)^2\overset{x_\gamma\ll 1}{\simeq} Q_\mathrm{min}^2\equiv \tfrac{m_\mathrm{N}^2x_\gamma^2}{1-x_\gamma}$, and defining $(1-x_\gamma)Q^2=k_\perp^2+\tfrac{E_\gamma^2}{\gamma_\mathrm{L}^2}$, thus $k_\perp^2=(1-x_\gamma)(Q^2-Q^2_\mathrm{min})$, which leads to
\begin{eqnarray}
\mathrm{d}N_{\gamma/Z}^{\mathrm{ChFF}}(x_\gamma, Q^2)&=&\frac{Z^2\alpha}{\pi}\frac{\mathrm{d}x_\gamma}{x_\gamma}\frac{\mathrm{d}Q^2}{Q^2} \left(1-\frac{Q^2_\mathrm{min}}{Q^2}\right)\left[F_{\mathrm{ch},A}\left(\sqrt{1-x_\gamma}Q\right)\right]^2.\label{eq:dnChFFForm2}
\end{eqnarray}
Both expressions are identical in the case $x_{\gamma}\ll 1$, which is fulfilled for most of the phenomenologically relevant phase space in $\gaga$ collisions at colliders. Indeed, a typical $\gaga\to X$ collision process with $m_X\approx 5$~GeV at midrapidity in PbPb collisions at the LHC probes $x_\gamma\approx 10^{-3}$ values, whereas the same production of a final state with $m_X\approx 100$~GeV (corresponding to $x_\gamma\approx 0.1$) has a cross section two orders-of-magnitude smaller~\cite{Shao:2022cly}.\\

As explained in Section~\ref{sec:th}, the matrix element calculations in \mgshort\ expect collinear (\ie\ $\kt$-independent) input photon distributions, and therefore when running \gammaUPC\,+\,\mgshort\ together to compute a given $\gaga$ process, the photon fluxes need to be fully integrated over $Q$ and $\kt$. In order to restore the full unintegrated dependencies of the photon fluxes, our \gammaUPC\ setup incorporates a small extra $\kt$ implemented by directly modifying the 4-momenta of the incoming photons and, correspondingly, outgoing produced particles in the Les Houches (\lhe) file output of the generated MC events. Such a ``$\kt$ smearing'' of the initial and final states is performed by running a python or Fortran script on the \lhe\ file that modifies the kinematics of each event as explained next. 

Let us consider a generic photon-photon collision process $\gamma_1\gamma_2\to X_3\cdots X_{n+2}$, with $n$ particles produced in the final state. First, the incoming photons transverse momentum $k_\perp \approx Q$ is sampled from the distribution given by Eq.~\eqref{eq:dnChFFForm1} or~\eqref{eq:dnChFFForm2} depending on the choice of the \verb|ion_Form| setting of the script, with uniform azimuthal plane dependence. Then, the kinematics of the final particles is appropriately reshuffled to respect momentum conservation and the on-mass-shell conditions. Technically, this is implemented as in Ref.~\cite{Frixione:2019fxg} where the 4-momenta of the two initial photons, in the \cm\ frame of the two initial colliding hadrons, are written as a function of the fractional momentum $x_\gamma\equiv x$ (for simplicity, we omit the $\gamma$ subindex in the following) carried out by each photon,
\begin{eqnarray}
k_1&=&\frac{\sqrt{\snn}}{2}x_1\left(1,0,0,1\right),\quad k_2=\frac{\sqrt{\snn}}{2}x_2\left(1,0,0,-1\right).
\end{eqnarray}
The 4-momenta of the final particles, $k_3,\ldots, k_{n+2}$, are then required to satisfy the momentum conservation $k_1+k_2=\sum_{j=3}^{n+2}{k_j}$ and onshell conditions $k_i^2=m_i^2$, where $m_1=m_2=0$. After the momentum reshuffling, we have
\begin{eqnarray}
k_i&\to&\bar{k}_i,\quad \bar{k}_1^2=\bar{m}_1^2=-Q_1^2,\quad \bar{k}_2^2=\bar{m}_2^2=-Q_2^2,\quad \bar{k}_j^2=m_j^2, j\geq 3.
\end{eqnarray}
The new momentum conservation is $\bar{k}_1+\bar{k}_2=\sum_{j=3}^{n+2}{\bar{k}_j}$.
There are two possible ways to proceed with the $k_\perp$ smearing of the collision process: either starting reshuffling the momenta of the initial photons, or modifying the kinematics of the final state:
\begin{itemize}
\item \textbf{Initial momentum reshuffling}:
This momentum reshuffling scheme applies for the $n\geq 1$ case, \ie\ for a photon-fusion process leading even to a single one-particle final state. The invariant mass of the final-state system is $\left(\sum_{j=3}^{n+2}{k_j}\right)^2=x_1x_2\snn$. Let us use
\begin{eqnarray}
\bar{k}_1&=&\left(\tfrac{\sqrt{\snn}}{2}\bar{x}_1,\vec{\bar{k}}_{\perp,1},\tfrac{\sqrt{\snn}}{2}\bar{x}_1\right),\quad \bar{k}_2=\left(\tfrac{\sqrt{\snn}}{2}\bar{x}_2,\vec{\bar{k}}_{\perp,2},-\tfrac{\sqrt{\snn}}{2}\bar{x}_2\right),
\end{eqnarray}
where $|\vec{\bar{k}}_{\perp,1}|=Q_1$ and $|\vec{\bar{k}}_{\perp,2}|=Q_2$. For simplicity, we denote $\vec{K}_{\perp}\equiv \vec{\bar{k}}_{\perp,1}+\vec{\bar{k}}_{\perp,2}$, where we know that $(Q_1-Q_2)^2\leq K_\perp^2\leq (Q_1+Q_2)^2$. Now, we must require $(\bar{k}_1+\bar{k}_2)^2=\snn\bar{x}_1\bar{x}_2-K_\perp^2=\snn x_1x_2$. Thus, $\bar{x}_1\bar{x}_2=x_1x_2+\tfrac{K_\perp^2}{\snn}$. If $K_\perp^2\leq \snn(1-x_1x_2)$, we can solve $\bar{x}_1$ and $\bar{x}_2$ by additionally requiring $\log{\tfrac{\bar{x}_1}{\bar{x}_2}}=\log{\tfrac{x_1}{x_2}}$. If there is a physical solution, this solution is $\bar{x}_1=\sqrt{\smash[b]{\tfrac{x_1}{x_2}\tfrac{K_\perp^2}{\snn}+x_1^2}}$ and $\bar{x}_2=\sqrt{\smash[b]{\tfrac{x_2}{x_1}\tfrac{K_\perp^2}{\snn}+x_2^2}}$. This gives us the final constraint on $K_\perp^2\leq \snn\left(\mathrm{min}(1,\tfrac{x_1}{x_2},\tfrac{x_2}{x_1})-x_1x_2\right)$. Otherwise, the values of $Q_1$ and $Q_2$ are regenerated. Lastly, the whole final system needs to be boosted from the frame $k_1+k_2$ to $\bar{k}_1+\bar{k}_2$. Since the fractional momentum $x$ changes, one can decide whether to apply the additional rescaling weight, $\frac{f_1(\bar{x}_1)f_2(\bar{x}_2)}{f_1(x_1)f_2(x_2)}$, or not.

\item \textbf{Final momentum reshuffling}:  We can only use this scheme for the $n\geq 2$ case, because we need to partition the final particles into two subsets. In this scheme, let us keep $x_1$ and $x_2$, \ie\,
\begin{eqnarray}
\bar{k}_1&=&\left(\tfrac{\sqrt{\snn}}{2}x_1,\vec{\bar{k}}_{\perp,1},\tfrac{\sqrt{\snn}}{2}x_1\right),\quad \bar{k}_2=\left(\tfrac{\sqrt{\snn}}{2}x_2,\vec{\bar{k}}_{\perp,2},-\tfrac{\sqrt{\snn}}{2}x_2\right)
\end{eqnarray}
with $|\vec{\bar{k}}_{\perp,1}|=Q_1$ and $|\vec{\bar{k}}_{\perp,2}|=Q_2$. Without loss of generality, we assume that the two subsets are $\left\{3,\ldots,{m+2}\right\}$ and $\left\{m+3,\ldots,n+2\right\}$, where $1\leq m\leq n-1$. Let us define $M_1^2\equiv K_1^2\equiv\left(\sum_{j=3}^{m+2}{k_j}\right)^2$ and $M_2^2\equiv K_2^2\equiv\left(\sum_{j=m+3}^{n+2}{k_j}\right)^2$, and keep the invariant masses $M_1$ and $M_2$ of the two final systems unmodified. In the case that $\bar{s}\equiv x_1x_2\snn-K_\perp^2\leq (M_1+M_2)^2$, we need to regenerate $Q_1$ and $Q_2$ and $\vec{K}_\perp=\vec{\bar{k}}_{\perp,1}+\vec{\bar{k}}_{\perp,2}$. Otherwise, in the partonic rest frame, the energies of the two subsystems are $\bar{K}_1^0=\varepsilon(\sqrt{\bar{s}},M_1,M_2)$ and $\bar{K}_2^0=\varepsilon(\sqrt{\bar{s}},M_2,M_1)$, where $\varepsilon(m,m_1,m_2)=\tfrac{m}{2}\left(1+\tfrac{m_1^2-m_2^2}{m^2}\right)$. The mass on-shell conditions enforce $\overrightarrow{\bar{K}}_1=\sqrt{\left(\bar{K}_1^0\right)^2-M_1^2}\tfrac{\overrightarrow{K}_1}{|\overrightarrow{K}_1|}$ and $\overrightarrow{\bar{K}}_2=\sqrt{\left(\bar{K}_2^0\right)^2-M_2^2}\tfrac{\overrightarrow{K}_2}{|\overrightarrow{K}_2|}$. Note that because $\sqrt{\left(\bar{K}_1^0\right)^2-M_1^2}=\sqrt{\left(\bar{K}_2^0\right)^2-M_2^2}$ and in the partonic rest frame $\tfrac{\overrightarrow{K}_1}{|\overrightarrow{K}_1|}=-\tfrac{\overrightarrow{K}_2}{|\overrightarrow{K}_2|}$, it follows that $\overrightarrow{\bar{K}}_1+\overrightarrow{\bar{K}}_2=\overrightarrow{0}$. For $k_j, 3\leq j\leq m+2$, the $\bar{k}_j$ is obtained by first boosting $k_j$ into the rest frame of $K_1$, and then boosting it back to the frame of $\bar{K}_1$. A similar method is applied to  obtain $\bar{k}_j$ (with $m+3\leq j\leq n+2$) from $k_j$. Finally, the momenta of the final particles are boosted from the $\left(\sqrt{\bar{s}},0,0,0\right)$ to the $\bar{k}_1+\bar{k}_2$ frame.
\end{itemize}
Since the initial-momentum reshuffling works out for the simplest production of a single resonance in two-photon collisions, $\gaga\to X_1$, this is the method currently implemented in our $\kt$ smearing script.\\

%\section{Merging of NLO and $\kt$ smearing}
%\label{sec:app}

The procedure to combine $\gaga\to\mathrm{X}$ kinematic distributions including LO$\,+\,\kt$ smearing %(LO+smear) 
and NLO corrections has some arbitrary freedom. The combined spectrum should follow closely the LO$\,+\,\kt$ result in the lowest values of the distributions, whereas the tails should match exactly the NLO result. We adopt a rather simple approach here to combine the intermediate region of the distributions where both contributions overlap. For a given kinematic distribution differential in $O$ generated in a given AA UPC, a weighted distribution is defined as follows
\begin{eqnarray}
\frac{\mathrm{d}\sigma^\mathrm{AA}_\mathrm{merged}}{\mathrm{d}O}&=&(1-w_\mathrm{AA}(O))\frac{\mathrm{d}\sigma^\mathrm{AA}_\mathrm{NLO}}{\mathrm{d}O}+w_\mathrm{AA}(O)\frac{\mathrm{d}\sigma^\mathrm{AA}_\mathrm{LO+sm}}{\mathrm{d}O},
\label{eq:A7}
\end{eqnarray}
where the weights for the dilepton variables $O = A_\phi,\,\pT^{\lele}$ are, respectively,
\begin{equation}
\begin{aligned}
w_\mathrm{PbPb}(A_{\phi})&=\frac{1}{1+(A_{\phi}/0.005)^4},\quad\quad\;\;\; w_\mathrm{pp}(A_{\phi})=\frac{1}{1+(A_{\phi}/0.03)^4},\\ %\nonumber\\
w_\mathrm{PbPb}(\pT^{\lele})&=\frac{1}{1+(\pT^{\lele}/0.01~\mathrm{GeV})^4},\quad w_\mathrm{pp}(\pT^{\lele})=\frac{1}{1+(\pT^{\lele}/0.7~\mathrm{GeV})^4}.
%w_\mathrm{PbPb}(A^{\mathrm{X}}_{\phi})&=&\frac{1}{1+(A^{\mathrm{X}}_{\phi}/0.005)^4},\quad w_\mathrm{pp}(A^{\mathrm{X}}_{\phi})=\frac{1}{1+(A^{\mathrm{X}}_{\phi}/0.03)^4},\nonumber\\
%w_\mathrm{PbPb}(\pT^{\lele}^{\mathrm{X}})&=&\frac{1}{1+(\pT^{\lele}^{\mathrm{X}}/0.01~\mathrm{GeV})^4},\quad w_\mathrm{pp}(\pT^{\lele}^{\mathrm{X}})=\frac{1}{1+(\pT^{\lele}^{\mathrm{X}}/0.7~\mathrm{GeV})^4}.
\label{eq:A8}
\end{aligned}
\end{equation}

Figures~\ref{fig:dimuonPTAco_pp13TeV} and~\ref{fig:dimuonPTAco_PbPb5TeV} show distributions for dimuon acoplanarity $A_{\phi}^{\mm}$ (left) and $\pT^{\mm}$ (right) in \pp\ UPCs at $\sqrts=13$ TeV and \PbPb\ UPCs at $\sqrtsnn=5.02$ TeV, respectively, in the fiducial phase space of Refs.~\cite{ATLAS:2017sfe,ATLAS:2020epq}, computed with \gammaUPC\ (ChFF photon flux) with LO$\,+\,\kt$-smearing (blue histograms) and NLO QED (red histograms) accuracy, as well as their corresponding combined distributions (filled grey histograms) following Eqs.~(\ref{eq:A7}) and (\ref{eq:A8}). The agreement with the acoplanarity distribution measured in \PbPb\ UPCs is good as shown in the bottom left panel of Fig.~\ref{fig:dimuonPTAco_PbPb5TeV}.

\end{document}